\documentclass[12pt]{article}
\usepackage{xcolor}
\usepackage{amsmath,amssymb,amsthm,bbm}
\usepackage[colorlinks,citecolor=blue,urlcolor=blue,linkcolor=blue]{hyperref}
\usepackage{fullpage}
\usepackage{natbib}
\usepackage[onehalfspacing]{setspace}
\usepackage{dsfont}
\usepackage{tikz}
\usepackage{pgfplots}
\AtBeginDocument{%
\setlength\abovedisplayskip{0.05in}
\setlength\belowdisplayskip{0.08in}}
\usepackage{sectsty}
\subsectionfont{\fontsize{13}{16}\selectfont}
\subsectionfont{\large}
\usepackage{fancyhdr}
\newtheorem{assumption}{Assumption}
\newtheorem{definition}{Definition}
\newtheorem{lemma}{Lemma}
\newtheorem{proposition}{Proposition}
\newtheorem{corollary}{Corollary}
\newtheorem{theorem}{Theorem}

\newtheorem{example}{Example}
\newtheorem*{examplecont}{Example 1 (continued)}
\newtheorem{obs}{Observation}
\newcommand{\Cov}{\operatorname{Cov}}
\newcommand{\Var}{\operatorname{Var}}
\newcommand{\E}{\mathbb{E}}
\newcommand{\R}{\mathbb{R}}

\newcommand{\m}[1]{\ensuremath{\boldsymbol {#1}}}

\title{\small\bf\MakeUppercase{Cognitive Constraints and Misspecification Concern}\thanks{This paper builds on and supersedes our previous paper, ``Complexity and Misspecification''. We thank Simone Cerreia-Vioglio, Piotr Dworczak, Giacomo Lanzani, Massimo Marinacci, Lasse Mononen, Pietro Ortoleva, Arthur Robson, Larry Samuelson, Jakub Steiner,  and Tomasz Strzalecki for helpful comments, and NSF grant SES-2417162 for financial support.}}
\author{Drew Fudenberg\thanks{Department of Economics, MIT, drew.fudenberg@gmail.com} 
\and Florian Mudekereza\thanks{Department of Economics, MIT, florianm@mit.edu}}
\date{\today}
\begin{document}

\maketitle
\thispagestyle{empty}

\begin{abstract}
We study repeated decision problems in which an agent has  cognitive constraints  and  is concerned that her models might be misspecified. She first uses data to weight candidate models, subject to a cognitive-capacity constraint on the  complexity she can process, and then she evaluates choices according to how well they perform under those models. More data need not overcome limited capacity: a statistically inferior but simpler model can remain relevant in the long run. When the constraint binds, evidence that favors models in proportion to their complexity can leave current behavior unchanged while still affecting later responses to data. This constraint eliminates cycles between risky and safe actions when simple models favor safety, and relaxing it can make risky play more attractive initially yet less frequent in the long run. It can also sustain misspecified model classes by suppressing choices that would reveal their errors.
\end{abstract}

\newpage
\setcounter{page}{1}

\section{Introduction}

Economic agents face two distinct problems when learning
from data. They may be unable to use the models that the
data favor because using those models requires more
cognitive resources than they have available. They may also doubt that any model they
entertain is a correct representation of the true data-generating process. An agent can therefore have
compelling evidence for a model without being able to rely
on it fully, and become more cautious without changing
which models she uses. We study how these two problems
interact when agents learn from the outcomes of their
own choices.

To fix ideas, consider a manager deciding whether to launch a product in a new market or maintain an established product line. She has two models of demand: a simple benchmark that extrapolates past sales and a more sophisticated model that accounts for differences across consumers. Each quarter, she observes sales and can assess how well each model predicted them. While the simple model is relatively easy to apply, using the sophisticated model to guide her strategy is cognitively demanding because it requires analyzing how different consumers will respond to each option. Thus, even if the accumulated evidence strongly favors the sophisticated model, limited cognitive capacity may prevent the manager from relying on it exclusively. Separately, if even the sophisticated model predicts sales poorly, she may worry that neither model adequately captures demand and become more cautious about launching. The first limitation determines how much weight she can place on each model; the second determines how much she trusts the predictions of those models. 

Our framework combines cognitive limitations with concern
about model misspecification. The agent repeatedly chooses among actions and observes their outcomes. She evaluates the statistical performance of a collection of candidate models using Bayesian updating, but faces a  constraint on the models she can use for decisions. Each model has a complexity cost, interpreted as the cognitive resources required to represent it and evaluate alternative actions under it. The expected complexity cost across candidate models must satisfy a cognitive \textit{capacity} constraint. Greater reliance on a demanding model consumes more cognitive resources, so the  agent chooses the distribution of model weights that comes closest to her Bayesian posterior among those she can implement. Adapting the terminology of \citet{samuelsonsteiner24},
we call this distribution of model weights the \emph{recognition distribution}.

The agent also recognizes that every candidate model may be misspecified.  Following \citet{hansen01}, she uses the multiplier criterion
to account for possible misspecification of each model. Building on \citet{lanzani25}, we let misspecification concern respond to both the statistical fit of the model class and the realized performance of the agent's actions. Our safety-adjusted likelihood-ratio rule provides a rationality benchmark for complexity-constrained agents: it guarantees approximately the maxmin payoff and, when sufficiently accurate predictions are cognitively feasible, ensures that limiting choices are approximately optimal under predictions that are approximately correct for the actions taken. 

Our first main result is that complexity filters evidence (Theorem \ref{thm:complexity_filters}). Consider again the manager with two models of demand.  If her cognitive-capacity constraint binds, she cannot increase her reliance on the sophisticated model beyond the level her resources permit. Additional evidence favoring that model therefore changes her Bayesian posterior without changing the weights used for choice. More generally, evidence that favors models in proportion to their complexity is absorbed by an increase in the shadow value of capacity rather than by a change in recognition. Only the component of evidence unrelated to complexity changes recognition, and its effect on choices depends on whether it favors models that rank the available actions differently. This result shows that stronger statistical evidence does not always induce a stronger behavioral response.

We also find that evidence that leaves current choices unchanged   can still affect how the agent responds to subsequent observations. When evidence favors sophisticated models while capacity binds, the shadow value of capacity rises. Later evidence against those models can initially reduce this shadow value without changing recognition. Only after sufficiently unfavorable evidence has accumulated does the agent begin to reduce her reliance on the sophisticated models. Proposition \ref{prop:recognition_invariance}  characterizes which changes in statistical evidence can leave recognition unchanged. This result suggests that two agents using the same models and same weights can respond differently to identical new evidence because their past observations provide different amounts of support for complex models. Moreover, behavior can change while recognition remains unchanged because the agent may switch actions if the new evidence changes her concern about misspecification. 

Limited capacity can continue to affect recognition in the long run.  Theorem \ref{thm:long_run_frontier} characterizes the separate limits of complexity-constrained recognition and misspecification concern when the empirical frequencies of actions converge. Concern about misspecification approaches a measure of the statistical error of the best candidate model, adjusted for the extent
to which the actions generating the data perform below the
maxmin payoff. Recognition, by contrast, approaches the set of  best-fitting distributions  that satisfy the capacity constraint. Consequently, the agent can continue to rely on a statistically inferior model even when unlimited data establish that a more sophisticated model predicts outcomes better: she learns which model fits the data, but cannot fully use it.

This distinction gives rise to a \emph{fit-complexity frontier}: the best statistical fit attainable with a given amount of capacity. Proposition \ref{prop:frontier}  shows that the shadow value of capacity measures the improvement in statistical fit made possible by additional resources. It also establishes that an efficient long-run recognition distribution can be supported on at most two models, one  that fits well  and one that is easier to use. 
Proposition \ref{prop:equilibrium}  incorporates this tradeoff into an equilibrium in which the actions generating the data are optimal given the resulting recognition distribution and concern about misspecification. Such an equilibrium exists, and joint convergence
of empirical action frequencies and recognition
selects such an equilibrium.

One implication is that correct statistical specification need not eliminate the behavioral effects of limited capacity. Suppose the uniquely best-fitting model correctly predicts the outcomes of the actions the agent continues to take, but requires more capacity than she has. If those
actions also perform at least as well as the maxmin payoff,
misspecification concern vanishes while the capacity constraint
continues to prevent exclusive reliance on that model.  Proposition \ref{thm:independent} shows that the best fit she can attain is then strictly worse than the fit of the best model, and that the long-run shadow price of capacity is strictly positive, even though that model predicts the outcomes of her chosen actions exactly. Corollary \ref{lem:eu_constrained} shows that her limiting
choices maximize expected utility under a recognition distribution that  places positive weight on models that fit the data strictly worse. 

We then investigate the implications for risky choice. Safe choices may produce outcomes that are relatively easy for the agent's models to explain, reducing her concern about misspecification and making risk-taking attractive. Risky choices may then generate outcomes that the models explain poorly, restoring caution and inducing a return to safety. The ensuing safe period reduces concern again, restarting the process. Such fluctuations can arise even when the environment itself is stationary. Theorem \ref{thm:complexity_cycles} characterizes the resulting long-run frequency of risky play in a two-model environment.

Limited capacity changes this feedback by restricting which models influence choices. Suppose the sophisticated model is statistically superior and favors risk when concern about misspecification is low, while the simple model favors safety even when model fit is relatively good. With sufficient capacity, the agent repeatedly returns to risk after periods of safe play. When capacity falls below a threshold, however, the sophisticated model receives too little weight to make risk attractive even after favorable safe-period observations. The cycle is eliminated, and the agent eventually chooses safety exclusively (Corollary \ref{cor:complexity_eliminates_cycles}). Whether limited capacity favors safety depends on which action the simpler model supports; here, the threshold identifies when constraints on model use prevent favorable safe-period evidence from restarting risky play.

Increasing capacity need not increase risk-taking over time. Section \ref{sec:capacity_risky_play}  constructs  a decision problem in which additional capacity makes the risky action more attractive before learning, yet strictly reduces its long-run frequency. Initially, additional capacity shifts recognition away from a strongly pessimistic simple model toward a more optimistic sophisticated model, encouraging risk-taking. As evidence accumulates, the pessimistic model becomes irrelevant, and the remaining simple and sophisticated models differ in their assessments of downside risk. Greater reliance on the sophisticated model then makes the agent more cautious. The reversal arises because learning changes which models determine the behavioral effect of capacity.

Finally, we extend our framework to study how limited capacity affects the survival of misspecified models when agents can compare and switch between competing model classes. As in \citet{choKasa2015}, \citet{schwartzsteinSunderam2021} and  \citet{ba26}, we suppose the agent evaluates model classes using their statistical performance;  we add the capacity constraint on decisions within whichever class she adopts.  We show  that sufficiently limited capacity can make a misspecified class persist with positive probability against every competing class, even when it would not persist under unconstrained Bayesian choice, and that it can  make a previously robust model class vulnerable to competition. The direction depends on whether the capacity constraint suppresses or encourages actions that reveal misspecification.

This application also identifies when the persistence of a misspecified model has welfare consequences and how it can be overcome. In environments involving risky and safe actions, the risky action is objectively superior, but sufficiently limited capacity can induce the agent to choose safety indefinitely and remain with a misspecified class. Corollary \ref{cor:ba_welfare} shows that, when the correct model class is available as a competitor, requiring risky experimentation at an increasingly sparse sequence of dates can restore eventual adoption of the correct class and optimal long-run payoffs without increasing cognitive capacity. The reason is that even a vanishing frequency of informative choices can generate decisive evidence, provided those choices occur infinitely often (Corollary \ref{cor:ba_information_rates}). A misspecified model class then persists because capacity constraints keep the agent from choosing the actions that would falsify it, not because she cannot process the resulting evidence.

\subsection{Related Work}

\paragraph{Variational inference and constrained data fitting.}
Our approach builds on variational Bayesian inference, which represents Bayesian updating as the solution to an optimization problem involving likelihoods and Shannon entropy \citep{zellner1988optimal}. Imposing a constraint on the feasible distributions provides a tractable way to model limitations on inference \citep{jordan99,blei2017variational}. In economics, \citet{samuelsonsteiner24} studies this constrained variational problem in environments with observable data and latent variables, showing how restrictions on feasible beliefs can capture cognitive limitations.\footnote{\citet{samuelson2025robust} builds on this approach to obtain an axiomatic characterization of a statistical procedure based on maximum-likelihood estimation and Bayes rule.} We combine  the variational approach with a capacity constraint based on the complexity of the  models used for decisions.

\paragraph{Cognitive resource constraints.}
Our complexity constraint follows the resource-bound approach of  \citet{liederGriffiths17} and \citet{oprea26}, in which procedures consume cognitive resources such as time, working memory, or attention, and the agent must operate within a fixed resource budget.  More broadly, \citet{liederGriffiths2020} surveys the resource-rational approach to cognition, which analyzes how limited computational resources constrain otherwise rational procedures. We apply this perspective to the use of statistical models, allowing the agent to allocate limited capacity across candidate models with different processing requirements. Experimental evidence that implementing decision rules and forming predictive models can be costly, and that agents favor simpler procedures when processing demands increase, supports model-specific complexity costs \citep{oprea2020, mandler24,
kendallOprea2024,banovetzOprea2023}.  \citet{halpern2015algorithmic} supposes that agents implement their strategies as Turing machines, and pay costs to use more complex machines.  Other  work supposes that complexity  
restricts the strategies agents can implement, for example by limiting them to finite automata \citep{abreuRubinstein88} or finite memory \citep{wilson14}. Those approaches constrain
strategies or memory; we constrain the average complexity of the models used to evaluate actions.

\paragraph{Complexity measures.}
Our expected complexity cost admits several interpretations. In statistical model averaging, the weighted average of model dimensions measures the effective number of parameters \citep{hansen07}. In the theory of technical languages, \citet{cremer07} measures expected diagnosis costs across representations that differ in the number of possibilities left unresolved. In sequential information processing, \citet{mandler24} evaluates question trees by their expected processing costs along realized paths. In rational metareasoning, \citet{liederGriffiths17} measures the cost of a cognitive strategy as the expected time required to execute it. Weighted averages of complexity measures also arise in mechanism design; for example, \citet{segal05} considers the ``average-case'' communication complexity as the expected number of bits transmitted in a protocol, also known as ``distributional communication complexity.'' These examples motivate representing the resource requirement of a collection of procedures by an average of their individual costs.

\paragraph{Misspecification concern.}
Our decision criterion for evaluating actions belongs to the average robust control (ARC) family of \citet{miss25}, which combines uncertainty over structured models with concern that those models may be misspecified. Within each structured model, the agent uses \citeauthor{hansen01}’s (\citeyear{hansen01}) multiplier criterion, axiomatized by \citet{strz11}, to protect against adverse departures from its predictions. \citet{lanzani25} further endogenizes misspecification concern in dynamic environments through the statistical fit of the model class and characterizes its implications for long-run behavior. We retain this treatment of misspecification and the ARC evaluation of actions, but restrict the model weights entering that evaluation according to the resources required to use the models. This restriction changes the appropriate adjustment of misspecification concern, because  statistical model fit no longer guarantees adequate performance. We therefore extend \citeauthor{lanzani25}'s (\citeyear{lanzani25}) likelihood-ratio rule to restore its normative properties in the presence of cognitive constraints.

\paragraph{Rational inattention.}
Our cognitive-capacity constraint differs from those in the rational-inattention literature \citep{Sims2003,matejkaMcKay2015,caplinDean2015,mackowiak2023rational}. In that literature, agents optimally acquire or process information about uncertain states, often subject to information costs or limits on the informativeness of signals. Here, the data are observed without cost, and the capacity constraint applies instead to the complexity of the models used to evaluate actions. \citet{ortoleva25} finds experimental evidence that people are averse to complexity even after information is gathered and argues that models of inattention fail to capture some key aspects of complexity.

\section{Static Framework}
\label{sec:static}
The agent's decision has two steps.  She first maps observed data into a distribution over candidate models, subject to a constraint on the complexity of the models she can use.  She then evaluates actions under these models with a robust-control criterion.

\subsection{Primitives and data}

Let $A$ be a finite set of actions and $Y$ a finite set of outcomes.  For each
$a\in A$ and $y\in Y$, the agent's per-period utility is $u(a,y)\in\R$.
A \emph{structured model} specifies action-dependent outcome distributions
$\m q:=(q_a)_{a\in A}\in\Delta(Y)^A$.\footnote{Throughout, we use bold font  for distributions in $\Delta(Y)^A$ to distinguish them from those in $\Delta(Y)$.}  The agent entertains a finite \emph{model class}
$Q\subset\Delta(Y)^A$, and begins with a full-support prior
$\pi_0\in\Delta(Q)$. The true model is denoted
$\m p^\star=(p^\star_a)_{a\in A}\in\Delta(Y)^A$; it need
not belong to $Q$.

\begin{assumption}[Full support]\label{ass:full_support}
For each $\m q\in Q$, $a\in A$, and $y\in Y$, $q_a(y)>0$.
\end{assumption}

Fix a finite sample
$\mathcal D_n=((a_i,y_i))_{i=1}^n\in(A\times Y)^n$.  An observation
$(a_i,y_i)$ records the outcome $y_i$ generated following action $a_i$.
For now the sample is taken as given in order to define the static map from data to recognition and choice.  In Section \ref{sec:dynamic}, the observations will be generated endogenously from the agent's sequential choices.

\subsection{Step 1: model complexity}
\label{subsec:constrained_fitter}

The first step translates observed data into weights on the structured models in $Q$, subject to the agent's capacity to process them.

\subsubsection{Constrained data fitting}
For $\m q\in Q$, let $L_n(\m q):=\prod_{i=1}^n q_{a_i}(y_i)$ be the likelihood of $\mathcal D_n$, and let $\pi_n(\m q)\propto\pi_0(\m q)L_n(\m q)$ be the Bayesian posterior. The posterior is our benchmark for how the sample ranks the structured models absent any cognitive limitations.
Bayesian updating has a standard variational characterization; see \citet{zellner1988optimal} and \citet{blei2017variational}. To state it formally, let $H(\nu):=-\sum_{x}\nu(x)\log\nu(x)$ denote Shannon entropy, and let $R(\nu\Vert\varphi):=\sum_{x}\nu(x)\log[\nu(x)/\varphi(x)]$ denote relative entropy when $\nu\ll\varphi$, with $R(\nu\Vert\varphi)=+\infty$ otherwise. We have
\begin{equation}\label{eq:bayes-variational}
\sum_{\m q\in Q}\eta(\m q)\log\big(\pi_0(\m q)L_n(\m q)\big)+H(\eta)
=
\log\Big(\sum_{\m q\in Q}\pi_0(\m q)L_n(\m q)\Big)
-R(\eta\Vert\pi_n).
\end{equation}
Since relative entropy is nonnegative and vanishes only when its arguments coincide, $\pi_n$ uniquely maximizes this objective over $\Delta(Q)$. Equivalently, it maximizes $\sum_{\m q\in Q}\eta(\m q)\log L_n(\m q)-R(\eta\Vert\pi_0)$, trading off expected log likelihood against divergence from the prior.

We model cognitive constraints by assuming that the agent maximizes the objective in \eqref{eq:bayes-variational} over a nonempty feasible set $\mathcal V\subseteq\Delta(Q)$. That is, the agent selects a solution to
\begin{equation}\label{eq:const}
\max_{\eta\in\mathcal V}
\left\{\sum_{\m q\in Q}\eta(\m q)\log\big(\pi_0(\m q)L_n(\m q)\big)+H(\eta)\right\}.
\end{equation} We call the solution to \eqref{eq:const} the \emph{recognition distribution}, adapting the recognition
concept of \citet{samuelsonsteiner24}, which provides a
microfoundation for constrained variational updating.
In our setting, recognition describes the distribution
of model weights that the agent can implement for choice.

\subsubsection{Complexity capacity}

We focus on model complexity as the cognitive limitation on the models that the agent can effectively use.  Each structured model $\m q\in Q$ has a procedural complexity $k(\m q)\in\R_+$, interpreted as the cognitive resources required to rely on $\m q$ when evaluating actions.

\par Under a recognition distribution $\eta\in\Delta(Q)$, the expected complexity of models used for choice is 
$
\E_\eta[k]
=
\sum_{\m q\in Q}\eta(\m q)k(\m q)$; weighting $k(\m q)$ by $\eta(\m q)$ captures the idea that cognitive resources scale with the extent to which the agent uses each model $\m q$.  Section \ref{subsec:complexity_examples} illustrates some examples of complexity measures that can be represented by $\E_\eta[k]$.  Now, let $K$ denote the available \textit{complexity capacity}. The  agent can  use only model weights restricted to the set 
\begin{equation}\label{eq:complexity_set}
\mathcal E_K
:=
\Big\{
\eta\in\Delta(Q):
\E_\eta[k]\le K
\Big\}.
\end{equation}
Unless otherwise stated, we assume
$\min_{\m q\in Q}k(\m q)<K<\max_{\m q\in Q}k(\m q)$.
This restricts attention to environments in which
capacity permits some but not all structured models
to receive full recognition. The constraint in \eqref{eq:complexity_set} is a hard resource bound, as in \citeauthor{oprea26}'s (\citeyear[][Problem (2)]{oprea26}) general model of complexity, where  feasible procedures must respect an upper bound on a cognitive resource such as time, working memory, or attention.  Applying this capacity restriction to the data-fitting problem in \eqref{eq:const} yields the following problem:
\begin{equation}\label{eq:recognition_problem}
\max_{\eta\in\mathcal E_K}
\left\{
\sum_{\m q\in Q}
\eta(\m q)
\log\!\big(\pi_0(\m q)L_n(\m q)\big)
+
H(\eta)
\right\}.\end{equation}
 We call this  maximization \emph{complexity-constrained data fitting}. The prior and likelihood determine the Bayesian posterior $\pi_n$, while the capacity set $\mathcal E_K$ restricts which model-weight distributions the agent can use for choice.  When $\pi_n$ is feasible, the agent implements the Bayesian benchmark; otherwise, the recognition distribution is the feasible distribution closest to $\pi_n$ in relative entropy.  The next result formalizes these insights. 
\begin{lemma}[Constrained posterior]
\label{prop:constrained_posterior}
The unique solution to  \eqref{eq:recognition_problem} is 
$\eta_n^K
=
\arg\min_{\eta\in\mathcal E_K}R(\eta\Vert\pi_n).$  If $\E_{\pi_n}[k]\le K$, then $\eta_n^K=\pi_n$.  If $\E_{\pi_n}[k]> K$, then there is a
unique $\mu_n>0$ such that
\[
\eta_n^K(\m q)
=
\frac{\pi_n(\m q)e^{-\mu_nk(\m q)}}
{\sum_{\m r\in Q}\pi_n(\m r)e^{-\mu_nk(\m r)}},
\quad\text{ and }\quad
\E_{\eta_n^K}[k]=K.
\]
\end{lemma}
Lemma \ref{prop:constrained_posterior} says that complexity is irrelevant when the exact Bayesian posterior is feasible. Otherwise, the exponential tilt reduces
the relative weight on more complex models. The multiplier $\mu_n$, which we set to zero whenever
$\E_{\pi_n}[k]\le K$, is the shadow value of capacity
and determines the size of this adjustment.
We refer to this multiplier as \emph{complexity pressure}.

\begin{example}[Running example]\label{ex:two_model}\normalfont
Consider an environment in which there are two structured models $Q=\{\m q^S,\m q^C\}$ with $k(\m q^S)=0$ and $k(\m q^C)=1$. 
For any $K\in(0,1)$, limited capacity requires
$\eta(\m q^C)\le K$, so  Lemma \ref{prop:constrained_posterior} gives
\[
\eta_n^K(\m q^C)
=
\min\big\{\pi_n(\m q^C),K\big\},
\qquad
\mu_n
=
\max\Big\{
0,\
\log\frac{\pi_n(\m q^C)}{\pi_n(\m q^S)}
-
\log\frac{K}{1-K}
\Big\}.
\]
Once the posterior weight on the complex model exceeds capacity,
additional evidence in its favor no longer changes the recognition weight;
it instead raises only the shadow value $\mu_n$ of additional capacity. To see this, suppose $K=0.4$.
For posterior weights $\pi_n(\m q^C)=0.3,0.6,0.9$, the corresponding recognition weights are $0.3,0.4,0.4$, while
$\mu_n=0,\log(2.25)\simeq0.81,\log(13.5)\simeq2.60$. Recognition is therefore constant for all $\pi_n(\m q^C)\ge0.4$, and the evidence distinguishing these posteriors is reflected only in $\mu_n$.\hfill$\bigtriangledown$
\end{example}

\subsubsection{Comparative statics}
Fix a full-support distribution $\pi\in\Delta(Q)$. For each capacity $K$, let $\eta^K$ and $\mu^K$ denote the associated recognition distribution and multiplier from Lemma \ref{prop:constrained_posterior}. The complexity constraint \emph{binds} if $\E_\pi[k]>K$; by Lemma \ref{prop:constrained_posterior}, this holds if and only if $\mu^K>0$, in which case $\E_{\eta^K}[k]=K$.

\begin{proposition}[The value of complexity capacity]
\label{prop:capacity}
For every $K<\E_\pi[k]$,
\[
\frac{d\mu^K}{dK}
=
-\frac{1}{\Var_{\eta^K}(k)},
\qquad
\frac{d}{dK}\log\eta^K(\m q)
=
\frac{k(\m q)-K}{\Var_{\eta^K}(k)},
\]
and, for every function $w:Q\to\R$,
\[
\frac{d}{dK}\E_{\eta^K}[w]
=
\frac{\Cov_{\eta^K}(k,w)}
{\Var_{\eta^K}(k)}.
\]
\end{proposition}
Additional capacity lowers its shadow value and shifts
recognition weight toward models whose complexity exceeds
the current average $K$. Its effect on any model-dependent
quantity is determined by that quantity's covariance
with complexity.

\begin{examplecont}\normalfont
In the two-model running example, whenever the complexity capacity binds, $\eta^K(\m q^C)=K$ and $\eta^K(\m q^S)=1-K$.
Thus, an additional unit of capacity transfers recognition weight one-for-one from the simple model to the complex model.  For any statistic $w:Q\to\R$, the last identity in
Proposition \ref{prop:capacity} becomes
$\frac{d}{dK}\E_{\eta^K}[w]=w(\m q^C)-w(\m q^S)$. Greater capacity therefore has no intrinsic tendency to make the agent more optimistic, more pessimistic, or more willing to take risk: its direction is determined by what the complex model predicts relative to the simple model.\hfill$\bigtriangledown$
\end{examplecont}

\subsubsection{Complexity measures}
\label{subsec:complexity_examples}
The function $k$ can represent different resource costs.
We give four interpretations.\footnote{Another interpretation is  the cost of ``forming mental models'' as in \citet{kendallOprea2024}.} 

\paragraph{Effective model dimension.}
Consider the least-squares model-averaging environment of \citet{hansen07}.  Let
$Q=\{\m q_1,\ldots,\m q_M\}$ index the approximating regression models, where model
$\m q_m$ uses $k_m$ regressors, and let $P_m$ denote its least-squares projection matrix.
Identifying our recognition weight $\eta(\m q_m)$ with the model-averaging weight
$w_m$, set
$k(\m q_m):=k_m.$
The projection matrix associated with the model average is
$P(\eta)=\sum_{m=1}^M\eta(\m q_m)P_m$.  Since the trace of $P_m$ satisfies $\operatorname{tr}(P_m)=k_m$, \citet[][Lemma 1(i)]{hansen07} shows that the effective number of parameters becomes
$
\operatorname{tr}(P(\eta))
=
\sum_{m=1}^M\eta(\m q_m)k_m
=
\E_\eta[k]$, which is used as  the linear  penalty term   $r\E_\eta[k]$ in \citeauthor{hansen07}'s (\citeyear[][eq. (11)]{hansen07}) criterion, for some $r>0$.  Thus, one specification of our complexity
function is the dimension of a statistical representation: $k(\m q)$ is the number of
coefficients that must be represented and estimated under model $\m q$, while
$\E_\eta[k]$ is the effective dimension of the collection of models actually used.
Under this interpretation, our capacity constraint \eqref{eq:complexity_set} places a hard bound on the
effective number of model parameters that the agent can process.
\paragraph{Coarse representations.}
\citet{cremer07} study technical languages in which a code partitions a finite set of events into words $C=\{W_1,\ldots,W_J\}$. Word $j$ has breadth $n_j:=|W_j|$ and is used with probability $p_j$. On receiving it, the receiver pays a diagnosis cost $d(n_j)$, with $d$ increasing, because a broader word leaves more events to be distinguished. Identify word $j$ with an internal representation $\m q_j$, set $\eta(\m q_j):=p_j$ and $k(\m q_j):=d(n_j)$. The expected diagnosis cost of $C$, denoted $D(C)$, is
$
D(C)=\sum_{j=1}^J p_j\,d(n_j)=\sum_{j=1}^J \eta(\m q_j)\,k(\m q_j)=\E_\eta[k]$ 
\citep[eq. (1)]{cremer07}. Under the normalization $d(n)=n$, the complexity of a representation is the number of possibilities it leaves unresolved, that is, the downstream effort needed to make it precise. The capacity constraint can therefore be read as a bound on expected interpretive effort.

\paragraph{Information acquisition.}
Another interpretation comes from the question trees of \citet{mandler24}.  An agent has a decision partition and acquires enough information to identify
its relevant cell by proceeding through a tree of questions, denoted $T$.  At every nonterminal
node $v$, the current question has $k_v$ possible answers and incurs a processing
cost $c_T(k_v)$.  Thus, if the realized state leads to leaf $l$, the total processing
cost is the sum of the costs of all questions encountered along the corresponding
path $p(l)$,
$C_T(l)
:=
\sum_{v\in p(l)}c_T(k_v).$
\citet{mandler24} evaluates a tree by its expected path cost: when the $n$
cells of the decision partition are equally likely, the  expectation criterion is
$\frac{1}{n}\sum_l C_T(l)$; with nonuniform probabilities the same criterion weights
the path costs by the probabilities of the corresponding leaves. 
\par To obtain a complexity measure for our framework, fix
a revealing recognition tree
$T$ whose leaves identify the structured models in $Q$, so that
each $\m q\in Q$ is associated with a leaf $l(\m q)$.  Define
$k(\m q)
:=
C_T(l(\m q))
=
\sum_{v\in p(l(\m q))}c_T(k_v).$
Weighting the model-specific path costs by $\eta$ gives  
\[
\sum_{\m q\in Q}\eta(\m q)
\sum_{v\in p(l(\m q))}c_T(k_v)=\sum_{\m q\in Q}\eta(\m q)
C_T(l(\m q))=\E_\eta[k].
\]
Under this interpretation, $k(\m q)$ is the amount of cognitive resource required to
process model $\m q$.  \citeauthor{mandler24}'s (\citeyear{mandler24}) specifications give
simple candidates for the node cost $c_T$.  If processing a question requires
considering all of its possible answers, $c_T(k_v)=k_v$; if the final answer can be
deduced once the other $k_v-1$ answers have been rejected, $c_T(k_v)=k_v-1$.
This environment suggests that the capacity constraint can  be interpreted as
an upper bound on the expected amount of sequential processing that the agent can
devote to model recognition.

\paragraph{Cognitive strategies.}
In the theory of rational metareasoning,
\citet{liederGriffiths17} study how agents select among cognitive strategies with different accuracy and execution times. There, the value of using a cognitive strategy is its expected payoff net of a cost proportional to its expected execution time. Suppose that each structured model $\m q\in Q$ is associated with a cognitive strategy $m_{\m q}$ for evaluating actions, and let $T$ denote the time required to execute the selected strategy.
Set $k(\m q):=\E[T|m=m_{\m q}]$, where $m$
denotes the strategy used, and identify
$\eta(\m q)$ with its utilization weight.
By the law of iterated expectations, the expected execution time is
$\E[T]=\sum_{\m q\in Q}\eta(\m q)k(\m q)=\E_\eta[k]$. In this context, the capacity constraint places a hard bound on the expected amount of time the agent can devote to evaluating actions using alternative models.
\subsection{Step 2: model misspecification}
\label{subsec:robust_choice}
The agent also allows for misspecification of the structured
models. We evaluate actions using the multiplier criterion
of \citet{hansen01}, axiomatized by \citet{strz11}.
For $a\in A$ and $\m q\in Q$, the multiplier criterion is 
\[
v_\lambda(a;\m q)
:=
\min_{p\in\Delta(Y)}
\left\{
\E_p[u(a,y)]
+
\frac{1}{\lambda}R(p\Vert q_a)
\right\}
=
-\frac{1}{\lambda}
\log\Big(\sum_{y\in Y}q_a(y)e^{-\lambda u(a,y)}\Big)
\]
for $\lambda >0$. We define $v_0(a;\m q):=\operatorname{lim}_{\lambda\downarrow0} v_{\lambda}(a;\m q)=\E_{q_a}[u(a,y)]$; this is  the subjective expected-utility (SEU) benchmark. The unique minimizing distribution is
$\widehat p_\lambda(a;\m q)(y)\propto
q_a(y)e^{-\lambda u(a,y)}$, and the parameter $\lambda$ determines how
far the agent is willing to depart from each model when evaluating an action.   Relative to $q_a$, the minimizer $\widehat p_\lambda(a;\m q)$ shifts probability toward outcomes with lower utility.  If $u(a,\cdot)$ is nonconstant, $v_\lambda(a;\m q)$ is strictly decreasing in $\lambda$, and, by Assumption \ref{ass:full_support}, $v_\lambda(a;\m q)\to\min_{y}u(a,y)$ as $\lambda\to\infty$; this is \citeauthor{gilboa89}'s (\citeyear{gilboa89}) maxmin criterion where the set of candidate models is the entire simplex $\Delta(Y)$.  If $u(a,\cdot)$ is constant, $v_\lambda(a;\m q)$ does not depend on $\lambda$.  An increase in $\lambda$ therefore lowers the value of actions with nonconstant payoffs relative to actions with constant payoffs.
\par For any recognition distribution
$\eta\in\Delta(Q)$, define the average robust value and
best-reply correspondence by
\begin{equation*}
V_\lambda(a;\eta)
:=
\sum_{\m{q}\in Q}\eta(\m{q})v_\lambda(a;\m{q}),
\qquad
\operatorname{BR}_\lambda(\eta)
:=
\arg\max_{a\in A}V_\lambda(a;\eta).
\end{equation*}

Given the data $\mathcal D_n$, the agent evaluates action $a$ by $V_\lambda(a;\eta_n^K)$ (Lemma \ref{prop:constrained_posterior}).  Complexity affects choice through the recognition
distribution $\eta_n^K$, while misspecification concern $\lambda$ affects the value assigned to each action under each model weighted by $\eta_n^K$. The average robust value was introduced by \citet{miss25} and axiomatized by \citet{lanzani25}.

\subsubsection{Effects of complexity on choices}

This section studies how evidence affects recognition and action
values, holding misspecification concern $\lambda$ fixed.
Fix a full-support distribution $\pi\in\Delta(Q)$ and an
\textit{evidence score} $g:Q\to\R$, and let
$\pi^s(\m q)\propto\pi(\m q)e^{sg(\m q)}$ for $s\in\R$.
The score $g$ specifies a direction of log-likelihood
change, such as the cumulative likelihood increment in Section \ref{sec:finite}, and the constant $s$ scales its magnitude:
$\pi^s$ is the posterior obtained from $\pi$ after observations that shift the log likelihood of each $\m q$ by $sg(\m q)$.
Let $(\eta^s,\mu^s)$ denote the recognition distribution and multiplier associated with $\pi^s$. Let $I\subset\R$ denote a \emph{binding region} for the family $(\pi^s)_{s\in\R}$, meaning an open interval on which  $K<\E_{\pi^s}[k]$ for all $s\in I$.

Define
\[
\beta^s
:=
\frac{\Cov_{\eta^s}(k,g)}{\Var_{\eta^s}(k)},
\qquad
g^s(\m q)
:=
g(\m q)-\E_{\eta^s}[g]
-\beta^s\big(k(\m q)-K\big).
\]
These  objects separate statistical evidence into a component associated with model complexity and a component orthogonal to complexity. Since
$\E_{\eta^s}[k]=K$, the pair
$\left(
\E_{\eta^s}[g]-\beta^sK,\,
\beta^s
\right)$
solves the weighted least-squares projection of $g$ on a constant and $k$, so $\beta^s$ is the regression coefficient of the evidence score on model complexity under the agent's current recognition distribution.  It measures how much more favorable the incoming evidence is, on average, to models that require one additional unit of complexity.  A positive $\beta^s$ means that the evidence systematically favors more complex models; a negative $\beta^s$ means that it systematically favors simpler
models. The function $g^s$ is the corresponding regression residual. It is the component of evidence not explained
by a linear relationship with complexity.

\begin{theorem}[Complexity filters evidence]
\label{thm:complexity_filters}
For every $s\in I$,
\[
\frac{d\mu^s}{ds}=\beta^s,
\qquad
\frac{d}{ds}\log\eta^s(\m q)=g^s(\m q),
\]
and, for every $a,b\in A$,
\[
\frac{d}{ds}
\left[
V_\lambda(a;\eta^s)-V_\lambda(b;\eta^s)
\right]
=
\Cov_{\eta^s}
\big(
v_\lambda(a;\cdot)-v_\lambda(b;\cdot),
g^s
\big).
\]
\end{theorem}
The first equality shows that evidence proportional to
complexity changes the multiplier, while the second shows
that only the residual changes recognition. Thus, evidence
can change the Bayesian posterior without changing the
weights used for choice. The third equality links changes
in recognition to action values: the relative value of
two actions changes according to the covariance of the
residual with their model-specific robust-value difference.

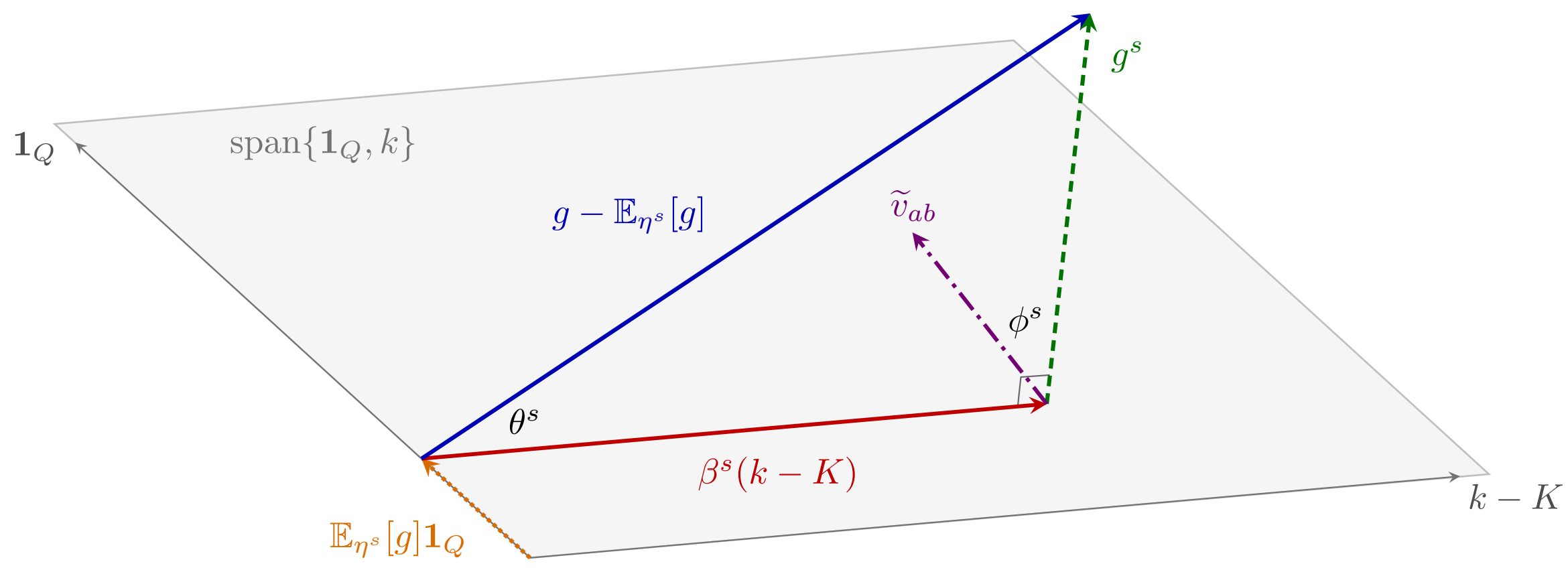
\begin{figure}[t]
\centering
\begin{tikzpicture}[scale=1.02,>=stealth]

\filldraw[
fill=black!4,
draw=black!25,
line width=0.5pt
]
(0,0) -- (9.28,0.81) -- (4.68,5.01) -- (-4.60,4.20) -- cycle;

\node[black!55, font=\footnotesize] at (-2.0,4)
{$\operatorname{span}\{\mathbf 1_Q,k\}$};

\coordinate (O) at (0,0);
\coordinate (M) at (-1.05,0.96);
\coordinate (B) at (5.00,1.49); 
\coordinate (G) at (5.42,5.27); 
\coordinate (H) at (3.70,3.15); 

\draw[->,black!55,thin]
(O) -- (-4.40,4.02)
node[above left,black!70,font=\footnotesize,inner sep=5pt] at (-4.40,3.62) {$\mathbf 1_Q$};

\draw[->,black!55,thin]
(O) -- (9.00,0.79)
node[below right,black!70,font=\footnotesize,inner sep=2pt] {$k-K$};

\draw[black!60,thin]
(4.72,1.47) -- (4.75,1.75) -- (5.03,1.77);

\draw[
->,
orange!85!black,
densely dotted,
line width=1.15pt
]
(O) -- (M);
\node[
below left,
orange!85!black,
font=\footnotesize,
inner sep=1.5pt
] at (-0.55,0.42)
{$\E_{\eta^s}[g]\mathbf 1_Q$};

\draw[
->,
red!75!black,
solid,
line width=1.1pt
]
(M) -- (B);
\node[
below,
red!75!black,
font=\footnotesize
] at (2.4,1.12)
{$\beta^s(k-K)$};

\draw[
->,
blue!70!black,
solid,
line width=1.15pt
]
(M) -- (G);
\node[
above left,
blue!70!black,
font=\footnotesize
] at (1.85,3.)
{$g-\E_{\eta^s}[g]$};

\draw[
->,
green!45!black,
densely dashed,
line width=1.15pt
]
(B) -- (G);
\node[
right,
green!45!black,
font=\footnotesize
] at (5.48,4.85)
{$g^s$};

\draw[
->,
violet!90!black,
dash pattern=on 6pt off 2pt on 1.2pt off 2pt,
line width=1.1pt
]
(B) -- (H);
\node[
above left,
violet!90!black,
font=\footnotesize,
inner sep=1.5pt
] at (4.,3.2)
{$\widetilde v_{ab}$};

\node[font=\footnotesize] at (-0.05,1.32) {$\theta^s$};
\node[font=\footnotesize] at (4.80,2.28) {$\phi^s$};

\end{tikzpicture}

\caption{The geometry of evidence filtering in Theorem \ref{thm:complexity_filters}.}
\label{fig:evidence_geometry}
\end{figure}

\paragraph{Geometric interpretation.} Theorem \ref{thm:complexity_filters} has a geometric interpretation: since the capacity constraint binds, the evidence score $g$, which records how strongly the new data favor each model, decomposes as in a regression into three parts:
an irrelevant  constant, a part proportional to complexity, which is 
absorbed by $\mu$,  and a residual, which is the only part that moves recognition.

Here, $\operatorname{span}\{\mathbf 1_Q,k\}$ is the set of evidence patterns in which models differ only through their complexity. To make this precise, equip functions on $Q$ with the $\eta^s$-weighted inner product $\langle f,h\rangle:=\E_{\eta^s}[fh]$ and induced norm $\lVert f\rVert:=\sqrt{\langle f,f\rangle}$, so that orthogonality to complexity means zero correlation with complexity under the agent's current weights. Since $\E_{\eta^s}[k]=K$, removing the common level gives the orthogonal decomposition
$g-\E_{\eta^s}[g]
=
\beta^s(k-K)+g^s$.
Figure \ref{fig:evidence_geometry} illustrates this decomposition and its implication for choice.

Let $\theta^s$ denote the angle between the centered evidence score $g-\E_{\eta^s}[g]$ and the complexity direction $k-K$, and let $\phi^s$ denote the angle between the residual $g^s$ and the centered relative robust-value vector
$\widetilde v_{ab}:=v_\lambda(a;\cdot)-v_\lambda(b;\cdot)
-\E_{\eta^s}[v_\lambda(a;\cdot)-v_\lambda(b;\cdot)]$.
Whenever the angles are well defined, the three identities in Theorem \ref{thm:complexity_filters} take the geometric form
$\beta^s
=
\frac{\lVert g-\E_{\eta^s}[g]\rVert}
 {\lVert k-K\rVert}
\cos\theta^s,$ 
$\lVert g^s\rVert
=
\lVert g-\E_{\eta^s}[g]\rVert
\sin\theta^s,$  and 
$\frac{d}{ds}
\big[V_\lambda(a;\eta^s)-V_\lambda(b;\eta^s)\big]
=
\lVert \widetilde v_{ab}\rVert
\lVert g^s\rVert
\cos\phi^s.$

The angle $\theta^s$ measures how closely the evidence resembles a complexity tilt. When it is close to zero or $\pi$, the evidence is mostly absorbed by $\mu^s$ and the weights barely move; when it is near a right angle, the evidence mostly reallocates weight across models. The relative value of $a$ increases when $\cos\phi^s>0$,
decreases when $\cos\phi^s<0$, and has zero derivative
when $\cos\phi^s=0$.
Thus, even holding fixed both the complexity projection $\beta^s(k-K)$ and the residual magnitude $\lVert g^s\rVert$, the same evidence can have opposite choice effects depending on which models receive the residual support.

\paragraph{Comparative statics.} An implication of Theorem \ref{thm:complexity_filters}
concerns the effect of evidence generated by an action
on complexity pressure. For $a\in A$, let
$D_a(\m q):=R(p^\star_a\Vert q_a)$.
The \emph{expected-evidence family} generated by action
$a$ is the family
$\pi^s(\m q)\propto\pi(\m q)e^{sg(\m q)}$
with $g(\m q)=\E_{p^\star_a}[\log q_a(y)]
=-H(p^\star_a)-D_a(\m q)$.
Thus, $s$ scales the expected log-likelihood
contribution of observations under action $a$.

\begin{corollary}[Complexity pressure]
\label{cor:evidence_complexity}
Let $I$ be a binding region for the expected-evidence
family generated by action $a$. Then, for every $s\in I$,
\[
\frac{d\mu^s}{ds}
=
-\frac{\Cov_{\eta^s}(k,D_a)}
{\Var_{\eta^s}(k)}.
\]
\end{corollary}

When evidence is measured by its expected log likelihood
under action $a$, complexity pressure rises if
$\Cov_{\eta^s}(k,D_a)<0$ and falls if this covariance
is positive.  Pressure depends on relative fit: poor fit need not raise
complexity pressure if simple and complex models perform
equally poorly, while evidence favoring complex models
can raise pressure even when the class fits well.
\section{Dynamic Framework}
\label{sec:dynamic}
We now let the data be generated endogenously by the agent's own actions.  The recognition distribution and misspecification concern are computed from the same history but depend on different features of it: the former on relative likelihoods within $Q$, the latter on the fit of $Q$ relative to an unrestricted class and the
realized performance of the agent's actions.  The static choice criterion applies at each history
with endogenous misspecification concern $\lambda_t$.

\subsection{Dynamic environment}
 At date
$t+1$, the agent chooses $a_{t+1}$ and then observes an outcome $y_{t+1}$ whose
conditional distribution, given the history up to date $t$ and the chosen action
$a_{t+1}$, is $p^\star_{a_{t+1}}$. For $t\ge0$, let $H_t:=(A\times Y)^t$ denote the
set of histories of length $t$, where
$h_t=((a_1,y_1),\ldots,(a_t,y_t))\in H_t$,
with $h_0=\varnothing$. The sample introduced in Section
\ref{sec:static} is now explicitly
$\mathcal D_t(h_t)=((a_i,y_i))_{i=1}^t$.
By Lemma \ref{prop:constrained_posterior}, the recognition distribution at $h_t$ is
\begin{equation}\label{eq:dynamic_recognition}
\eta_t
=
\arg\min_{\eta\in\mathcal E_K}R(\eta\Vert\pi_t),
\end{equation}
where we suppress the dependence on $K$. The likelihood and Bayesian posterior are   $L_t(\m q)=\prod_{i=1}^tq_{a_i}(y_i)$ and $\pi_t(\m q)\propto\pi_0(\m q)L_t(\m q)$, respectively. 

\paragraph{Misspecification concern.}

Following \citet{lanzani25}, we assess the statistical fit
of $Q$ by comparing its maximized likelihood with that of
the unrestricted model class $\Delta(Y)^A$:
\[
\operatorname{LLR}(h_t;Q)
:=
-\log
\frac{\max_{\m q\in Q}L_t(\m q)}
{\max_{\m p\in\Delta(Y)^A}\prod_{s=1}^tp_{a_s}(y_s)},
\]
for all $t$ and all $h_t\in H_t$. This log-likelihood ratio (LLR) is nonnegative and is large
when the structured models fit the data poorly. In our setting, though, limited capacity can prevent the agent from
relying on a model that explains the data well, so  good statistical fit need not ensure good performance.  

Let $\underline u := \max_{a\in A}\min_{y\in Y} u(a,y)$ denote the maxmin payoff,
and let $B := \max_{a,y} u(a,y) - \min_{a,y} u(a,y)$ denote the payoff range.
We assume that the agent's misspecification concern $\lambda_t$ evolves according to the
\emph{safety-adjusted LLR} (S-LLR) rule:
\begin{equation}\label{eq:dynamic_rule}
  \lambda_0 := 0, \qquad
  \lambda_t := \frac{\operatorname{LLR}(h_t;Q) + \psi_t}{ct} \quad \text{for } t \ge 1,
\end{equation}
where $c>0$ is a sensitivity parameter and $\psi_t$ is the cumulative
\emph{safety adjustment}. For $t \ge 1$ and $B>0$,
\[
  \psi_t := \frac{1}{B} \sum_{a\in A}
  \Big( \sum_{i=1}^{t} \mathbbm{1}\{a_i = a\}\big[\underline u - u(a_i,y_i)\big] \Big)_+ ,
\]
with $\psi_t := 0$ if $B=0$ or $t=0$.\footnote{We write $(x)_+ := \max\{x,0\}$.}
The adjustment increases misspecification concern when
the average realized payoff of an action falls below the
maxmin payoff, with greater weight on more frequently
chosen actions. S-LLR coincides with LLR  whenever
every observed action has delivered an average payoff
of at least $\underline u$; this holds at every history,
for instance, when all actions have the same minimum
payoff, even if their outcome distributions differ
and are misspecified.
Appendix \ref{app:safety} shows that, although LLR need not
guarantee approximately the maxmin payoff under limited
capacity, S-LLR does so for sufficiently small $c$.
For every $c>0$, sufficiently accurate feasible predictions
ensure that all sufficiently late choices are approximately
optimal under feasible predictions that approximately match
their outcomes (Theorem \ref{prop:safety_normative}).

\paragraph{Decision criterion.}

The agent's date-$t$ criterion and choice are
\begin{equation}\label{eq:dynamic_criterion}
V_{\lambda_t}(a;\eta_t)=\sum_{\m{q}\in Q}\eta_t(\m{q})v_{\lambda_t}(a;\m{q}),
\qquad
a_{t+1}\in\operatorname{BR}_{\lambda_t}(\eta_t)=\arg\max_{a\in A}V_{\lambda_t}(a;\eta_t),
\end{equation}
with $v_0(a;\m q)=\E_{q_a}[u(a,y)]$ as in Section \ref{subsec:robust_choice}.  The agent is myopic: she does not value the information generated by her choices.  A \emph{policy} is a selection from the correspondence in \eqref{eq:dynamic_criterion}; unless stated otherwise, results hold for every policy.
Let $\mu_t\ge0$ denote the complexity pressure
associated with history $h_t$, with $\mu_t=0$
whenever the capacity constraint is slack.

\subsection{Finite-sample dynamics}\label{sec:finite}
Theorem \ref{thm:complexity_filters} concerns marginal changes in evidence.  We now
consider the evidence that arrives between two dates $t<T$ and ask which evidence
leaves the recognition distribution unchanged. Such evidence can still affect behavior, because it changes how the agent responds to later observations and how cautiously she evaluates current actions. 

Fix $1\le t<T$ and define the cumulative log-likelihood
increment
\[
g_{t,T}(\m q)
:=
\log\frac{L_T(\m q)}{L_t(\m q)}
=
\sum_{i=t+1}^T\log q_{a_i}(y_i).
\]
Write $\mathbf 1_Q$ for the constant-one function on $Q$.  Since
$\pi_T(\m q)\propto\pi_t(\m q)e^{g_{t,T}(\m q)}$, the function $g_{t,T}$
contains precisely the new cross-model likelihood evidence between the two histories.

\begin{proposition}[Recognition-invariant evidence]
\label{prop:recognition_invariance}
Fix $1\le t<T$.
\begin{enumerate}
\item[\textnormal{(i)}]
If $\E_{\eta_t}[k]<K$, then $\eta_T=\eta_t$ if and only if $g_{t,T}$ is
constant on $Q$.
\item[\textnormal{(ii)}]
If $\E_{\eta_t}[k]=K$, then
$\eta_T=\eta_t$ if and only if $g_{t,T}=\zeta\mathbf 1_Q+\beta k$ for some $\zeta\in\R, \beta\ge-\mu_t.$
\end{enumerate}
Whenever these equivalent conditions hold, $\mu_T=\mu_t+\beta$, with $\beta=0$ in case (i).
\end{proposition}

Part \textnormal{(i)} gives the unconstrained benchmark.  If the recognition distribution
uses strictly less than the available capacity, it coincides with the Bayesian posterior, so
only a common multiplicative change in every model's likelihood can leave recognition
unchanged.  

Part \textnormal{(ii)} identifies what changes once all available capacity is used. In this case, recognition remains unchanged exactly when the log-likelihood increment is affine in complexity with slope at least \(-\mu_t\).  Thus, capacity adds  one
direction of evidence that can be absorbed without reallocating model weights: evidence
whose relative support for two models is proportional to their complexity difference.  The
coefficient $\beta$ is absorbed one-for-one by the shadow price.  Positive $\beta$ raises
$\mu$, while negative $\beta$ runs down the accumulated shadow price.  The restriction
$\beta\ge-\mu_t$ is therefore substantive: evidence favoring simpler
models can remain hidden from recognition only until the shadow price reaches zero; at
that point, further evidence must change the recognition distribution.

With three or more models, 
any component of $g_{t,T}$ outside $\operatorname{span}\{\mathbf 1_Q,k\}$ necessarily
changes recognition.  With two models of different complexities, every increment
belongs to this span, so recognition changes only when
evidence against the complex model exceeds the threshold
set by $\mu_t$.

\begin{corollary}
\label{prop:lag}
Suppose $Q=\{\m q^S,\m q^C\}$ with
$k(\m q^S)<K<k(\m q^C)$ and $\E_{\eta_t}[k]=K$. Define
$B_{t,T}
:=
\frac{1}{k(\m q^C)-k(\m q^S)}
\log
\frac{L_T(\m q^C)L_t(\m q^S)}
{L_t(\m q^C)L_T(\m q^S)}.$  If $B_{t,T}\ge-\mu_t$, then $\eta_T=\eta_t$ and
$\mu_T=\mu_t+B_{t,T}$. If $B_{t,T}<-\mu_t$, then the complexity
constraint is slack at $T$, $\mu_T=0$, and
$\eta_T(\m q^C)=\pi_T(\m q^C)<\eta_t(\m q^C)$.
\end{corollary}

\begin{examplecont}\normalfont
Suppose $K=0.4$ and
$\pi_t(\m q^C)=0.9$ in the two-model benchmark.  Then,
$\eta_t(\m q^C)=0.4$ and $\mu_t\simeq2.6$.
Corollary \ref{prop:lag} implies that subsequent evidence against the complex model holds its recognition weight at $0.4$ until the cumulative log Bayes factor against it exceeds approximately $2.6$.  A Bayesian posterior begins to fall immediately; by contrast, complexity-constrained recognition does not change until the previously accumulated evidence favoring the complex model has been exhausted.\hfill$\bigtriangledown$
\end{examplecont}

The preceding results concern model recognition, not directly choice.  Holding the recognition distribution
fixed need not hold behavior fixed because misspecification concern is generated by a
different statistic.  For $\eta\in\Delta(Q)$, the \emph{strict best-reply region} of action $a$ is $\mathcal C_a(\eta):=\{\lambda\ge0:\operatorname{BR}_\lambda(\eta)=\{a\}\}$.
Also, let
$U_t:=\max_{\m p\in\Delta(Y)^A}\prod_{s=1}^tp_{a_s}(y_s)$.  Whenever
$g_{t,T}=\zeta\mathbf 1_Q+\beta k$, define
$\Delta^U_{t,T}:=\log(U_T/U_t)$ and
$\mathcal M_t(\beta):=\log\big(\max_{\m q}L_t(\m q)e^{\beta k(\m q)}
/\max_{\m q}L_t(\m q)\big)$.  Then, the misspecification concern associated with this
likelihood increment is
\[
\widetilde\lambda_{t,T}(\zeta,\beta)
:=
\frac{ct\lambda_t+\Delta^U_{t,T}-\zeta-\mathcal M_t(\beta)+\psi_T-\psi_t}{cT}.
\]

\begin{corollary}[Action switching with frozen recognition]
\label{cor:frozen_recognition_switch}
Suppose $\E_{\eta_t}[k]=K$ and the best replies at histories $h_t$ and $h_T$ are
unique.  Then, the agent switches actions while $\eta_T=\eta_t$ if and only if there are
$\zeta\in\R$ and $\beta\ge-\mu_t$ such that
$g_{t,T}=\zeta\mathbf 1_Q+\beta k$,
$\lambda_t\in\mathcal C_a(\eta_t)$, and
$\widetilde\lambda_{t,T}(\zeta,\beta)\in\mathcal C_b(\eta_t)$ for two distinct
actions $a,b\in A$.  Under these conditions,
$\lambda_T=\widetilde\lambda_{t,T}(\zeta,\beta)$.
\end{corollary}

Corollary \ref{cor:frozen_recognition_switch} shows how misspecification concern can change choice while recognition remains fixed: recognition depends on relative likelihoods within $Q$ after filtering out the complexity direction, while misspecification concern depends on the absolute fit of the best model in $Q$ and on realized payoffs. 

\subsection{The fit-complexity frontier}\label{sec:fit}
The finite-sample analysis shows how scarce complexity capacity separates statistical
evidence from the model distribution used for choice.  We now analyze the
long-run version of this tradeoff: the fit-complexity frontier identifies the best
statistical fit attainable at capacity $K$ and the marginal statistical value of relaxing
that capacity.
\par For any distribution over actions $\alpha\in\Delta(A)$, define
\[
D(\m q;\alpha) := \sum_{a\in A}\alpha(a)\,R(p^\star_a\Vert q_a),
\qquad
d(\alpha):=\min_{\m q\in Q}D(\m q;\alpha),
\qquad
Q(\alpha):=\arg\min_{\m q\in Q}D(\m q;\alpha).
\]
Thus, $d(\alpha)$ is the absolute misspecification of the best structured model
when actions are drawn from $\alpha$. Next, let the \emph{expected normalized safety shortfall} of $\alpha$ be
\[
\psi(\alpha):=\frac{1}{B}\sum_{a\in A}\alpha(a)\big(\underline u-\E_{p^\star_a}[u(a,y)]\big)_+
\quad\text{if } B>0,
\qquad \psi(\alpha):=0 \ \text{ if } B=0,
\]
and define the \emph{misspecification level} of $\alpha$ by
\begin{equation}\label{eq:population_misspecification}
\tau(\alpha):=\frac{d(\alpha)+\psi(\alpha)}{c}.
\end{equation}
Thus, $\tau(\alpha)$ combines the statistical misspecification of the best structured
model with the payoff shortfall incurred by playing $\alpha$. Observation \ref{lem:convergence}
in Appendix \ref{app:safety} verifies that $\tau(\alpha)$ is the level toward which the S-LLR concern $\lambda_t$
drifts when play settles at $\alpha$.  

The corresponding complexity-constrained fit is summarized by
\begin{equation*}
F_K(\alpha)
:=
\min_{\eta\in\mathcal E_K}\E_\eta[D(\cdot;\alpha)],
\qquad
\mathcal H_K(\alpha)
:=
\arg\min_{\eta\in\mathcal E_K}\E_\eta[D(\cdot;\alpha)],
\qquad
G_K(\alpha):=F_K(\alpha)-d(\alpha).
\end{equation*}
The function $F_K(\alpha)$ is the \emph{fit-complexity frontier},
$\mathcal H_K(\alpha)$ is the set of recognition distributions that attain the frontier, and $G_K(\alpha)$ is the additional statistical error induced by the
complexity constraint. Define the set of shadow prices supporting the frontier by
\[
\mathcal R_K(\alpha)
:=
\arg\max_{\rho\ge0}
\left\{
\min_{\m q\in Q}
\big[D(\m q;\alpha)+\rho k(\m q)\big]
-\rho K
\right\}.
\]

\begin{proposition}[Properties of fit-complexity frontier]
\label{prop:frontier}
For every $\alpha\in\Delta(A)$:
\begin{enumerate}
\item[\textnormal{(i)}]
$F_K(\alpha)
=
\max_{\rho\ge0}
\big\{\min_{\m q}[D(\m q;\alpha)+\rho k(\m q)]-\rho K\big\}$.

\item[\textnormal{(ii)}]
$G_K(\alpha)=0$ if and only if
$K\ge\min_{\m q\in Q(\alpha)}k(\m q)$.

\item[\textnormal{(iii)}]
If $\eta\in\mathcal H_K(\alpha)$ and
$\rho\in\mathcal R_K(\alpha)$, then
$\operatorname{supp}(\eta)\subset
\arg\min_{\m q}\big\{D(\m q;\alpha)+\rho k(\m q)\big\},$
$\rho(\E_\eta[k]-K)=0$, and
$-\rho$ is a subgradient of the map $K\mapsto F_K(\alpha)$ at $K$.

\item[\textnormal{(iv)}]
There is an element of $\mathcal H_K(\alpha)$ supported on at most two
structured models.  If $\mathcal H_K(\alpha)$ is a singleton, its element has support
of size at most two.
\end{enumerate}
\end{proposition}

Proposition \ref{prop:frontier} gives a geometric characterization of how complexity
capacity changes model selection.  For fixed $\alpha$, each structured model $\m q$
can be viewed as a point whose two coordinates are its complexity $k(\m q)$ and its
statistical misspecification $D(\m q;\alpha)$.  Allowing recognition distributions
convexifies this menu: the agent can trade greater reliance on statistically better models
against the resources required to use them.  The frontier $F_K(\alpha)$ is the lowest
misspecification attainable from this tradeoff at capacity $K$.

Parts (i) and (iii) show that models receiving positive
weight minimize $D(\m q;\alpha)+\rho k(\m q)$.
A statistically inferior model can remain in recognition
because its lower complexity compensates for its worse
fit. The supporting price $\rho$ measures the marginal
value of additional capacity. Part (ii) identifies when capacity worsens attainable
fit. If one best-fitting model has complexity at most
$K$, the agent can attain $d(\alpha)$. Otherwise,
$G_K(\alpha)>0$: even unlimited data cannot remove
the restriction on model use. Part (iv) shows that some optimal recognition distribution
uses at most two models. If the optimum is unique,
it therefore has support of size at most two.

\subsubsection{Convergence to fit-complexity frontier}
The next result connects the fit-complexity frontier to the endogenous learning process. For every $t>0$, denote empirical action frequencies as
$\alpha_t(a):=\frac{1}{t}\sum_{i=1}^t\mathbbm{1}\{a_i=a\}$.

\begin{theorem}
\label{thm:long_run_frontier}
For every policy, there is an event of probability one such that, on every realized path in that event, if $\alpha_t\to\alpha$ for some
$\alpha\in\Delta(A)$, then
\begin{enumerate}
\item[\textnormal{(i)}]
$\lambda_t\to \tau(\alpha)$.

\item[\textnormal{(ii)}]
Every accumulation point of $\eta_t$ belongs to $\mathcal H_K(\alpha)$.

\item[\textnormal{(iii)}]
The sequence $\mu_t/t$ is bounded, and every accumulation point of $\mu_t/t$ belongs to $\mathcal R_K(\alpha)$.
\end{enumerate}
Consequently, on the same path, if
$\mathcal H_K(\alpha)=\{\eta^\star\}$ and
$\mathcal R_K(\alpha)=\{\rho^\star\}$, then
\[
\eta_t\longrightarrow\eta^\star,
\qquad
\lambda_t\longrightarrow\tau(\alpha),
\qquad
\frac{\mu_t}{t}\longrightarrow\rho^\star.
\]
\end{theorem}

Theorem \ref{thm:long_run_frontier} characterizes recognition
and misspecification concern when empirical action frequencies
converge to $\alpha$. By part \textnormal{(i)},
misspecification concern converges to  $\tau(\alpha)$ in \eqref{eq:population_misspecification},
which combines the statistical error of the best-fitting
model with the payoff shortfall associated with those
action frequencies. For fixed $\alpha$, the limiting misspecification concern does
not depend on capacity.  Limited capacity affects misspecification concern
through the choices it induces and the resulting statistical
and payoff evidence.

Part \textnormal{(ii)} shows that complexity  restricts which models the agent uses in the long run. As data accumulate, relative log likelihoods grow at rate $t$, so the prior and the entropy term become negligible, but the capacity constraint does not.  The recognition distribution $\eta_t$ therefore approaches the
fit-complexity solution set $\mathcal H_K(\alpha)$, rather than  the set of distributions supported on the unconstrained
best-fitting set $Q(\alpha)$ as in \citet{lanzani25}.   This result indicates that persistent reliance on a statistically inferior model is the long-run implication of a binding resource constraint, rather than a failure to learn the available evidence.

Part \textnormal{(iii)} identifies what accumulates when the evidence persistently favors models that are too costly to use.  The finite-sample multiplier $\mu_t$ is the  complexity pressure generated by past evidence. Every accumulation point of $\mu_t/t$ is a supporting
price $\rho\in\mathcal R_K(\alpha)$ of the fit-complexity
frontier.   When the supporting price is unique, $\mu_t/t\to\rho^\star$.

\subsection{Complexity-constrained robust equilibrium}

Theorem \ref{thm:long_run_frontier} motivates the following definition, which we call \textit{complexity-constrained $c$-robust equilibrium}.  

\begin{definition}[Equilibrium]
\label{def:ccre}
A tuple $(\alpha,\eta,\tau,\rho)\in
\Delta(A)\times\mathcal E_K\times\R_+^2$ is a
\emph{complexity-constrained $c$-robust equilibrium} if
\begin{enumerate}
\item[\textnormal{(i)}] $\eta\in\mathcal H_K(\alpha)$;
\item[\textnormal{(ii)}] $\operatorname{supp}(\alpha)\subseteq
\operatorname{BR}_\tau(\eta)$;
\item[\textnormal{(iii)}] $\tau=\tau(\alpha)$ and
$\rho\in\mathcal R_K(\alpha)$.
\end{enumerate}
\end{definition}

The three conditions combine the economic forces identified in Theorem \ref{thm:long_run_frontier}.  The  action
distribution $\alpha$ determines the statistical fit of every structured model.
The recognition distribution $\eta$ attains the best fit that is feasible at capacity
$K$, and every action receiving positive long-run frequency is optimal under the
resulting robust evaluation.  The scalar $\tau$  combines the statistical misspecification
of the best-fitting structured model with the safety
adjustment induced by the action distribution, while $\rho$
measures the marginal statistical value of additional
complexity capacity.

\begin{proposition}[Equilibrium existence and dynamic selection]
\label{prop:equilibrium}\hfill
\begin{enumerate}
\item[\textnormal{(i)}]
A complexity-constrained $c$-robust equilibrium exists.

\item[\textnormal{(ii)}]
For every policy, along almost every path on which
$\alpha_t\to\alpha$ and $\eta_t\to\eta$, every accumulation
point $\rho$ of $\mu_t/t$ yields a complexity-constrained
$c$-robust equilibrium $(\alpha,\eta,\tau(\alpha),\rho)$.
Moreover, $a_t\in\operatorname{BR}_{\tau(\alpha)}(\eta)$
at every sufficiently large date.

\item[\textnormal{(iii)}]
For every policy, along almost every path on which
$\alpha_t\to\alpha$ and
$\mathcal H_K(\alpha)=\{\eta^\star\}$,
we have $\eta_t\to\eta^\star$, and the conclusions
in part \textnormal{(ii)} hold with $\eta=\eta^\star$.
If, in addition, $\mathcal R_K(\alpha)=\{\rho^\star\}$,
then
$(\alpha_t,\eta_t,\lambda_t,\mu_t/t)
\longrightarrow
\big(\alpha,\eta^\star,\tau(\alpha),\rho^\star\big)$.
\end{enumerate}
\end{proposition}
Appendix \ref{app:action_convergence} gives conditions under which actions are eventually confined to a smaller set, including conditions under which  empirical frequencies converge. It shows that, in two-action problems, there are  mild conditions that guarantee frequency convergence; it also gives a condition under which the agent eventually stops switching actions.

\subsection{Separating complexity and misspecification concern}

The distinction between misspecification and complexity is sharpest when the statistically best-fitting structured model is unique.  Away from the knife-edge case in which its complexity equals capacity, whether that model is strictly feasible or infeasible determines which long-run concern remains active.

\begin{proposition}
\label{thm:independent}
Fix $\alpha\in\Delta(A)$ and suppose
$Q(\alpha)=\{\m q^\star\}$.  Then,

\begin{enumerate}
\item[\textnormal{(i)}]
If $k(\m q^\star)<K$, then
$G_K(\alpha)=0,$ $\mathcal H_K(\alpha)=\{\delta_{\m q^\star}\},$ and $\mathcal R_K(\alpha)=\{0\}.$

\item[\textnormal{(ii)}]
If $k(\m q^\star)>K$, then
$G_K(\alpha)>0,$ $\delta_{\m q^\star}\notin\mathcal H_K(\alpha),$ and $\mathcal R_K(\alpha)\subset\R_{++}.$
\end{enumerate}
\end{proposition}

Recall that
$G_K(\alpha)=F_K(\alpha)-d(\alpha)$ is the additional statistical error induced by
the complexity constraint. In part (i), when the uniquely best-fitting model $\m q^\star$ is strictly feasible,
the agent attains the unconstrained statistical optimum, assigns it all recognition
weight, and additional complexity capacity has zero marginal value.  Misspecification
concern may nevertheless remain positive because the model itself can still fit the true
model imperfectly.

In part (ii), when the uniquely best-fitting model exceeds capacity, the opposite configuration
arises.  The fit-complexity wedge is strictly positive, the statistically best model
cannot receive all recognition weight, and relaxing capacity has strictly positive
marginal statistical value.  This can occur even when
$d(\alpha)=0$, so that the structured model class $Q$ contains a model $\m q$ that, in the limit, fits the  data perfectly, i.e., $q_a=p^\star_a$ for every $a\in\operatorname{supp}(\alpha)$.  Statistical fit and complexity scarcity are therefore separate
long-run properties of the same learning problem.

\begin{corollary}
\label{lem:eu_constrained}
Suppose $Q(\alpha)=\{\m q^\star\}$, $k(\m q^\star)>K$, $\alpha_t\to\alpha$ a.s., $d(\alpha)=\psi(\alpha)=0$, and $\mathcal H_K(\alpha)=\{\eta^\star\}$.
Then, the following hold a.s.:
$\eta_t\rightarrow\eta^\star,$  $\lambda_t\rightarrow0$,
and, for every action $a\in A$,
\[
V_{\lambda_t}(a;\eta_t)
\longrightarrow
U(a;\eta^\star)
:=
\sum_{\m q\in Q}
\eta^\star(\m q)\E_{q_a}[u(a,y)].
\]
Moreover,
$\operatorname{supp}(\alpha)
\subseteq
\arg\max_{a\in A}U(a;\eta^\star).$
\end{corollary}

In Corollary \ref{lem:eu_constrained}, the conditions
$d(\alpha)=\psi(\alpha)=0$ imply that $\m q^\star$
correctly predicts the outcomes of actions chosen with
positive long-run frequency, and that each such action has
expected payoff at least equal to the maxmin payoff.
Misspecification concern therefore vanishes, and these
actions maximize expected utility under $\eta^\star$. 
Capacity nevertheless prevents the agent from placing
all weight on $\m q^\star$. Proposition
\ref{thm:independent}\textnormal{(ii)} and Theorem
\ref{thm:long_run_frontier}\textnormal{(iii)} imply that
every asymptotic complexity shadow price belongs to
$\mathcal R_K(\alpha)\subset\R_{++}$.
Thus, the limiting criterion is expected utility, but
the model weights remain constrained by capacity.
In the corresponding environment of \citet{lanzani25},
a unique best-fitting model receives all long-run
model weight.

The separation between complexity and misspecification is sharper with two models.

\begin{corollary}[Two-model decomposition]
\label{cor:two_model}
Suppose
$Q=\{\m q^S,\m q^C\}$,
$k(\m q^S)<K<k(\m q^C)$, and
$D(\m q^C;\alpha)<D(\m q^S;\alpha)$.  Then,
$\mathcal H_K(\alpha)=\{\eta^\star\}$ and $\mathcal R_K(\alpha)=\{\rho^\star\}$, where
\[
\eta^\star(\m q^C)
=
\frac{K-k(\m q^S)}
{k(\m q^C)-k(\m q^S)},
\qquad
\rho^\star
=
\frac{
D(\m q^S;\alpha)-D(\m q^C;\alpha)}
{k(\m q^C)-k(\m q^S)},
\qquad
\tau^\star
=\frac{D(\m q^C;\alpha)+\psi(\alpha)}{c}.
\]
\end{corollary}

Corollary \ref{cor:two_model} decomposes the three long-run objects.  The recognition weight on the complex model is the fraction of the complexity gap that available capacity can support.  The misspecification concern $\tau^\star$ depends on the absolute statistical error of the better-fitting model and the payoff shortfall associated with the
action distribution.  In contrast, $\rho^\star$ is the complex model's improvement in statistical fit divided by its additional complexity requirement.  It therefore measures the statistical return to one
extra unit of complexity capacity.

\begin{examplecont}\normalfont
Consider the two-model benchmark, where
$A=\{\mathfrak{s},\mathfrak{r}\}$ and $Y=\{0,1\}$, with
$u(\mathfrak{r},y)=y$ and $u(\mathfrak{s},y)=0.477$ for all $y$. An interpretation is that $\mathfrak{s}$ is a safe action and $\mathfrak{r}$ risky.
Suppose the simple model satisfies $q^S_{\mathfrak{s}}=q^S_{\mathfrak{r}}=\mathrm{Bern}(0.52)$,
while the complex model allows their success probabilities to differ, $q^C_{\mathfrak{s}}=\mathrm{Bern}(0.2)$ and $q^C_{\mathfrak{r}}=\mathrm{Bern}(0.6)$. Let $p^\star_{\mathfrak{s}}=\mathrm{Bern}(0.298)$,
$p^\star_{\mathfrak{r}}=\mathrm{Bern}(0.75)$, and $c=0.038$. 
\par Define $\xi:=\alpha(\mathfrak{r})$. Then, $D(\m q^S;\alpha)\simeq0.1+0.011\xi$ and $D(\m q^C;\alpha)\simeq0.027+0.023\xi$, so the complex model fits better
for every $\xi\in[0,1]$. Since the safe action yields $\bar u=0.477$ and the
objective expected payoff of the risky action is $0.75$,
$\psi(\alpha)=0$ for every $\alpha\in\Delta(A)$. Corollary \ref{cor:two_model} implies $\eta^\star(\m q^C)=K\in(0,1),$ $\rho^\star\simeq0.074-0.012\xi,$ and $\tau^\star\simeq0.71+0.599\xi.$
More frequent risky play  therefore raises the absolute misspecification of the best-fitting model while lowering the marginal statistical value of additional complexity capacity, even though the recognition weight on the complex model remains fixed at $K$.  \hfill$\bigtriangledown$
\end{examplecont}

\section{Long-run cycles and risky play}\label{sec:risky}
We now examine how complexity capacity and misspecification concern jointly govern the willingness to bear risk. Throughout
this section, we consider the following class of decision problems.

\begin{definition}[Risky-safe problem]
\label{def:risky_safe}
A \emph{risky-safe problem} has two actions,
$A=\{\mathfrak{s},\mathfrak{r}\}$, with
$u(\mathfrak{s},y)=\bar u$ for every $y\in Y$ and
$u(\mathfrak{r},\cdot)$ nonconstant. We refer to
$\mathfrak{s}$ as the safe action and $\mathfrak{r}$ as
the risky action.
\end{definition}

These problems capture choices between a guaranteed payoff
and an alternative whose profitability depends on uncertain
outcomes, as in investment and technology adoption.
The distribution of outcomes can depend on the
chosen action even when its payoff is constant.
\subsection{Endogenous cycles}\label{sec:cycle}
Since the agent's choices affect the data she observes, they can 
change her concern about misspecification and hence her subsequent
choices. This feedback can generate persistent switching even
in a stationary environment. In a risky-safe problem, safe observations can reduce
$\lambda_t$ enough to make risky play  attractive, while risky observations can
subsequently worsen model fit or reveal inadequate realized performance and therefore raise $\lambda_t$ enough to restore the safe action. We study how the complexity constraint changes this
switching and when it leads to eventual safe play.

Consider a risky-safe problem with two structured models,
$Q=\{\m q^S,\m q^C\}$, with
$k_S:=k(\m q^S)<K<k(\m q^C)=:k_C$.  For
$j\in\{S,C\}$ and
$a\in\{\mathfrak{s},\mathfrak{r}\}$, let
$d_a^j:=R(p_a^\star\Vert q_a^j).$
Suppose
$d_a^C<d_a^S$ 
$\text{for }a=\mathfrak{s},\mathfrak{r},$ and 
$d_{\mathfrak{s}}^C<d_{\mathfrak{r}}^C$, so that the complex model $\m q^C$ is statistically better under either action, while risky observations are
harder for even this best-fitting model to explain.\footnote{For example, a manager's models may predict sales in an
established market more accurately than sales in a new
market, where demand is less well understood.} Let
$\lambda_{\mathfrak{s}}
:=\tau(\delta_{\mathfrak s})
=[d_{\mathfrak s}^C+\psi(\delta_{\mathfrak s})]/c$ and 
$\lambda_{\mathfrak{r}}
:=\tau(\delta_{\mathfrak r})
=[d_{\mathfrak r}^C+\psi(\delta_{\mathfrak r})]/c$; these are the limiting levels of misspecification concern
under exclusive safe and risky play, respectively. When $\min_yu(\mathfrak r,y)<\bar u$,
the maxmin payoff is $\underline u=\bar u$,
so $\psi(\delta_{\mathfrak s})=0$ and
$\lambda_{\mathfrak r}>\lambda_{\mathfrak s}$. Also, let  
$\gamma_a:=d_a^S-d_a^C>0, $ and for $j=S,C$, define
$
\Delta_j(\lambda)
:=
v_\lambda(\mathfrak{r};\m q^j)-\bar u$ and $\Delta_K(\lambda)
:=
(1-\chi_K)\Delta_S(\lambda)
+
\chi_K\Delta_C(\lambda)$ where $\chi_K
:=
\frac{K-k_S}{k_C-k_S}.$ 
\begin{lemma}
\label{lem:delta_monotone}
For each $j\in\{S,C\}$,
$\Delta_j$ is continuous and strictly decreasing on $\R_+$.
Thus, $\Delta_K$ is continuous and strictly decreasing on $\R_+$ for every
$K\in(k_S,k_C)$.
\end{lemma}

Lemma \ref{lem:delta_monotone}  shows that greater
misspecification concern makes risky play less attractive.  A larger $\lambda$ permits a more adverse distortion of the risky model
and therefore strictly lowers the robust value of the risky action.  The safe action has
state-independent utility, so its value remains $\bar u$ at every $\lambda$. Following \citet{lanzani25}, we say that the environment exhibits a
\emph{$\lambda$-cycle} when
$\alpha_t(\mathfrak{r})$ converges to a value in $(0,1)$, so that both actions  occur with
positive asymptotic frequency. As we show below, misspecification concern nevertheless converges.

\begin{theorem}[Robustness cycles]
\label{thm:complexity_cycles}
For every $K\in(k_S,k_C)$ and every policy satisfying
\eqref{eq:dynamic_criterion},
$\eta_t(\m q^C)\rightarrow\chi_K$ a.s.,
and
$\lambda_t-
\left[
\lambda_{\mathfrak{s}}
+
\alpha_t(\mathfrak{r})
\big(
\lambda_{\mathfrak{r}}
-\lambda_{\mathfrak{s}}
\big)
\right]
\rightarrow0$ a.s.
Moreover,
$\alpha_t(\mathfrak{r})\to x_K$ a.s., where
\[
x_K=
\begin{cases}
0,
& \Delta_K(\lambda_{\mathfrak{s}})\le0,\\[3pt]
1,
& \Delta_K(\lambda_{\mathfrak{r}})\ge0,\\[3pt]
x\in(0,1) \text{ \normalfont s.t. }
\Delta_K\!\left(
\lambda_{\mathfrak{s}}
+x(\lambda_{\mathfrak{r}}-\lambda_{\mathfrak{s}})
\right)=0,
\quad & \Delta_K(\lambda_{\mathfrak{s}})>0>
\Delta_K(\lambda_{\mathfrak{r}}),
\end{cases}
\]
and $\mu_t/t
\rightarrow
\dfrac{
(1-x_K)\gamma_{\mathfrak{s}}
+x_K\gamma_{\mathfrak{r}}
}{
k_C-k_S
}
>0$ a.s., so a $\lambda$-cycle occurs if and only if
$\Delta_K(\lambda_{\mathfrak{s}})
>0>
\Delta_K(\lambda_{\mathfrak{r}}).$
\end{theorem}

This result shows that statistical
evidence eventually favors $\m q^C$, but its recognition weight
converges to the maximal feasible level $\chi_K$. Misspecification concern converges to \(\lambda_{\mathfrak s}+x_K(\lambda_{\mathfrak r}-\lambda_{\mathfrak s})\), where \(x_K\) is the limiting frequency of risky play.  Since the risky-safe value difference is strictly decreasing in
$\lambda$, an interior frequency is selected  when risky behavior is preferred at
the low endpoint and safe behavior is preferred at the high endpoint. 

  In the limit of $\mu_t/t$, the quantity $\gamma_a=d_a^S-d_a^C$ is the expected per-observation
log-likelihood advantage of the complex model over the simple model when action $a$ is
played, so 
$(1-x_K)\gamma_{\mathfrak{s}}+x_K\gamma_{\mathfrak{r}}$
is the average long-run statistical advantage of the complex model under the endogenous
action frequency.  Dividing by $k_C-k_S$ converts that statistical advantage into a
return per additional unit of complexity.  

Theorem \ref{thm:complexity_cycles} also clarifies the difference from \citet{lanzani25}.  Since $\m q^C$ is
the unique statistically best model under both actions, unconstrained long-run model
selection places all weight on $\m q^C$.  Complexity-constrained recognition instead
fixes its long-run recognition weight at $\chi_K<1$.  The endogenous law governing $\lambda_t$ responds to
both absolute statistical fit and payoff
shortfalls. Capacity can therefore alter or eliminate the cycle by changing both the model weights used to compare the two actions and, through the resulting choices, the observations that drive misspecification concern.

\subsubsection{Elimination of cycles}

To characterize when complexity eliminates $\lambda$-cycles, consider the endpoint comparisons:
\[\Delta_C(\lambda_{\mathfrak{s}})>0>
\Delta_C(\lambda_{\mathfrak{r}}),\quad 
\Delta_S(\lambda_{\mathfrak{s}})<0.
\]
The first two inequalities characterize  the cycle under full recognition of the statistically superior model  $\m q^C$: the risky action is preferred
after safe observations have reduced misspecification concern to
$\lambda_{\mathfrak{s}}$, but the safe action is preferred after risky observations have
raised it to $\lambda_{\mathfrak{r}}$.  

The last inequality says that the simple model $\m q^S$
favors safety even at $\lambda_{\mathfrak{s}}$, the limiting
level of misspecification concern under exclusively safe play.  This condition is necessary for eventual safe play at
some $K\in(k_S,k_C)$: otherwise,
$\Delta_K(\lambda_{\mathfrak s})>0$ at every such capacity,
so continued safe play would eventually make risk attractive.
Let 
$K^\ast:=k_S+(k_C-k_S)\chi^\ast$, where
$\chi^\ast:=
-\Delta_S(\lambda_{\mathfrak{s}})/
[\Delta_C(\lambda_{\mathfrak{s}})
-\Delta_S(\lambda_{\mathfrak{s}})]$.

\begin{corollary}[Complexity eliminates $\lambda$-cycles]\label{cor:complexity_eliminates_cycles}
If $\Delta_C(\lambda_{\mathfrak{s}})>0>
\Delta_C(\lambda_{\mathfrak{r}})$ and
$\Delta_S(\lambda_{\mathfrak{s}})<0$, then 
$\chi^\ast\in(0,1)$ and
$K^\ast\in(k_S,k_C)$.  Moreover,

\begin{enumerate}
\item[\textnormal{(i)}]
if $K<K^\ast$, then
$a_t=\mathfrak{s}$ for all sufficiently large $t$ a.s.;

\item[\textnormal{(ii)}]
if $K=K^\ast$, then
$\alpha_t(\mathfrak{r})\to0$ a.s.;

\item[\textnormal{(iii)}]
if $K\in(K^\ast,k_C)$, then
$\alpha_t(\mathfrak{r})\to x_K\in(0,1)$ a.s. and the environment exhibits a
$\lambda$-cycle.
\end{enumerate}
\end{corollary}

The three cases in Corollary \ref{cor:complexity_eliminates_cycles} describe whether complexity prevents risky behavior from restarting after a
safe phase.  If $K<K^\ast$, recognition of the complex model is sufficiently restricted that
even at $\lambda_{\mathfrak{s}}$ the safe action is strictly preferred; since the risky--safe
value difference decreases with $\lambda$, safety is then optimal throughout the endogenous
range and eventually persists.  At $K=K^\ast$, the agent is  indifferent at the
lower endpoint, so the long-run risky frequency vanishes, although  this does not establish that risky choices
eventually cease. If $K>K^\ast$, risky behavior remains attractive at the
lower endpoint and safety at the upper endpoint, so the robustness feedback continues to
support a cycle.

These comparisons arise naturally when the structure that makes $\m q^C$ more complex is what makes
the risky action attractive when model fit is good.  The complex model then supports risky
behavior at low $\lambda$, while robustness overturns that ranking after sufficiently poor
fit.  If the simpler model cannot represent the mechanism
that generates the risky upside, it can instead favor safety already at
$\lambda_{\mathfrak{s}}$.  A simple sufficient condition is $\E_{q_{\mathfrak{r}}^S}
[u(\mathfrak{r},y)]
<
\bar u.$
Indeed, the multiplier criterion implies
$\Delta_S(\lambda)\le\Delta_S(0)
=\E_{q_{\mathfrak{r}}^S}[u(\mathfrak{r},y)]-\bar u$
for every $\lambda\ge0$.  This sufficient condition is not necessary because
robustness can make the simple model favor safety at
$\lambda_{\mathfrak{s}}$ even when its expected payoff from the risky action exceeds
$\bar u$.  The running example illustrates this possibility: the simple model's
expected risky payoff is $0.52>\bar u=0.477$, but
$\Delta_S(\lambda_{\mathfrak{s}})<0$. 

\begin{figure}[hbt!]
\centering

\def\pzero{0.52}
\def\pone{0.6}
\def\ubar{0.477}

\def\lamlow{0.7131342250139462}
\def\lamhigh{1.3120198993216696}
\def\Kstar{0.5617753931072901}

\begin{tikzpicture}
\begin{axis}[
    width=0.94\textwidth,
    height=0.58\textwidth,
    xmin=0.10, xmax=1.50,
    ymin=-0.12, ymax=0.12,
    axis on top,
    xlabel={Misspecification concern $\lambda$},
    ylabel={Value difference $\Delta_K(\lambda)$},
    xtick={0.2,0.4,0.6,0.8,1.0,1.2,1.4},
    xticklabels={$0.2$, $0.4$, $0.6$, $0.8$, $1.0$, $1.2$, $1.4$},
    ytick={-0.12,-0.08,-0.04,0,0.04,0.08,0.12},
    yticklabels={$-0.12$, $-0.08$, $-0.04$, $0$, $0.04$, $0.08$, $0.12$},
    ticklabel style={font=\footnotesize},
    label style={font=\footnotesize},
    legend style={
        at={(0.03,0.03)},
        anchor=south west,
        draw=black!20,
        fill=white,
        fill opacity=0.92,
        text opacity=1,
        font=\scriptsize,
        legend image post style={sharp plot}
    },
    samples=450,
    domain=0.10:1.50,
    unbounded coords=jump,
    clip=false,
    declare function={
        vr(\p)= -(1/x)*ln((1-\p)+\p*exp(-x));
        Delta(\chi)= (1-\chi)*vr(\pzero)+\chi*vr(\pone)-\ubar;
    }
]

\path[fill=black!7]
(axis cs:\lamlow,-0.12) rectangle
(axis cs:\lamhigh,0.12);

\addplot[black!45, dashed, thin, forget plot]
coordinates {(\lamlow,-0.12) (\lamlow,0.12)};
\addplot[black!45, dashed, thin, forget plot]
coordinates {(\lamhigh,-0.12) (\lamhigh,0.12)};

\addplot[black, thin, forget plot]
coordinates {(0.10,0) (1.50,0)};

\node[font=\footnotesize, anchor=north west]
at (axis cs:0.8,0.113)
{endogenous range
$[\lambda_{\mathfrak{s}},\lambda_{\mathfrak{r}}]$};

\node[font=\scriptsize, text=gray, anchor=south west]
at (axis cs:0.09,0.0006)
{$\Delta_K>0$: risky action preferred};

\node[font=\scriptsize, text=gray, anchor=north west]
at (axis cs:0.09,-0.0006)
{$\Delta_K<0$: safe action preferred};

\addplot[very thick, blue!75!black, densely dashed]
{Delta(1.00)};
\addlegendentry{$K=1$}

\addplot[very thick, orange!90!black, densely dotted]
{Delta(0.80)};
\addlegendentry{$K=0.80$}

\addplot[very thick, violet!80!black, dashdotted]
{Delta(\Kstar)};
\addlegendentry{$K^\ast\simeq0.56$}

\addplot[very thick, green!60!black, solid]
{Delta(0.40)};
\addlegendentry{$K=0.40$}

\addplot[
    only marks,
    mark=*,
    mark size=2.2pt,
    blue!75!black,
    forget plot
] coordinates {(0.9988689905052638,0)};

\addplot[
    only marks,
    mark=*,
    mark size=2.2pt,
    orange!90!black,
    forget plot
] coordinates {(0.8690508863651075,0)};

\addplot[
    only marks,
    mark=*,
    mark size=2.2pt,
    violet!80!black,
    forget plot
] coordinates {(0.7131342250139457,0)};

\addplot[
    only marks,
    mark=*,
    mark size=2.2pt,
    green!60!black,
    forget plot
] coordinates {(0.6068478574210411,0)};

\end{axis}
\end{tikzpicture}

\caption{Tighter complexity capacity moves the risky-safe indifference point toward
the lower boundary of the misspecification concern.  At
$K=K^\ast$ the indifference point reaches
$\lambda_{\mathfrak{s}}$; below $K^\ast$ the safe action is preferred throughout the
entire range.}
\label{fig:complexity_cycles}
\end{figure}
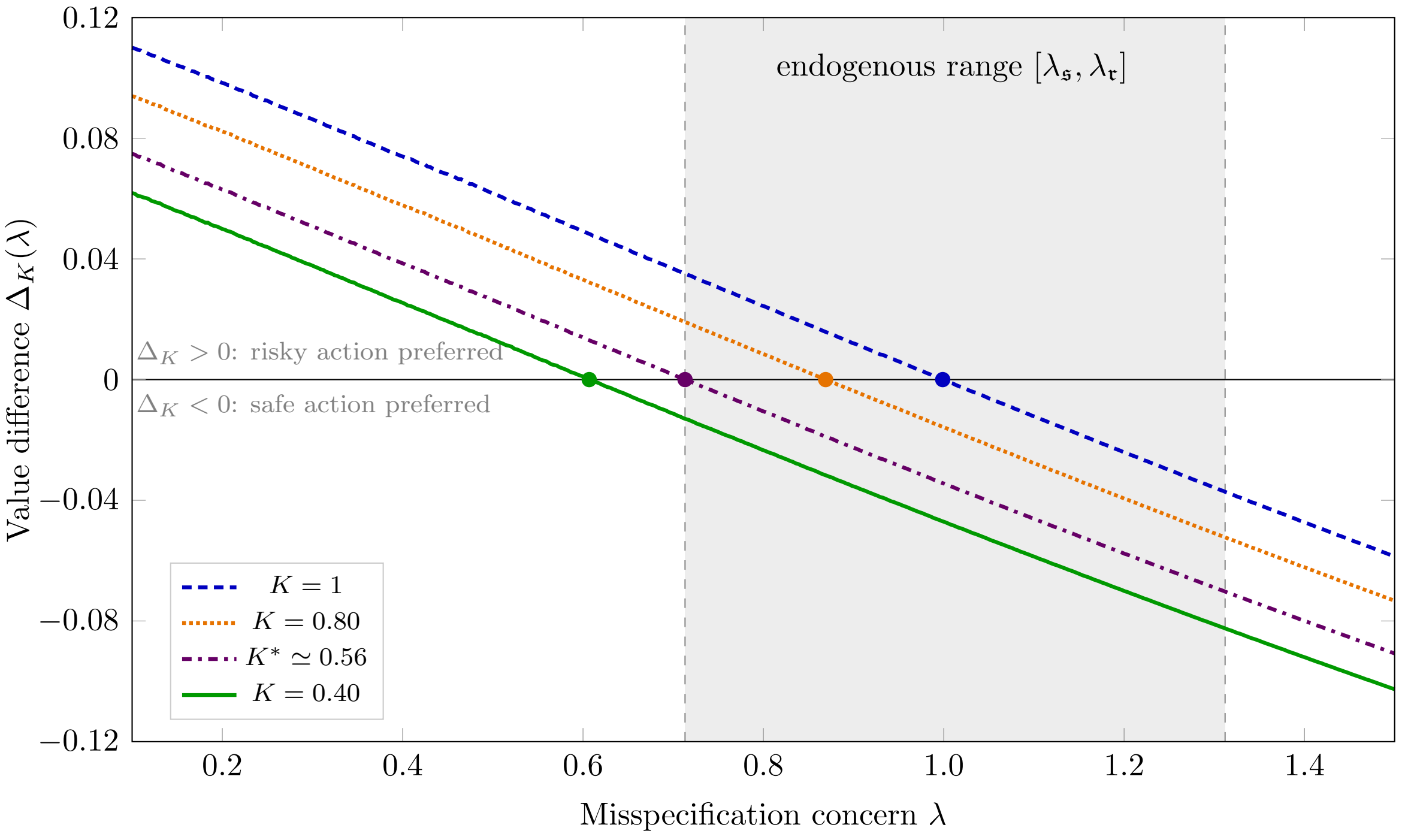

\begin{examplecont}\normalfont
Figure \ref{fig:complexity_cycles} illustrates the threshold in the running example. Here 
$\lambda_{\mathfrak{s}}\simeq0.71$, $\lambda_{\mathfrak{r}}\simeq1.31$, and
$K^\ast\simeq0.56$. The curve $K=1$ corresponds to full recognition of $\m q^C$ 
and has an indifference point in
$(\lambda_{\mathfrak{s}},\lambda_{\mathfrak{r}})$, so robustness cycles between safe and risky actions can be sustained.  Tightening capacity shifts recognition toward the simple model and moves the indifference point to the
left.  At $K=K^\ast$, it reaches the left boundary of the range $\lambda_{\mathfrak{s}}$, so below $K^\ast$, even the most favorable level of endogenous misspecification concern cannot restart risky play.\hfill$\bigtriangledown$
\end{examplecont}

\subsection{Risky play}
\label{sec:capacity_risky_play}
The following example shows that greater capacity can make risk initially attractive
while reducing its long-run frequency. Consider a risky-safe problem
with $Y=\{-1,1,2\}$, $\bar u=0$, and
$u(\mathfrak r,y)=y$. Let $Q=\{\m q^P,\m q^S,\m q^C\}$, with
$k(\m q^P)=k(\m q^S)=0$ and $k(\m q^C)=1$; as we show below, having three structured models instead of two will allow learning
to change which model comparison governs the effect of
capacity.
Their predictions under the risky action and the true outcome distribution
are
\[
\begin{array}{c|ccc}
& -1 & 1 & 2 \\ \hline
q_{\mathfrak r}^P & 0.94 & 0.05 & 0.01 \\
q_{\mathfrak r}^S & 0.10 & 0.80 & 0.10 \\
q_{\mathfrak r}^C & 0.30 & 0.10 & 0.60 \\
p_{\mathfrak r}^\star & 0.40 & 0.01 & 0.59
\end{array}
\]
For the safe action,
$p_{\mathfrak s}^\star
=q_{\mathfrak s}^S
=q_{\mathfrak s}^C
=(1/3,1/3,1/3)$, while
$q_{\mathfrak s}^P=(0.80,0.10,0.10)$.
All distributions have full support.
The prior is
$(\pi_0(\m q^P),\pi_0(\m q^S),\pi_0(\m q^C))
=(0.09,0.01,0.90)$, and the agent follows
the decision criterion \eqref{eq:dynamic_criterion} with misspecification sensitivity $c=0.03$.

At date $0$, $\lambda_0=0$, so choice is governed by expected utility.
The expected risky payoffs under
$\m q^P$, $\m q^S$, and $\m q^C$ are respectively
$-0.87$, $0.90$, and $1$.
For $K<0.90$, the capacity constraint binds, where
$\eta_0(\m q^P)=0.9(1-K),$ $\eta_0(\m q^S)=0.1(1-K),$ and $\eta_0(\m q^C)=K.$
Hence, the date-$0$ risky value is
$1.693K-0.693$.
Define
$\bar K:=\frac{0.693}{1.693}\simeq0.409.$
Increasing capacity through $\bar K$ therefore changes the date-$0$ optimal
action from safe to risky.

The long-run comparison is different because the recognition
weight on $\m q^P$ converges to zero. 
The remaining comparison is therefore between
$\m q^S$ and $\m q^C$.
For these two models, $k_S=0$ and $k_C=1$,
so $\chi_K=K$, and the limiting recognition weights
are $1-K$ and $K$, respectively.
Let $\Delta_K$ be the corresponding risky-safe value
difference defined in Section \ref{sec:cycle}. By Lemma \ref{lem:delta_monotone}, $\Delta_K$ is continuous and strictly
decreasing in $\lambda$.  In this environment,
$\Delta_K(0)>0$ and
$\Delta_K(\lambda)\to-1$ as $\lambda\to\infty$, so let $\tau_K$ denote its unique zero. All proofs of results from this section are in Appendix \ref{app:olarisky}.

\begin{proposition}[Reversal in the effect of capacity]
\label{prop:capacity_risky_reversal}
In this environment, for every $K\in(0,1)$:

\begin{enumerate}
\item[\textnormal{(i)}]
At date $0$, the safe action is strictly optimal if $K<\bar K$, the risky
action is strictly optimal if $K>\bar K$, and the two actions are
indifferent if $K=\bar K$.

\item[\textnormal{(ii)}]
The following hold a.s.,
\[
\eta_t(\m q^P)\to0,
\quad
\eta_t(\m q^S)\to1-K,
\quad
\eta_t(\m q^C)\to K, \quad \lambda_t\to\tau_K,
\quad
\alpha_t(\mathfrak r)
\to
\frac{c\tau_K}{D_{\mathfrak r}(\m q^C)}.
\]

\item[\textnormal{(iii)}]
Both $\tau_K$ and the limiting risky frequency in
\textnormal{(ii)} are strictly decreasing in $K$.
\end{enumerate}
\end{proposition}

Part \textnormal{(i)} identifies the short-run effect.
Before any outcomes have been observed, additional capacity transfers
recognition weight from the two simple models toward $\m q^C$.
Since $\m q^P$ assigns the risky action a strongly negative expected
payoff, this raises the risky value and changes the optimal action once
$K$ crosses $\bar K$.

In part \textnormal{(ii)}, learning changes the relevant model comparison: the recognition weight on $\m q^P$ converges to zero, while
statistical evidence increasingly favors $\m q^C$ over $\m q^S$.
Capacity nevertheless limits recognition of $\m q^C$ to $K$, leaving
weights $1-K$ and $K$ on the two models that remain relevant for choice.
Since $\m q^C$ correctly specifies the safe-action distribution  and $\psi(\alpha)=0$ for every
$\alpha\in\Delta(A)$, safe
observations make no asymptotic contribution to misspecification concern.
Risky observations do, so the long-run risky frequency is the
frequency required to sustain the indifference level $\tau_K$.

Part \textnormal{(iii)} then gives the opposite long-run comparative static.
Although $\m q^C$ gives the risky action a higher expected payoff than
$\m q^S$, it assigns substantially more probability to the loss state. For these distributions, the robust risky value under $\m q^C$
is lower than under $\m q^S$ at every $\tau_K$.
Greater capacity consequently lowers $\Delta_K$ at the
long-run indifference point and lowers its zero $\tau_K$.  Since
misspecification concern is asymptotically generated by risky
observations, the limiting risky frequency falls as well.

The next result shows why changing the relevant model
comparison matters.

\begin{obs}
\label{prop:one_simple}
Consider a risky-safe problem with
$Q=\{\m q^S,\m q^C\}$,
$k(\m q^S)<k(\m q^C)$, and 
$\min_yu(\mathfrak r,y)<\bar u$.
For $j\in\{S,C\}$, define
$\bar\lambda_j
:=
\inf\{
\lambda\ge0:
\Delta_j(\lambda)\le0
\}.$ Fix capacities
$K<K'$ in
$(k(\m q^S),k(\m q^C))$.
Consider two copies of the dynamic decision problem with identical
primitives, differing only in their
capacities $K$ and $K'$.
For each action $a$, couple the two copies using a common sequence
$(Y_n^a)_{n\ge1}$ of outcomes with distribution $p_a^\star$: whenever
either copy chooses action $a$ for the $n$th time, it observes $Y_n^a$.
If both copies are indifferent between the two actions at the same
decision problem, select the same action.

Let
$R_t^\ell
:=
\sum_{\tau=1}^t
\mathbbm 1\{a_\tau^\ell=\mathfrak r\}$ for $\ell\in\{K,K'\}$.
Then, on every coupled outcome path and for every $t$,
$R_t^{K'}\ge R_t^K$ if
$\bar\lambda_C>\bar\lambda_S$;
$R_t^{K'}\le R_t^K$ if
$\bar\lambda_C<\bar\lambda_S$; and
$R_t^{K'}=R_t^K$ if
$\bar\lambda_C=\bar\lambda_S$.
\end{obs}

Observation \ref{prop:one_simple} rules out the mechanism behind
Proposition \ref{prop:capacity_risky_reversal} when there are only two
models. If the complex model favors risk at higher
levels of misspecification concern than the simple model, greater capacity
weakly raises the number of risky choices at every date.  If the ordering
of the model-specific thresholds is reversed, greater capacity weakly
lowers that number at every date.  With equal thresholds, capacity does
not affect the realized risky-choice count under the coupling.  In this setting, a reversal
of the kind in Proposition \ref{prop:capacity_risky_reversal} therefore
requires learning to change which model comparison governs the effect of
capacity.

\section{Application: Model Selection}
\label{sec:ba}

So far, the agent has learned and chosen actions within a
fixed model class. We now allow the agent to compare
classes and select which one to use. Averaging across competing explanatory frameworks may
be more demanding than assessing uncertainty within one,
particularly when the frameworks use different variables
or causal mechanisms. At the same time, selecting a class does not resolve uncertainty within it: the agent must still assess its competing predictions and use them to choose an action.

In \citet{ba26} and \citet{schwartzsteinSunderam2021},
the agent selects a statistical specification and then updates beliefs within it.\footnote{Model comparison guides choice in \citet{choKasa2015}, although
the testing and learning rules differ.} We combine this idea with our capacity constraint,
distinguishing the statistical evidence for selecting
a class from the capacity required to use its predictions. This section 
extends the switching rule in \citet{ba26}, where the
agent repeatedly compares given classes using data
generated by past choices. Capacity can therefore affect
which classes survive by changing the observations.

A model $\theta$ in \citet{ba26} corresponds to a model
class $Q$: each parameter $\omega\in\Omega_\theta$ indexes
a structured model $\m q^\omega$, with
$q^\omega_a=q_\theta(\cdot|a,\omega)$, so we continue to denote structured models by suppressing the $\omega$ index.
The agent updates beliefs within two classes using every
observation and chooses actions using the Bayesian
posterior within the adopted class. Classes are compared
by their marginal likelihoods,
$\ell_t(Q):=\sum_{\m q\in Q}\pi_0(\m q)L_t(\m q)$.
We connect these rules by
allowing scoring weights to exceed prior weights by
a factor of at most $h\ge1$:
\begin{equation}\label{eq:class_score}
S_{h,t}(Q):=
\max_{\substack{\nu\in\Delta(Q)\\
\nu(\m q)\le h\pi_0(\m q)\hspace{0.03in} \forall\m q\in Q}}
\ \sum_{\m q\in Q}\nu(\m q)L_t(\m q).
\end{equation}
The parameter $h$ bounds the increase in each scoring weight relative
to its initial prior weight. At $h=1$, the  score is the marginal likelihood. As $h\to\infty$, it approaches maximized likelihood; we include this
endpoint by defining $S_{\infty,t}(Q):=\max_{\m q\in Q}L_t(\m q)$.
For finite $h$,  
\begin{equation}\label{eq:class_score_bounds}
\ell_t(Q)\le S_{h,t}(Q)\le h\ell_t(Q).
\end{equation}

The score in \eqref{eq:class_score} is the upper-tail
conditional value-at-risk (CVaR) of model likelihoods,
an operator used in
stochastic optimization
\citep[see,][]{rockafellar2000optimization,rockafellar2002cvar}.\footnote{Distributions that attain the maximum in the score \eqref{eq:class_score} are called ``type II maximum likelihood priors'' in robust statistics \citep[see,][]{type20}.} All the proofs of results from this section are in Appendix \ref{app:generalized_selection_proofs}.

\begin{obs}[Tail-likelihood representation]
\label{lem:ba_cvar}
For every finite $h\ge1$,
\begin{equation}\label{eq:ba_cvar}
S_{h,t}(Q)=
\min_{\zeta\in\R}
\Big\{
\zeta+h\sum_{\m q\in Q}\pi_0(\m q)
\bigl(L_t(\m q)-\zeta\bigr)_+
\Big\}.
\end{equation}
\end{obs}
Observation \ref{lem:ba_cvar} shows that $h$ governs how much weight can be shifted toward
the best-fitting models. Unlike Bayesian marginal
likelihood, the score emphasizes the upper tail of
model fit, without changing the models used for choice.

Use the same $h$ for both classes and set
$B_{h,t}:=S_{h,t}(Q')/S_{h,t}(Q)$.
Starting from $Q$, the agent switches to $Q'$ when $B_{h,t}>\gamma$,
switches back when $B_{h,t}<1/\gamma$, and otherwise retains the current
class, where $\gamma>1$. 

\citet{ba26} says that a  class \emph{persists} against a competitor if it is eventually used
forever with positive probability. At a given prior, it is
\emph{globally robust} if it persists against every admissible
competitor and prior, and \emph{locally robust} if it persists against
all sufficiently small perturbations of its predictions and prior.
Competitors are finite nonempty classes with full-support predictions
and priors. The distances between classes and priors are defined in
Appendix \ref{app:ba_definitions}; the switching rule is now determined
by $B_{h,t}$.

\paragraph{The risky-safe problem without a capacity constraint.}
Consider a risky-safe problem as in Definition
\ref{def:risky_safe}, with $Y=\{0,1\}$,
$\bar u\in(0,1)$, and $u(\mathfrak r,y)=y$. Let $Q=\{\m q^S,\m q^C\}$, with
$w:=\pi_0(\m q^C)\in(0,1)$.
Assume
$q^C_{\mathfrak s}=p^\star_{\mathfrak s}\ne q^S_{\mathfrak s}$,
$q^S_{\mathfrak r}=\operatorname{Bern}(x)$,
$q^C_{\mathfrak r}=\operatorname{Bern}(z)$, and
$p^\star_{\mathfrak r}=\operatorname{Bern}(p_R)$,
where all predictions have full support,
$x,z,p_R\in(0,1)$, $p_R>\bar u$, and $p_R\notin\{x,z\}$.
Ties favor safety. Only $\m q^C$ predicts safe observations correctly,
and neither model predicts risky observations correctly.
Risk is objectively optimal.

In the Bayesian benchmark, the agent chooses risk under $Q$
exactly when $x+(z-x)\pi_t(\m q^C)>\bar u$.
The parameter $h$ affects which class is adopted, but does not change
this action rule at a given history. Let $Q^\star:=\{\m p^\star\}$
be the true singleton competitor. Its score is
$S_{h,t}(Q^\star)=L_t(\m p^\star)$ for every $h$.
For finite $h$, write $a_h(w):=\min\{hw,1\}$, and set
$a_\infty(w):=1$. Whenever $z<\bar u<x$ or $x<\bar u<z$, define $K_\star:=(x-\bar u)/(x-z)$.

\begin{obs}[Unconstrained benchmark]
\label{prop:ba_unconstrained}
For every $h\in[1,\infty]$:
\begin{enumerate}
\item[\textnormal{(i)}]
If $x<\bar u<z$, $Q$ is not locally robust. Against $Q^\star$, the agent
eventually adopts $Q^\star$ permanently and chooses risk forever,
almost surely.
\item[\textnormal{(ii)}]
If $z<\bar u<x$, then against $Q^\star$ there is positive
probability that the agent never leaves $Q$ and always
chooses safety whenever $w>K_\star$ and
$a_h(w)\ge1/\gamma$.
Moreover, $Q$ is globally robust if $h<\infty$, $w>K_\star$, and
\begin{equation}\label{eq:ba_global_prior}
\frac{(1-w)K_\star}{w(1-K_\star)}
+\frac{h}{\gamma a_h(w)}<1.
\end{equation}
\end{enumerate}
\end{obs}

In part (i), safe observations eventually concentrate the posterior
on $\m q^C$, which favors risk. Continued use of $Q$ therefore generates
infinitely many risky observations, exposing both models' errors.
This argument applies to every selection rule in the family.
In part (ii), the same posterior concentration instead favors safety,
so it can preserve the observations on which $\m q^C$ is correct. Along histories with only safe actions,
$S_{h,t}(Q)\ge a_h(w)L_t(\m p^\star)$, so
$a_h(w)\ge1/\gamma$ prevents switching to $Q^\star$.
Increasing $h$ relaxes this condition, but need not
improve global robustness because it also increases
competitors' scores. Condition \eqref{eq:ba_global_prior}
holds for $w$ sufficiently close to one whenever
$h<\gamma$.

\paragraph{Capacity and misspecification concern.}
We now apply the decision rule of Section \ref{sec:dynamic} within
whichever class is adopted. Assign costs $k(\m q^S)=0$ and
$k(\m q^C)=1$, with $K\in(0,1)$, so
$\eta_t(\m q^C)=\min\{\pi_t(\m q^C),K\}$.
For each class, compute misspecification concern using the entire
history and the S-LLR rule in \eqref{eq:dynamic_rule}. Write
$\lambda_t^Q:=\big[\operatorname{LLR}(h_t;Q)+\psi_t\big]/(ct)$ for $t\ge1$, with
$\lambda_0^Q=0$, and define $\lambda_t^{Q'}$ analogously.
The same sensitivity parameter $c>0$ applies to both classes.
Under $Q$, risk is chosen exactly when
\begin{equation}\label{eq:ba_capacity_action}
\eta_t(\m q^C)v_{\lambda_t^Q}(\mathfrak r;\m q^C)
+\bigl(1-\eta_t(\m q^C)\bigr)
v_{\lambda_t^Q}(\mathfrak r;\m q^S)>\bar u.
\end{equation}
Capacity restricts recognition weights, while $h$
restricts scoring weights. Misspecification concern
continues to follow \eqref{eq:dynamic_rule}. Since $p_R>\bar u$, the expected safety adjustment is zero under every action distribution, i.e., $\psi(\alpha)=0$ for every $\alpha\in\Delta(A)$.
Misspecification concern leaves the safe value at $\bar u$ and weakly lowers the risky value. Along eventual exclusive safe play, $\lambda_t^Q\to0$
because $\m q^C$ predicts safe observations correctly
and the safe payoff equals the maxmin payoff.

The scoring and switching rules are unchanged.
Competing classes have nonempty feasible recognition
sets, so recognition is defined by
\eqref{eq:dynamic_recognition} even at boundary capacities. 
The following results hold for every $c>0$.

\begin{proposition}[Capacity and model robustness]
\label{prop:ba_capacity}
Fix $\gamma>1$ and $h\in[1,\infty]$.
\begin{enumerate}
\item[\textnormal{(i)}]
Suppose $x<\bar u<z$. For $K>K_\star$, $Q$ is not locally robust.
For $K\le K_\star$, it is globally robust if
$h<\gamma$ and $w>1/\gamma$, although it has no Bayesian self-confirming equilibrium (SCE; Definition \ref{def:ba_sce}).
\item[\textnormal{(ii)}]
Suppose $z<\bar u<x$. For $K<K_\star$, $Q$ is not locally robust.
For $K\ge K_\star$, it is globally robust if $h<\infty$,
$w>K_\star$, and \eqref{eq:ba_global_prior} holds.
\end{enumerate}
\end{proposition}
The failure of local robustness in part (i) holds even against arbitrarily small perturbations that preserve the original complexity costs. By contrast, sufficiently low capacity makes safety optimal at every history while $Q$ is adopted.

In part (ii), low capacity instead limits reliance on
the prediction that favors safety. Eventual exclusive
safe play would make misspecification concern vanish
and risk strictly preferable, so permanent use of $Q$
requires infinitely many risky observations, exposing
its errors. At higher capacity, a sufficiently
concentrated prior can sustain safety. Along safe histories, increasing $h$ lowers the prior
weight on $\m q^C$ sufficient to prevent a switch to
the true class from $1/\gamma$ to $1/(h\gamma)$. Global robustness need not
increase, since the competing rules  also gain scoring flexibility.

\begin{corollary}[Welfare and informative choices]
\label{cor:ba_welfare}
Consider competition against $Q^\star$, for any $h\in[1,\infty]$.
\begin{enumerate}
\item[\textnormal{(i)}]
If $x+(z-x)K>\bar u$, the agent eventually adopts $Q^\star$ permanently,
chooses risk forever, and obtains average realized payoff $p_R$,
almost surely.
\item[\textnormal{(ii)}]
If $x<\bar u<z$, $K\le K_\star$, and $a_h(w)\ge1/\gamma$,
the agent never leaves $Q$, always chooses safety, and receives $\bar u$
in every period. If $a_h(w)<1/\gamma$, the agent makes a first switch
to $Q^\star$ in finite time almost surely.
\item[\textnormal{(iii)}]
If risk is required at dates $t=j^2$, $j=1,2,\ldots$, the agent
eventually adopts $Q^\star$ permanently and obtains average realized
payoff $p_R$, almost surely, for every $K\in(0,1)$.
\end{enumerate}
\end{corollary}
Part (i) shows that when continued safe play would
eventually make risk attractive, the agent learns to
use the true class and attains the optimal payoff.
In part (ii), capacity sustains safety, while increasing
$h$ relaxes the prior condition preventing adoption of
the true class. At $h=\infty$, the agent remains with
$Q$ for every full-support prior, losing $p_R-\bar u$ per
period. When $a_h(w)<1/\gamma$, the first switch need
not be permanent. Part (iii) shows that sparse
experimentation restores optimal long-run behavior
without increasing capacity.

\paragraph{Model persistence.}
A misspecified model class may persist because the switching rule
underweights evidence against it, or because the agent's
choices prevent informative evidence from being generated.
These mechanisms call for different remedies: changing how
evidence is evaluated versus changing which observations
are collected. We distinguish them by characterizing how
scoring flexibility affects statistical discrimination
conditional on the frequency of informative choices.
Define $N_t(a):=\sum_{i\le t}\mathbbm 1\{a_i=a\}$.

\begin{proposition}[Information rates]
\label{prop:ba_information_rates}
Consider competition against $Q^\star$. For every
$t\ge0$ and $h,h'\in[1,\infty]$,
$|\log B_{h,t}-\log B_{h',t}|
\le\log(1/\min\{w,1-w\})$.
Under any history-dependent policy with
$N_t(\mathfrak r)/t\to\xi\in[0,1]$ a.s.,
\begin{equation}\label{eq:ba_information_rate}
\lim_{t\to\infty}\frac{\log B_{h,t}}{t}
=
\min_{\m q\in Q}
\left\{
(1-\xi)R(p^\star_{\mathfrak s}\Vert q_{\mathfrak s})
+\xi R(p^\star_{\mathfrak r}\Vert q_{\mathfrak r})
\right\}\quad \text{a.s.}
\end{equation}
for every fixed $h\in[1,\infty]$.
Moreover, the limit is strictly positive if and only if
$\xi>0$.
\end{proposition}

Proposition \ref{prop:ba_information_rates} shows that
scoring flexibility cannot change the asymptotic rate
of discrimination along a given sequence of action
frequencies. In particular, any positive frequency of
risk generates exponentially growing evidence against
$Q$, regardless of $h$. But a vanishing frequency need
not imply that informative observations cease altogether.

\begin{corollary}[Informative observations]
\label{cor:ba_information_rates}
For any history-dependent action policy and
$h\in[1,\infty]$, almost surely on
$\{N_t(\mathfrak r)\to\infty\}$,
$B_{h,t}\to\infty$, so $Q^\star$ is eventually
adopted permanently. Moreover, on
$\big\{N_t(\mathfrak r)\to\infty,\
N_t(\mathfrak r)/t\to0\big\}$,
$\log B_{h,t}=o(t)$ a.s.
\end{corollary}

Corollary \ref{cor:ba_information_rates} shows that informative observations need
not occur frequently to be decisive: infinitely many
risky observations suffice to overturn the misspecified
class, even when their frequency vanishes and evidence
grows only subexponentially. Thus, persistence against
$Q^\star$ requires that risky observations eventually
cease. The square-date intervention in Corollary
\ref{cor:ba_welfare}(iii) is one way to prevent this
informational trap. The result shows that capacity  affects model persistence by changing whether informative actions are taken infinitely often, rather than by slowing the rate at which those actions discriminate between classes.

\section{Conclusion}
\label{sec:conclusion}

We study the interaction between model complexity and model misspecification in repeated decision problems. Even when extensive data favor a particular model, limited cognitive capacity can prevent the agent from relying on it fully. Since choices also determine which evidence is generated, such constraints can alter risk-taking and allow misspecified models to survive by preventing the observations that would discredit them. 

We restrict attention to myopic choice to avoid the complications of combining misspecification concern with dynamic consistency. Extending the framework to forward-looking agents is a question for future research.

\appendix

\section{Appendix: Proofs of main results}
\label{app:new_proofs}
This appendix contains the proofs of the main results. The proofs of all lemmas, corollaries, and observations are in  Appendix \ref{app:oproof}.

\subsection{Proof of Proposition \ref{prop:capacity}}

Fix $K<\E_\pi[k]$ and suppress $\pi$ from the notation. On the open interval $(\min_{\m q\in Q}k(\m q),\E_\pi[k])$, Lemma \ref{prop:constrained_posterior} gives
$
\eta^K(\m q)
=
\frac{\pi(\m q)e^{-\mu^Kk(\m q)}}
{\sum_{\m r}\pi(\m r)e^{-\mu^Kk(\m r)}},$
for $\E_{\eta^K}[k]=K.$ 
For $\mu\ge0$, let $C(\mu):=\E_{\eta^\mu}[k]$.  As in the proof of
Lemma \ref{prop:constrained_posterior},
$C'(\mu)=-\Var_{\eta^\mu}(k)<0$.  Since
$C(\mu^K)=K$, implicit differentiation gives
$
\frac{d\mu^K}{dK}
=
-\frac{1}{\Var_{\eta^K}(k)}.$ Next,
$
\log\eta^K(\m q)
=
\log\pi(\m q)-\mu^Kk(\m q)
-\log\!\big(\sum_{\m r}\pi(\m r)e^{-\mu^Kk(\m r)}\big).$ 
Differentiating and using $\E_{\eta^K}[k]=K$ gives
$\frac{d}{dK}\log\eta^K(\m q)
=
-\frac{d\mu^K}{dK}\big(k(\m q)-K\big)
=
\frac{k(\m q)-K}{\Var_{\eta^K}(k)}.$ Finally, for any $w:Q\to\R$,
$\frac{d}{dK}\E_{\eta^K}[w]
=
\sum_{\m q}
w(\m q)\eta^K(\m q)
\frac{d}{dK}\log\eta^K(\m q)=
\frac{
\E_{\eta^K}[wk]
-K\E_{\eta^K}[w]
}{
\Var_{\eta^K}(k)}=
\frac{\Cov_{\eta^K}(k,w)}
{\Var_{\eta^K}(k)}$. \qed

\subsection{Proof of Theorem \ref{thm:complexity_filters}}

Fix $s\in I$.  Since $I$ is a binding region, Lemma 
\ref{prop:constrained_posterior} gives
$
\eta^s(\m q)
=
\frac{
\pi(\m q)e^{sg(\m q)-\mu^sk(\m q)}
}{
\sum_{\m r}
\pi(\m r)e^{sg(\m r)-\mu^sk(\m r)}
}$ and $\E_{\eta^s}[k]=K.$  
For arbitrary $(s,\mu)$, define the corresponding exponential-family distribution
$\eta_{\mu}^s$ and let
$C(s,\mu):=\E_{\eta_{\mu}^s}[k]-K$.  Finite summation permits
term-by-term differentiation.  Since
$
\frac{\partial}{\partial s}\eta_{\mu}^s(\m q)
=
\eta_{\mu}^s(\m q)
\big(g(\m q)-\E_{\eta_{\mu}^s}[g]\big),$ we obtain
$
\frac{\partial C}{\partial s}
=
\Cov_{\eta_{\mu}^s}(k,g).
$ 
Similarly,
$
\frac{\partial C}{\partial\mu}
=
-\Var_{\eta_{\mu}^s}(k).$ 
The variance is strictly positive because $\eta_{\mu}^s$ has full support and $k$ is
nonconstant.  The implicit-function theorem applied to
$C(s,\mu^s)=0$ therefore yields
$\frac{d\mu^s}{ds}
=
-\frac{\partial C/\partial s}{\partial C/\partial\mu}
=
\frac{\Cov_{\eta^s}(k,g)}
{\Var_{\eta^s}(k)}
=
\beta^s.$ Taking logarithms gives $\frac{d}{ds}\log\eta^s(\m q)
=
g(\m q)-\frac{d\mu^s}{ds}k(\m q)
-\E_{\eta^s}[g-\frac{d\mu^s}{ds}k].$ 
Since $\E_{\eta^s}[k]=K$ and $\frac{d\mu^s}{ds}=\beta^s$,
$
\frac{d}{ds}\log\eta^s(\m q)
=
g(\m q)-\E_{\eta^s}[g]
-\beta^s(k(\m q)-K)
=
g^s(\m q).
$

Fix $a,b\in A$ and let
$\Delta v(\m q):=v_\lambda(a;\m q)-v_\lambda(b;\m q)$.  Since
$\Delta v$ does not depend on $s$, $
\frac{d}{ds}
\left[
V_\lambda(a;\eta^s)-V_\lambda(b;\eta^s)
\right]
=
\sum_{\m q}
\Delta v(\m q)\eta^s(\m q)g^s(\m q).$ By construction, $\E_{\eta^s}[g^s]=0$, so the right-hand side is
$\Cov_{\eta^s}(\Delta v,g^s)$.
 \qed

\subsection{Proof of Proposition \ref{prop:recognition_invariance}}

For $m\in\{t,T\}$, Lemma \ref{prop:constrained_posterior} implies that there is a
multiplier $\mu_m\ge0$ such that
$\eta_m(\m q)
=
\frac{\pi_m(\m q)e^{-\mu_mk(\m q)}}
{\sum_{\m r}\pi_m(\m r)e^{-\mu_mk(\m r)}}.$
This formula also covers the slack case, in which $\mu_m=0$ and $\eta_m=\pi_m$.
Moreover, if $\mu_m>0$, then $\E_{\eta_m}[k]=K$.  Bayes rule gives
$\pi_T(\m q)
=
\frac{\pi_t(\m q)e^{g_{t,T}(\m q)}}
{\sum_{\m r}\pi_t(\m r)e^{g_{t,T}(\m r)}}.$

We first prove part \textnormal{(i)}.  Suppose $\E_{\eta_t}[k]<K$ and
$\eta_T=\eta_t$.  Since a positive multiplier would imply expected complexity $K$, we must have $\mu_t=\mu_T=0$.  Hence, $\eta_t=\pi_t$ and $\eta_T=\pi_T$, so $\pi_T=\pi_t$.  Since $\pi_t$ has full support, the preceding Bayes formula implies that $e^{g_{t,T}(\m q)}$ is constant in $\m q$, and therefore $g_{t,T}$ is constant on $Q$.  Conversely, if $g_{t,T}$ is constant, then $\pi_T=\pi_t$, and uniqueness of the recognition distribution implies $\eta_T=\eta_t$.

We next prove necessity in part \textnormal{(ii)}.  Suppose
$\E_{\eta_t}[k]=K$ and $\eta_T=\eta_t$.  Combining the two exponential-tilt representations above, there is a constant $C\in\R$ such that, for every $\m q\in Q$,
$g_{t,T}(\m q)
=
C+(\mu_T-\mu_t)k(\m q).$
Set $\zeta:=C$ and $\beta:=\mu_T-\mu_t$.  Since $\mu_T\ge0$,
$\beta\ge-\mu_t$.  This proves the required affine representation and also shows that whenever recognition is unchanged its slope must satisfy $\mu_T=\mu_t+\beta$.

For sufficiency, suppose $g_{t,T}=\zeta\mathbf 1_Q+\beta k$ with $\beta\ge-\mu_t$, and set
$\widetilde\mu:=\mu_t+\beta\ge0$.  The representation of $\eta_t$ above implies
$\pi_t(\m q)\propto\eta_t(\m q)e^{\mu_tk(\m q)}$, and therefore Bayes rule gives
$\pi_T(\m q)
\propto
\eta_t(\m q)e^{\widetilde\mu k(\m q)}.$
If $\widetilde\mu=0$, then $\pi_T=\eta_t$.  Since
$\E_{\eta_t}[k]=K$, the Bayesian target is feasible, so Lemma
\ref{prop:constrained_posterior} gives $\eta_T=\pi_T=\eta_t$ and $\mu_T=0$.

Suppose instead that $\widetilde\mu>0$.  We claim that
$\E_{\pi_T}[k]>K$.  For any full-support distribution $\nu$ on $Q$ and any $r>0$,
let $\nu^r(\m q)\propto\nu(\m q)e^{rk(\m q)}$.  Then,
$\E_{\nu^r}[k]-\E_\nu[k]=
\frac{\Cov_\nu(k,e^{rk})}{\E_\nu[e^{rk}]}.$
Furthermore,
$2\Cov_\nu(k,e^{rk})
=
\sum_{\m q,\m r}
\nu(\m q)\nu(\m r)
\big(k(\m q)-k(\m r)\big)
\big(e^{rk(\m q)}-e^{rk(\m r)}\big)>0,$
because $k$ is nonconstant and $\nu$ has full support.  Applying this with
$\nu=\eta_t$ and $r=\widetilde\mu$ yields
$\E_{\pi_T}[k]>\E_{\eta_t}[k]=K$. Thus, the complexity constraint binds for the target $\pi_T$.  Tilting $\pi_T$ by
$e^{-\widetilde\mu k}$ recovers $\eta_t$, which induces expected complexity $K$.
The uniqueness statement in Lemma \ref{prop:constrained_posterior} therefore gives
$\eta_T=\eta_t$ and $\mu_T=\widetilde\mu=\mu_t+\beta$.  This completes the proof.
\qed

\subsection{Proof of Proposition \ref{prop:frontier}}

Fix $\alpha$ and set
$D_{\m q}:=D(\m q;\alpha)$.  For any capacity
$x\in[k_{\min},k_{\max}]$, where
$k_{\min}:=\min_{\m q}k(\m q)$ and
$k_{\max}:=\max_{\m q}k(\m q)$, define
$F(x)
:=
\min_{\eta\in\Delta(Q):\E_\eta[k]\le x}\E_\eta[D].$
The feasible set is nonempty and compact, so the minimum exists.

We first establish two elementary properties of $F$.  If $x_2\ge x_1$, the feasible
set at $x_1$ is contained in that at $x_2$, so $F(x_2)\le F(x_1)$.
Thus $F$ is nonincreasing.  Next, if $\eta_i$ attains $F(x_i)$ for
$i=1,2$ and $\theta\in[0,1]$, then
$\eta_\theta:=\theta\eta_1+(1-\theta)\eta_2$ satisfies
$\E_{\eta_\theta}[k]
\le
\theta x_1+(1-\theta)x_2.$
Therefore,
$F(\theta x_1+(1-\theta)x_2)
\le
\E_{\eta_\theta}[D]
=
\theta F(x_1)+(1-\theta)F(x_2),$
so $F$ is convex.

Because $K\in(k_{\min},k_{\max})$, a finite convex function on this interval has a
supporting slope $s$ at $K$.  For completeness, let
$s_-:=\sup_{x<K}\frac{F(K)-F(x)}{K-x}$ and $
s_+:=\inf_{x>K}\frac{F(x)-F(K)}{x-K}.$
Convexity implies $s_-\le s_+$.  Any
$s\in[s_-,s_+]$ therefore satisfies
$F(x)\ge F(K)+s(x-K)$ for every $x$.  Since $F$ is nonincreasing,
$s\le0$.  Set $\rho:=-s\ge0$.

For each $\m q$, $\delta_{\m q}$ is feasible at capacity $k(\m q)$, so
$F(k(\m q))\le D_{\m q}$.  The supporting inequality therefore implies
$D_{\m q}
\ge
F(K)-\rho(k(\m q)-K),$
and hence
$\min_{\m q}[D_{\m q}+\rho k(\m q)]-\rho K
\ge F(K).$
Conversely, for every $\rho'\ge0$ and every feasible $\eta$,
\begin{align*}
\E_\eta[D]\ge
\E_\eta[D]+\rho'(\E_\eta[k]-K)=
\E_\eta[D+\rho'k]-\rho'K\ge
\min_{\m q}[D_{\m q}+\rho'k(\m q)]-\rho'K.
\end{align*}
Taking the minimum over feasible $\eta$ gives the reverse inequality for every
$\rho'$.  Thus,
$F_K(\alpha)
=
\max_{\rho\ge0}
\left\{
\min_{\m q}
[D(\m q;\alpha)+\rho k(\m q)]
-\rho K
\right\},$
proving part \textnormal{(i)}.  The maximizing set is nonempty.  It is also bounded:
if $\m q^0$ minimizes $k$, the dual objective is no larger than
$D_{\m q^0}+\rho(k_{\min}-K)$, which tends to $-\infty$ as
$\rho\to\infty$.

For part \textnormal{(ii)}, every $D_{\m q}\ge d(\alpha)$, so
$F_K(\alpha)\ge d(\alpha)$.  Suppose
$K\ge\min_{\m q\in Q(\alpha)}k(\m q)$.  Choose
$\m q^\star\in Q(\alpha)$ satisfying $k(\m q^\star)\le K$.
Then, $\delta_{\m q^\star}$ is feasible and achieves
$d(\alpha)$, so $F_K(\alpha)=d(\alpha)$.

Conversely, suppose $F_K(\alpha)=d(\alpha)$ and let
$\eta\in\mathcal H_K(\alpha)$.  Then,
$0
=
\E_\eta[D]-d(\alpha)
=
\sum_{\m q}
\eta(\m q)\big(D_{\m q}-d(\alpha)\big).$
Every summand is nonnegative, so
$\operatorname{supp}(\eta)\subset Q(\alpha)$.  Consequently,
$K
\ge
\E_\eta[k]
\ge
\min_{\m q\in Q(\alpha)}k(\m q).$
This proves part \textnormal{(ii)}.

For part \textnormal{(iii)}, fix
$\eta\in\mathcal H_K(\alpha)$ and
$\rho\in\mathcal R_K(\alpha)$. By weak duality and primal-dual optimality:
\begin{align*}
F_K(\alpha)
=
\E_\eta[D]\ge
\E_\eta[D]+\rho(\E_\eta[k]-K)=
\E_\eta[D+\rho k]-\rho K\ge
\min_{\m q}[D_{\m q}+\rho k(\m q)]-\rho K=
F_K(\alpha).
\end{align*}
All inequalities are equalities.  Equality in the first yields
$\rho(\E_\eta[k]-K)=0$.  Equality in the second implies
$\eta(\m q)>0\Longrightarrow
\m q\in
\arg\min_{\m r}[D_{\m r}+\rho k(\m r)].$
For an alternative capacity $x$, weak duality at the same $\rho$ gives
$F(x)
\ge
\min_{\m q}[D_{\m q}+\rho k(\m q)]-\rho x
=
F(K)-\rho(x-K).$
Thus, $-\rho$ is a subgradient of $F$ at $K$.

For part \textnormal{(iv)}, a linear objective attains its minimum at an extreme point
of the polytope $\mathcal E_K$.  We show that every extreme point has support at most
two.  Suppose a feasible $\eta$ has three positive coordinates corresponding to
$\m q_1,\m q_2,\m q_3$.  The homogeneous system
$h_1+h_2+h_3=0$ and $
k(\m q_1)h_1+k(\m q_2)h_2+k(\m q_3)h_3=0$
has a nonzero solution because it has three unknowns and at most two independent
equations.  Extend $h$ by zeros on all other coordinates.  For sufficiently small
$\varepsilon>0$, both $\eta+\varepsilon h$ and $\eta-\varepsilon h$ are
nonnegative, have total mass one, and have the same expected complexity as $\eta$.
Therefore, both are feasible and
$\eta
=
\frac12(\eta+\varepsilon h)
+
\frac12(\eta-\varepsilon h),$
so $\eta$ is not extreme.  Thus, every extreme point has support at most two, proving
existence of an optimal solution with support at most two. If the optimizer is unique, it must be extreme.  Otherwise it could be expressed as a
nontrivial convex combination of two distinct feasible points; linearity of the objective
would then imply that both points are also optimal, contradicting uniqueness. \qed

\subsection{Proof of Theorem \ref{thm:long_run_frontier}}

Fix a history-dependent action rule and let
$\mathcal F_t$ be the sigma-field generated by the
realized history through date $t$. For every $\m q\in Q$, define
$X_s^{\m q}
:=
\log q_{a_s}(y_s)
-
\E_{p^\star_{a_s}}[\log q_{a_s}(Y)].$
Because $a_s$ is $\mathcal F_{s-1}$-measurable and
$y_s\sim p^\star_{a_s}$ conditional on $\mathcal F_{s-1}$,
$\E[X_s^{\m q}|\mathcal F_{s-1}]=0.$
For every $(a,y)\in A\times Y$, also define
$M_t^{a,y}
:=
\sum_{s=1}^t
\mathbbm{1}\{a_s=a\}
\big(
\mathbbm{1}\{y_s=y\}-p^\star_a(y)
\big).$
The same conditional property implies that $(M_t^{a,y})_{t\ge0}$ is a
martingale.  Since $A$, $Y$, and $Q$ are finite and every structured model
has full support, the increments of all of these martingales are uniformly
bounded.  Azuma-Hoeffding therefore gives summable tail bounds for
$t^{-1}\sum_{s=1}^tX_s^{\m q}$ and $t^{-1}M_t^{a,y}$.  Applying
Borel-Cantelli for $\varepsilon=1/m$, $m\in\mathbb N$, and using finiteness
of $Q$, $A$, and $Y$, there is an event $\Omega_0$ with
$\mathbb P(\Omega_0)=1$ such that, simultaneously for every
$\m q\in Q$ and $(a,y)\in A\times Y$,
$\frac1t\sum_{s=1}^tX_s^{\m q}\rightarrow0$ and $
\frac{M_t^{a,y}}{t}\rightarrow0.$ Intersect $\Omega_0$ with the probability-one event
of Lemma \ref{lem:safety_statistic}, continuing to
denote the resulting event by $\Omega_0$. Fix an arbitrary realized path in $\Omega_0$ on which $\alpha_t$ converges,
write $\alpha$ for its limit, and suppress dependence on this path. 

\paragraph{Step 1: asymptotic likelihoods.}
For every $\m q\in Q$,
$\frac1t\sum_{s=1}^tX_s^{\m q}\rightarrow0.$
Thus,
$\frac1t\log L_t(\m q)=
\sum_{a\in A}
\alpha_t(a)
\E_{p^\star_a}[\log q_a(Y)]
+o(1)\rightarrow
-\sum_a\alpha(a)H(p^\star_a)-D(\m q;\alpha).$
Let $\mathcal H^\star(\alpha):=\sum_a\alpha(a)H(p^\star_a)$.  Then,
$\frac1t\log L_t(\m q)
\rightarrow
-\mathcal H^\star(\alpha)-D(\m q;\alpha).$

\paragraph{Step 2: posterior rates.}
Since $Q$ is finite,
$\frac1t
\log\left(
\sum_{\m r}\pi_0(\m r)L_t(\m r)
\right)
-
\max_{\m r}\frac1t\log L_t(\m r)
\rightarrow0.$
Therefore,
$-\frac1t\log\pi_t(\m q)
\rightarrow
D(\m q;\alpha)-d(\alpha)$
for every $\m q$, uniformly over the finite set $Q$.

Define
$f_t(\eta):=\frac1tR(\eta\Vert\pi_t)$ and $f(\eta)
:=
\E_\eta[D(\cdot;\alpha)]-d(\alpha).$
Then, we can rewrite $f_t$ as $f_t(\eta)
=
\frac1t\sum_{\m q}\eta(\m q)\log\eta(\m q)
-
\sum_{\m q}\eta(\m q)\frac1t\log\pi_t(\m q).$
The first term is uniformly bounded in absolute value by $\log|Q|/t$, while the
second converges uniformly by the preceding posterior-rate result.  Hence,
$f_t\to f$ uniformly on $\Delta(Q)$,
and in particular on $\mathcal E_K$.

Since $\eta_t$ minimizes $f_t$ over $\mathcal E_K$, every accumulation point
$\bar\eta$ of $\eta_t$ minimizes $f$.  Indeed, along any subsequence
$\eta_{t_j}\to\bar\eta$ and for every fixed $\eta\in\mathcal E_K$,
$f_{t_j}(\eta_{t_j})\le f_{t_j}(\eta).$
Uniform convergence and continuity give
$f(\bar\eta)\le f(\eta)$ for every feasible $\eta$.  The minimizers of $f$ are
 $\mathcal H_K(\alpha)$.  This proves part \textnormal{(ii)}.

\paragraph{Step 3: misspecification concern.}
By Lemma \ref{lem:safety_statistic},
$\lambda_t-\tau(\alpha_t)\to0$ a.s.
Since $Q$ is finite and $D(\m q;\alpha)$ is linear
in $\alpha$ for every $\m q\in Q$, the function
$d(\alpha)=\min_{\m q\in Q}D(\m q;\alpha)$
is continuous. Since $\psi(\alpha)$ is linear
in $\alpha$, \eqref{eq:population_misspecification}
implies that $\tau(\alpha)$ is continuous.
Hence, $\alpha_t\to\alpha$ implies
$\lambda_t\to\tau(\alpha)$, proving part
\textnormal{(i)}.

\paragraph{Step 4: the asymptotic complexity price.}
The number $\mu_t$ is the multiplier for the constraint when minimizing
$R(\eta\Vert\pi_t)$.  Therefore $\mu_t/t$ is the multiplier for the same
constraint when minimizing $f_t(\eta)=t^{-1}R(\eta\Vert\pi_t)$.

For $r\ge0$, define
$g_t(r)
:=
\inf_{\eta\in\Delta(Q)}
\left\{
f_t(\eta)+r(\E_\eta[k]-K)
\right\}.$ The Gibbs variational identity gives
$g_t(r)
=
-\frac1t
\log
\big(
\sum_{\m q}
\pi_t(\m q)e^{-trk(\m q)}
\big)
-rK.$
If the capacity constraint binds, the first-order condition for the concave dual problem
is  $\E_{\eta_t}[k]=K$, and its maximizer is $r=\mu_t/t$.  If the
constraint is slack, $\mu_t=0$, which is the dual maximizer.  Hence,
$g_t(\mu_t/t)
=
f_t(\eta_t)
=
\max_{r\ge0}g_t(r).$

We first prove boundedness.  Choose $\m q^0$ with
$k(\m q^0)=k_{\min}<K$.  Since $\delta_{\m q^0}$ is feasible,
$0
\le
f_t(\eta_t)
=
g_t(\mu_t/t)
\le
f_t(\delta_{\m q^0})
+\frac{\mu_t}{t}(k_{\min}-K).$
Thus,
$0\le\frac{\mu_t}{t}
\le
\frac{f_t(\delta_{\m q^0})}{K-k_{\min}}.$
Step 2 implies
$f_t(\delta_{\m q^0})
=-(1/t)\log\pi_t(\m q^0)$ converges to the finite value
$D(\m q^0;\alpha)-d(\alpha)$. Thus, $\mu_t/t$ is bounded. Uniform convergence $f_t\to f$ implies uniform convergence of $g_t$ on every bounded
interval to
$g(r)
=
\inf_{\eta\in\Delta(Q)}
\left\{
f(\eta)+r(\E_\eta[k]-K)
\right\}
=
\min_{\m q}
\left[
D(\m q;\alpha)-d(\alpha)
+r(k(\m q)-K)
\right].$
Let $t_j$ be a subsequence such that
$\mu_{t_j}/t_j\to\bar\rho$.  For every fixed $r\ge0$,
$g_{t_j}(\mu_{t_j}/t_j)\ge g_{t_j}(r).$
Passing to the limit gives $g(\bar\rho)\ge g(r)$.  Thus $\bar\rho$ maximizes
$g$.  Adding the constant $d(\alpha)$ shows that the maximizers of $g$ are 
$\mathcal R_K(\alpha)$.  Every accumulation point of $\mu_t/t$ therefore belongs
to $\mathcal R_K(\alpha)$, proving part \textnormal{(iii)}.

If $\mathcal H_K(\alpha)$ is a singleton, part \textnormal{(ii)} implies
$\eta_t\to\eta^\star$.  If $\mathcal R_K(\alpha)$ is also a singleton, boundedness
and part \textnormal{(iii)} imply $\mu_t/t\to\rho^\star$.  With part
\textnormal{(i)}, this proves the final statement. \qed

\subsection{Proof of Proposition \ref{prop:equilibrium}}

Since $Q$ is finite and $D(\m q;\alpha)$ is linear
in $\alpha$ for every $\m q\in Q$,
$d(\alpha)=\min_{\m q\in Q}D(\m q;\alpha)$ is
continuous. Since $\psi(\alpha)$ is linear in $\alpha$,
the function $\tau(\alpha)$ defined in
\eqref{eq:population_misspecification} is continuous.

We first prove existence.  For every $\alpha\in\Delta(A)$,
$\mathcal H_K(\alpha)$ is nonempty, compact, and convex: it is the set of minimizers
of the continuous function $\eta\mapsto\E_\eta[D(\cdot;\alpha)]$, which is linear
in $\eta$, over the nonempty compact convex set $\mathcal E_K$.  Moreover,
$\mathcal H_K$ has closed graph.  To see this, suppose
$\alpha_n\to\alpha$, $\eta_n\to\eta$, and
$\eta_n\in\mathcal H_K(\alpha_n)$.  For every
$\widetilde\eta\in\mathcal E_K$,
$\E_{\eta_n}[D(\cdot;\alpha_n)]
\le
\E_{\widetilde\eta}[D(\cdot;\alpha_n)].$
Since $Q$ is finite, the map
$(\alpha,\eta)\mapsto\E_\eta[D(\cdot;\alpha)]$ is continuous, so taking limits gives
$\E_\eta[D(\cdot;\alpha)]\le
\E_{\widetilde\eta}[D(\cdot;\alpha)]$ for every
$\widetilde\eta\in\mathcal E_K$.  Thus,
$\eta\in\mathcal H_K(\alpha)$.  Because the domain and range are compact metric
spaces and the values are compact, the closed-graph property implies upper
hemicontinuity of $\mathcal H_K$.

For $(\alpha,\eta)\in\Delta(A)\times\mathcal E_K$, define
$\mathcal B(\alpha,\eta)
:=
\Delta(\operatorname{BR}_{\tau(\alpha)}(\eta)),$
where $\Delta(B)$ denotes the set of distributions in $\Delta(A)$ supported on
$B\subseteq A$.  The map $(\lambda,\eta)\mapsto V_\lambda(a;\eta)$ is continuous
for every $a$.  For $\lambda>0$, this follows directly from the multiplier formula;
at $\lambda=0$, finiteness of $Y$ gives
$\sum_yq_a(y)e^{-\lambda u(a,y)}
=1-\lambda\E_{q_a}[u(a,y)]+O(\lambda^2)$, and therefore
$v_\lambda(a;\m q)=\E_{q_a}[u(a,y)]+O(\lambda)$ as
$\lambda\downarrow0$.  Since $A$ and $Q$ are finite, the extension is continuous
jointly in all arguments.  Together with continuity of $\tau$, this implies continuity
of $(\alpha,\eta)\mapsto V_{\tau(\alpha)}(a;\eta)$.

The correspondence $\mathcal B$ has nonempty, compact, convex values.  It also has
closed graph.  Indeed, suppose
$(\alpha_n,\eta_n)\to(\alpha,\eta)$,
$\beta_n\to\beta$, and
$\beta_n\in\mathcal B(\alpha_n,\eta_n)$.  If $\beta(a)>0$, then
$\beta_n(a)>0$ for all sufficiently large $n$, so
$a\in\operatorname{BR}_{\tau(\alpha_n)}(\eta_n)$.  Hence, for every $b\in A$,
$V_{\tau(\alpha_n)}(a;\eta_n)\ge
V_{\tau(\alpha_n)}(b;\eta_n)$.  Passing to the limit yields
$V_{\tau(\alpha)}(a;\eta)\ge V_{\tau(\alpha)}(b;\eta)$ for every $b \in A$, so
$a\in\operatorname{BR}_{\tau(\alpha)}(\eta)$.  Thus,
$\operatorname{supp}(\beta)\subseteq
\operatorname{BR}_{\tau(\alpha)}(\eta)$, which means
$\beta\in\mathcal B(\alpha,\eta)$.  As above, compact-valuedness and closed graph
imply upper hemicontinuity.

Define the correspondence
$\Phi:\Delta(A)\times\mathcal E_K\rightrightarrows
\Delta(A)\times\mathcal E_K$ by
$\Phi(\alpha,\eta):=\mathcal B(\alpha,\eta)\times
\mathcal H_K(\alpha)$.  Its domain is nonempty, compact, and convex, and $\Phi$ has
nonempty, compact, convex values and is upper hemicontinuous.  Kakutani's fixed-point
theorem therefore yields $(\alpha^\star,\eta^\star)$ such that
$\alpha^\star\in\mathcal B(\alpha^\star,\eta^\star)$ and
$\eta^\star\in\mathcal H_K(\alpha^\star)$.  Proposition
\ref{prop:frontier}\textnormal{(i)} implies that
$\mathcal R_K(\alpha^\star)$ is nonempty.  Choose any
$\rho^\star\in\mathcal R_K(\alpha^\star)$ and set
$\tau^\star=\tau(\alpha^\star)$.  Then,
$(\alpha^\star,\eta^\star,\tau^\star,\rho^\star)$ satisfies all three clauses of
Definition \ref{def:ccre}.  This proves part \textnormal{(i)}.

For part \textnormal{(ii)}, work on the probability-one event on which Theorem
\ref{thm:long_run_frontier} holds, and suppose
$\alpha_t\to\alpha$ and $\eta_t\to\eta$.  Theorem
\ref{thm:long_run_frontier}\textnormal{(i)} gives
$\lambda_t\to\tau:=\tau(\alpha)$.  Theorem
\ref{thm:long_run_frontier}\textnormal{(ii)} says that
every accumulation point of $\eta_t$ belongs to $\mathcal H_K(\alpha)$; since
$\eta_t\to\eta$, this gives $\eta\in\mathcal H_K(\alpha)$.  Part
\textnormal{(iii)} says that $\mu_t/t$ is bounded and every one of its accumulation
points belongs to $\mathcal R_K(\alpha)$.

For any action $a$ chosen infinitely often, there is
a sequence $t_j\to\infty$ such that $a_{t_j+1}=a$.
By \eqref{eq:dynamic_criterion},
$V_{\lambda_{t_j}}(a;\eta_{t_j})
\ge V_{\lambda_{t_j}}(b;\eta_{t_j})$ for every $b\in A$.
Taking limits and using continuity of $V$ gives
$a\in\operatorname{BR}_{\tau(\alpha)}(\eta)$.
Since $A$ is finite, $a_{t} \in \operatorname{BR}_{\tau(\alpha)}(\eta)$ for  all sufficiently large $t$.  In particular,
$\operatorname{supp}(\alpha)\subseteq
\operatorname{BR}_{\tau(\alpha)}(\eta)$.
Therefore, for every accumulation point $\rho$ of
$\mu_t/t$, $(\alpha,\eta,\tau(\alpha),\rho)$ satisfies
Definition \ref{def:ccre}, proving part \textnormal{(ii)}.

For part \textnormal{(iii)}, suppose $\alpha_t\to\alpha$
and $\mathcal H_K(\alpha)=\{\eta^\star\}$. By Theorem \ref{thm:long_run_frontier}\textnormal{(ii)}
and $\mathcal H_K(\alpha)=\{\eta^\star\}$, every
accumulation point of $\eta_t$ equals $\eta^\star$.
Since $\mathcal E_K$ is compact, $\eta_t\to\eta^\star$,
so part \textnormal{(ii)} applies.
If also $\mathcal R_K(\alpha)=\{\rho^\star\}$,
Theorem \ref{thm:long_run_frontier}\textnormal{(iii)}
implies $\mu_t/t\to\rho^\star$, since the sequence
is bounded and has a unique accumulation point.
Together with $\alpha_t\to\alpha$ and
$\lambda_t\to\tau(\alpha)$, 
$(\alpha_t,\eta_t,\lambda_t,\mu_t/t)
\to(\alpha,\eta^\star,\tau(\alpha),\rho^\star)$,
proving part \textnormal{(iii)}. \qed

\subsection{Proof of Proposition \ref{thm:independent}}

Fix $\alpha\in\Delta(A)$ and suppose
$Q(\alpha)=\{\m q^\star\}$.  Then,
$D(\m q;\alpha)>D(\m q^\star;\alpha)=d(\alpha)$ $\forall \m q\neq\m q^\star.$

For part \textnormal{(i)}, suppose $k(\m q^\star)<K$.  Then,
$\delta_{\m q^\star}\in\mathcal E_K$ and
$\E_{\delta_{\m q^\star}}[D(\cdot;\alpha)]
=d(\alpha).$
Since $d(\alpha)$ is the unconstrained minimum of $D(\cdot;\alpha)$, it follows that
$F_K(\alpha)=d(\alpha)$ and hence $G_K(\alpha)=0$.

Moreover, if $\eta\neq\delta_{\m q^\star}$, then $\eta$ assigns positive
probability to at least one $\m q\neq\m q^\star$.  Therefore,
$\E_\eta[D(\cdot;\alpha)]
=
\eta(\m q^\star)d(\alpha)
+
\sum_{\m q\neq\m q^\star}
\eta(\m q)D(\m q;\alpha)
>d(\alpha).$
Thus,
$\mathcal H_K(\alpha)=\{\delta_{\m q^\star}\}$. Let $\rho\in\mathcal R_K(\alpha)$.  Proposition
\ref{prop:frontier}\textnormal{(iii)}, applied to the primal optimizer
$\delta_{\m q^\star}$, gives
$\rho\big(k(\m q^\star)-K\big)=0.$
Since $k(\m q^\star)<K$, we obtain $\rho=0$.  Proposition
\ref{prop:frontier}\textnormal{(i)} guarantees that
$\mathcal R_K(\alpha)$ is nonempty, so
$\mathcal R_K(\alpha)=\{0\}$.

For part \textnormal{(ii)}, suppose $k(\m q^\star)>K$.  Since
$Q(\alpha)=\{\m q^\star\}$,
$K<\min_{\m q\in Q(\alpha)}k(\m q).$
Proposition \ref{prop:frontier}\textnormal{(ii)} therefore implies
$G_K(\alpha)>0$.  Also
$\delta_{\m q^\star}\notin\mathcal E_K$, and hence
$\delta_{\m q^\star}\notin\mathcal H_K(\alpha)$.

Finally, suppose by contradiction that
$0\in\mathcal R_K(\alpha)$.  The dual objective at $\rho=0$ equals
$\min_{\m q\in Q}D(\m q;\alpha)=d(\alpha).$
But duality gives the optimal dual value
$F_K(\alpha)=d(\alpha)+G_K(\alpha)>d(\alpha)$.
Thus, $\rho=0$ cannot be dual optimal.  Hence,
$\mathcal R_K(\alpha)\subset\R_{++}$. \qed

\subsection{Proof of Theorem \ref{thm:complexity_cycles}}

Let
$Z_t
:=
\log\frac{\pi_t(\m q^C)}{\pi_t(\m q^S)}.$
Bayes rule gives
$Z_t
=
Z_0+
\sum_{i=1}^t
\log
\frac{q^C_{a_i}(y_i)}
{q^S_{a_i}(y_i)}.$
For each $a\in\{\mathfrak{s},\mathfrak{r}\}$,
\begin{align*}
\E_{p_a^\star}
\left[
\log\frac{q_a^C(y)}{q_a^S(y)}
\right]
=
\sum_y
p_a^\star(y)
\log\frac{q_a^C(y)}{q_a^S(y)}=
R(p_a^\star\Vert q_a^S)
-
R(p_a^\star\Vert q_a^C)=
\gamma_a
>
0.
\end{align*}
Define
$M_t
:=
\sum_{i=1}^t
\big[
\log
\frac{q^C_{a_i}(y_i)}
{q^S_{a_i}(y_i)}
-
\gamma_{a_i}
\big].$
Conditional on the history before date $i$, the action $a_i$ is known and
$y_i$ has distribution $p^\star_{a_i}$.  Hence, $(M_t)$ is a martingale.
Its increments are uniformly bounded because $A$, $Y$, and $Q$ are finite and every
structured model has full support.  Thus, for every $\varepsilon>0$,
Azuma-Hoeffding implies
$\mathbb P(|M_t|\ge\varepsilon t)
\le
2e^{-C_\varepsilon t}$
for some $C_\varepsilon>0$.  The right-hand side is summable, so
Borel-Cantelli yields
$\frac{M_t}{t}\longrightarrow0$ 
a.s. Consequently,
$\frac{Z_t}{t}
=
\alpha_t(\mathfrak{s})\gamma_{\mathfrak{s}}
+
\alpha_t(\mathfrak{r})\gamma_{\mathfrak{r}}
+
o(1)
\ge
\min\{\gamma_{\mathfrak{s}},\gamma_{\mathfrak{r}}\}+o(1).$
Hence,
$Z_t\to+\infty$ and
$\pi_t(\m q^C)\to1$ a.s.

For $\eta\in\Delta(Q)$ write
$w:=\eta(\m q^C)$.  Feasibility is equivalent to $
k_S+w(k_C-k_S)\le K,$ or equivalently $
w\le\chi_K.$  Writing $p_t:=\pi_t(\m q^C)$, the objective
$R(\eta\Vert\pi_t)$ is
$r_{p_t}(w)
=
w\log\frac{w}{p_t}
+
(1-w)\log\frac{1-w}{1-p_t}.$
For $w\in(0,p_t)$,
$r_{p_t}'(w)
=
\log
\frac{w(1-p_t)}
{p_t(1-w)}
<0.$
Since $p_t\to1$ and $\chi_K<1$, almost surely there is a finite random date after
which $p_t>\chi_K$.  From that date onward the unique constrained minimizer is
$w=\chi_K$.  Thus,
$\eta_t(\m q^C)=\chi_K$
for all sufficiently large $t$ a.s.

We next characterize $\lambda_t$.
Since $d_a^C<d_a^S$ for $a=\mathfrak s,\mathfrak r$,
$\m q^C$ uniquely minimizes $D(\m q;\alpha)$ for
every $\alpha\in\Delta(A)$.
Hence, $d(\alpha)=
\alpha(\mathfrak s)d_{\mathfrak s}^C+
\alpha(\mathfrak r)d_{\mathfrak r}^C$.
Since $\psi(\alpha)$ is linear in $\alpha$,
$\tau(\alpha)=
\alpha(\mathfrak s)\lambda_{\mathfrak s}+
\alpha(\mathfrak r)\lambda_{\mathfrak r}$.
Lemma \ref{lem:safety_statistic} therefore gives
$\lambda_t-[\lambda_{\mathfrak s}+
\alpha_t(\mathfrak r)
(\lambda_{\mathfrak r}-\lambda_{\mathfrak s})]
\to0$ a.s.

If $\min_yu(\mathfrak r,y)\ge\bar u$, then
$v_\lambda(\mathfrak r;\m q)>\bar u$ for every
$\m q\in Q$ and every finite $\lambda\ge0$:
the risky payoff is nonconstant and the structured
models have full support, so the strict inequality
follows from expected utility at $\lambda=0$ and
from the exponential representation of $v_\lambda$
at $\lambda>0$.
Hence, risk is strictly optimal at every history,
$x_K=1$, and $\alpha_t(\mathfrak r)=1$ for every
$t\ge1$. Moreover,
$\Delta_K(\lambda_{\mathfrak r})>0$,
$\lambda_t\to\lambda_{\mathfrak r}$, and,
using the eventual identity
$\eta_t(\m q^C)=\chi_K$ and
Lemma \ref{prop:constrained_posterior},
$\mu_t/t=
[Z_t/t-t^{-1}\log(\chi_K/(1-\chi_K))]/
(k_C-k_S)
\to\gamma_{\mathfrak r}/(k_C-k_S)>0$.
This proves all the conclusions in this case.

Suppose hereafter that
$\min_yu(\mathfrak r,y)<\bar u$.
Then, $\underline u=\bar u$,
$\psi(\delta_{\mathfrak s})=0$, and
$\psi(\delta_{\mathfrak r})\ge0$.
Since $d_{\mathfrak r}^C>d_{\mathfrak s}^C$,
we have $\lambda_{\mathfrak r}>\lambda_{\mathfrak s}$. Set $x_t:=\alpha_t(\mathfrak{r})$ and define
$\Phi_K(x)
:=
\Delta_K\!\left(
\lambda_{\mathfrak{s}}
+
x(\lambda_{\mathfrak{r}}-\lambda_{\mathfrak{s}})
\right),$ for $x\in[0,1].$
By Lemma \ref{lem:delta_monotone} and
$\lambda_{\mathfrak{r}}>\lambda_{\mathfrak{s}}$,
$\Phi_K$ is continuous and strictly decreasing.

Once $\eta_t(\m q^C)=\chi_K$, the risky--safe value difference is
$\Delta_K(\lambda_t)$.  The preceding approximation implies that $\lambda_t$
is eventually contained in a compact neighborhood of
$[\lambda_{\mathfrak s},\lambda_{\mathfrak r}]$.
Uniform continuity of $\Delta_K$ on this neighborhood
therefore gives
$\Delta_K(\lambda_t)=\Phi_K(x_t)+e_t,$ $e_t\to0$.
Let
$b_{t+1}:=\mathbbm 1\{a_{t+1}=\mathfrak{r}\}$.  Optimality implies
$b_{t+1}=1$ if $\Phi_K(x_t)+e_t>0,$ and $b_{t+1}=0$ if $\Phi_K(x_t)+e_t<0$, with arbitrary tie breaking at equality.  Moreover,
$x_{t+1}
=
x_t+
\frac{b_{t+1}-x_t}{t+1}.$ 

Suppose first that $\Phi_K(0)<0$.  Then
$\Phi_K(x)\le\Phi_K(0)<0$ for every $x\in[0,1]$.  Since $e_t\to0$,
$b_{t+1}=0$ for all sufficiently large $t$.  Hence,
$a_t=\mathfrak{s}$ eventually and $x_t\to0$.

Suppose next that $\Phi_K(0)=0$.  Fix $\varepsilon\in(0,1)$.  Strict decrease gives
$\Phi_K(\varepsilon)<0$.  Choose $T$ such that, for every $t\ge T$,
$|e_t|<-\frac12\Phi_K(\varepsilon)$ and $
\frac1{t+1}<\varepsilon.$
Then, $x_t\ge\varepsilon$ implies $b_{t+1}=0$.  The process must eventually fall
below $\varepsilon$: otherwise $b_{t+1}=0$ forever after $T$, which would imply
$x_t\to0$, a contradiction.  Once $x_t<\varepsilon$, one update can raise it by
less than $\varepsilon$; whenever it is at least $\varepsilon$, the next update
moves it downward.  Hence, $\limsup_{t\to\infty}x_t\le2\varepsilon.$
As $\varepsilon$ is arbitrary, $x_t\to0$. The case $\Phi_K(1)\ge0$ is symmetric and yields $x_t\to1$; when
$\Phi_K(1)>0$, risky play is eventually permanent.

Finally, suppose
$\Phi_K(0)>0>\Phi_K(1).$
There is a unique $x_K\in(0,1)$ with $\Phi_K(x_K)=0$.
Fix $\varepsilon>0$ such that
$[x_K-2\varepsilon,x_K+2\varepsilon]\subset(0,1),$
and define
$m_\varepsilon
:=
\min\{
\Phi_K(x_K-\varepsilon),
-\Phi_K(x_K+\varepsilon)
\}>0.$
For all sufficiently large $t$,
$|e_t|<\frac12m_\varepsilon$ and $\frac1{t+1}<\varepsilon.$
Hence,
$x_t\le x_K-\varepsilon$ implies $b_{t+1}=1$, whereas
$x_t\ge x_K+\varepsilon$ implies $b_{t+1}=0$.

The process must eventually enter
$[x_K-\varepsilon,x_K+\varepsilon]$.  Otherwise, after some date it would
remain on one side of this interval: repeated risky choices below it would force
$x_t\to1$, while repeated safe choices above it would force $x_t\to0$, both
contradictions.  A transition from one side to the other without entering the interval
is impossible once the one-period change is smaller than $\varepsilon$.
After the process enters the inner interval, a single update can move it by less than
$\varepsilon$, so it remains in
$[x_K-2\varepsilon,x_K+2\varepsilon]$; outside the inner interval the choice rule
points back toward it.  Therefore,
$\limsup_{t\to\infty}|x_t-x_K|
\le2\varepsilon.$
Letting $\varepsilon\downarrow0$ gives $x_t\to x_K$.

It remains to identify the limiting complexity pressure.  Once capacity binds,
the constrained-posterior formula gives
$\frac{\eta_t(\m q^C)}
{\eta_t(\m q^S)}
=
\frac{\pi_t(\m q^C)}
{\pi_t(\m q^S)}
e^{-\mu_t(k_C-k_S)}.$
Using the eventual identity $\eta_t(\m q^C)=\chi_K$,
$\log\frac{\chi_K}{1-\chi_K}
=
Z_t-\mu_t(k_C-k_S).$
Hence,
$\frac{\mu_t}{t}
=
\frac{1}{k_C-k_S}
\big[
\frac{Z_t}{t}
-
\frac1t
\log\frac{\chi_K}{1-\chi_K}
\big].$
Since $x_t\to x_K$ and $M_t/t\to0$,
$\frac{Z_t}{t}
\longrightarrow
(1-x_K)\gamma_{\mathfrak{s}}
+
x_K\gamma_{\mathfrak{r}}.$
Therefore,
$\frac{\mu_t}{t}
\longrightarrow
\frac{
(1-x_K)\gamma_{\mathfrak{s}}
+
x_K\gamma_{\mathfrak{r}}
}{
k_C-k_S
}
>0.$
By the definition of a $\lambda$-cycle and the three cases above, a cycle occurs if and
only if
$\Delta_K(\lambda_{\mathfrak{s}})
>
0
>
\Delta_K(\lambda_{\mathfrak{r}}).$
\qed

\bibliography{ref}
\bibliographystyle{apalike}
\newpage \renewcommand{\thepage}{OA-\arabic{page}} \setcounter{page}{1}

{\noindent\LARGE\bf For Online Publication}

\section{Normative Benchmark}\label{app:safety}

We extend the normative analysis of \citet[Section 3.1]{lanzani25}
to complexity-constrained agents. Throughout this appendix, a decision
problem is a tuple $(u,A,Y,Q,k,K,\pi_0)$ satisfying
Assumption \ref{ass:full_support}, with a full-support prior
and $\mathcal E_K\ne\varnothing$. Let $H:=\bigcup_{t\ge0}H_t$ and consider a rule
$\Lambda:H\to\R_+$ determining misspecification concern.
A policy is \emph{$\Lambda$-optimal} if, at every history
$h_t$, it selects an action from
$\operatorname{BR}_{\Lambda(h_t)}(\eta_t)$,
where $\eta_t$ is given by \eqref{eq:dynamic_recognition}.
The policy and the true model $\m p^\star$ induce a
probability measure on infinite histories.

\begin{definition}
\label{def:epsilon_safety}
Let $\varepsilon>0$. A rule $\Lambda$ is
\emph{$\varepsilon$-safe} for a decision problem if,
for every $\Lambda$-optimal policy and every
$\m p^\star\in\Delta(Y)^A$,
$\liminf_{t\to\infty}\frac{1}{t}\sum_{s=1}^tu(a_s,y_s)
\ge\underline u-\varepsilon$ a.s.
\end{definition}

This is the safety requirement in \citet[][Definition 2]{lanzani25}:
the agent approximately attains the maxmin payoff
regardless of the extent of misspecification.
Unlike an unconstrained agent, a complexity-constrained
agent can fail this requirement even when
$\m p^\star\in Q$, because the accurate model need
not receive enough recognition weight to govern choice.

Our second requirement adapts \citeauthor{lanzani25}'s
(\citeyear{lanzani25}) notion of self-confirmation to
complexity-constrained recognition. For $\varepsilon\ge0$,
let $\operatorname{SCE}_{\varepsilon,K}(\m p^\star)$
be the set of actions $a\in A$ for which there exists
$\eta\in\mathcal E_K$ satisfying
$\E_\eta[D_a]\le\varepsilon$ and
$U(a;\eta)+\varepsilon\ge\max_{b\in A}U(b;\eta)$.\footnote{Recall from Section \ref{sec:dynamic} that $D_a(\m q)=R(p^\star_a\Vert q_a)$ and $U(a;\eta)=\sum_{\m q\in Q}\eta(\m q)
\E_{q_a}[u(a,y)]$.}
When this set is nonempty, define
$G_{\varepsilon,K}(\m p^\star)=
\min_{a\in\operatorname{SCE}_{\varepsilon,K}(\m p^\star)}
\E_{p^\star_a}[u(a,y)]$.

An action in $\operatorname{SCE}_{\varepsilon,K}(\m p^\star)$
is approximately optimal given a feasible recognition
distribution that approximately predicts its outcomes.
At $\varepsilon=0$, every structured model receiving
positive recognition weight must predict these outcomes
correctly.

\begin{definition}
\label{def:epsilon_self_confirming}
Let $\varepsilon>0$. A rule $\Lambda$ is
\emph{$\varepsilon$-self-confirming under cognitively
feasible correct specification} for a decision problem if
there exists $\delta>0$ such that, for every
$\m p^\star\in\Delta(Y)^A$ satisfying
$\min_{\eta\in\mathcal E_K}\max_{a\in A}
\E_\eta[D_a]<\delta$,
the set $\operatorname{SCE}_{\varepsilon,K}(\m p^\star)$
is nonempty and, for every $\Lambda$-optimal policy,
$\liminf_{t\to\infty}\frac{1}{t}\sum_{s=1}^tu(a_s,y_s)
\ge G_{\varepsilon,K}(\m p^\star)-\varepsilon$ a.s.
\end{definition}

This requirement bounds the losses induced by
misspecification concern relative to the possible
self-confirming learning traps when accurate predictions
are cognitively feasible. The antecedent is satisfied,
in particular, when a sufficiently accurate structured
model has complexity no greater than capacity.
The feasibility qualification is essential: the presence
of an accurate but cognitively infeasible model in $Q$ is 
 not sufficient. 

\begin{theorem}[Normative benchmark]
\label{prop:safety_normative}
\hfill
\begin{enumerate}
\item[\textnormal{(i)}]
For every $\varepsilon>0$ and every decision problem,
there exists $\bar c>0$ such that, for every
$c\in(0,\bar c]$, S-LLR is $\varepsilon$-safe.
The bound $\bar c$ can be chosen independently of
$K$, $k$, and $\pi_0$.

\item[\textnormal{(ii)}]
There exist $\varepsilon>0$ and a decision problem
for which no rule $\Lambda$ is both $\varepsilon$-safe
and $\varepsilon$-self-confirming under cognitively
feasible correct specification if either\footnote{We use
the conventions $0/0:=0$, $0\log0:=0$, and
$z/0:=+\infty$ for $z>0$.}
\[
\lim_{t\to\infty}
\frac{\Lambda(h_t)}
{[\operatorname{LLR}(h_t;Q)+\psi_t]/t}=0
\quad\text{or}\quad
\lim_{t\to\infty}
\frac{[\operatorname{LLR}(h_t;Q)+\psi_t]/t}
{\Lambda(h_t)}=0
\]
holds for every $(h_t)_{t\ge1}\in\prod_{t\ge1}H_t$.
The same conclusion holds when the two rate conditions
use $\operatorname{LLR}(h_t;Q)/t$ in place of
$[\operatorname{LLR}(h_t;Q)+\psi_t]/t$.

\item[\textnormal{(iii)}]
Suppose $\m p^\star\in Q$ and $k(\m p^\star)\le K$.
Then, for every $c>0$ and every S-LLR-optimal policy,
$\lambda_t\to0$ a.s. Moreover, a.s.,
$\limsup_{t\to\infty}\alpha_t(a)>0
\implies a\in\operatorname{SCE}_{0,K}(\m p^\star)$
for every $a\in A$.
\item[\textnormal{(iv)}]
For every $\varepsilon>0$, every decision problem,
and every $c>0$, there exists $\delta>0$ such that,
for every $\m p^\star$ satisfying
$\min_{\eta\in\mathcal E_K}\max_{a\in A}
\E_\eta[D_a]<\delta$,
every S-LLR-optimal policy almost surely chooses only
actions in $\operatorname{SCE}_{\varepsilon,K}(\m p^\star)$
at every sufficiently large date.
In particular, S-LLR is $\varepsilon$-self-confirming
under cognitively feasible correct specification.
\end{enumerate}
\end{theorem}
The proofs of all results from this appendix are in Appendix \ref{app:sproof}. Parts \textnormal{(i)}, \textnormal{(ii)}, and \textnormal{(iv)} extend the normative analysis of \citet[][Theorem 1]{lanzani25}. Part \textnormal{(i)} establishes approximate safety even when capacity is severely limited. Part \textnormal{(ii)} shows that both excessively lenient and excessively demanding adjustments of misspecification concern can defeat the normative properties. Part \textnormal{(iii)} shows that under cognitively feasible correct specification, misspecification concern vanishes for every sensitivity parameter, and every action chosen with positive asymptotic frequency along some subsequence is self-confirming given a feasible recognition distribution. Part \textnormal{(iv)} establishes eventual approximate self-confirmation whenever sufficiently accurate predictions are cognitively feasible. Together, these results provide a normative justification for adjusting misspecification concern in proportion to the safety-adjusted average likelihood shortfall.

To relate these results more directly to \citet{lanzani25},
let $\operatorname{SCE}^L_\varepsilon(\m p^\star)$
denote the unconstrained $\varepsilon$-self-confirming correspondence:
the set of actions $a$ for which there exists
$\nu\in\Delta(Q)$ supported on
$\{\m q\in Q:D_a(\m q)\le\varepsilon\}$
such that $U(a;\nu)+\varepsilon\ge\max_bU(b;\nu)$.
For every $\varepsilon>0$, a sufficiently small
$\varepsilon'>0$ satisfies
$\operatorname{SCE}_{\varepsilon',K}(\m p^\star)
\subseteq\operatorname{SCE}^L_\varepsilon(\m p^\star)$
uniformly over true models and capacities. To see this, choose $\varepsilon'<\varepsilon$ with
$\varepsilon'(1+2B/\varepsilon)\le\varepsilon$.
A supporting recognition distribution assigns mass at most
$\varepsilon'/\varepsilon$ to models with
$D_a(\m q)>\varepsilon$. Conditioning on the remaining
models increases each payoff difference
$U(b;\cdot)-U(a;\cdot)$ by at most
$2B\varepsilon'/\varepsilon$, so $a$ remains an
$\varepsilon$-best reply.
Thus, when $K\ge\max_{\m q\in Q}k(\m q)$,\footnote{Note that at the boundary $K=\min_{\boldsymbol q\in Q}k(\boldsymbol q)$, the recognition distribution in Lemma \ref{prop:constrained_posterior} equals the Bayesian posterior conditional on the minimum-complexity models. The exponential-tilt representation then holds in the limit as the multiplier tends to infinity, unless all models have minimum complexity.}
part \textnormal{(iv)} also delivers
\citeauthor{lanzani25}'s (\citeyear{lanzani25}) original
self-confirming guarantee under the original
approximate-specification condition.

In contrast, the LLR rule need not be $\varepsilon$-safe
under a binding capacity constraint, even under correct specification. S-LLR addresses this failure by tying additional caution to realized payoff shortfalls, rather than to statistical errors induced by limited capacity alone.
\begin{obs}\label{lem:convergence}
Suppose outcomes satisfy $y_i\sim p^\star_{a_i}$ independently of the past given $a_i$,
and the empirical action frequencies converge to $\alpha$ a.s. Then,
$\psi_t/t\to\psi(\alpha)$ a.s. If in addition
$\operatorname{LLR}(h_t;Q)/t\to d(\alpha)$ a.s., then $\lambda_t\to\tau(\alpha)$.
\end{obs}

\section{Convergence of action frequencies and actions}
\label{app:action_convergence}

Proposition \ref{prop:equilibrium} characterizes
equilibrium outcomes conditional on convergence.
Here we provide sufficient conditions for convergence.
Following \citet{lanzani25}, we distinguish convergence
of empirical action frequencies from convergence of
actions, since frequencies can converge even when
the agent continues to switch between actions.

For $\alpha\in\Delta(A)$, let $\mathcal A_K(\alpha):=
\bigcup_{\eta\in\mathcal H_K(\alpha)}
\operatorname{BR}_{\tau(\alpha)}(\eta)$ denote the
\emph{asymptotic best-reply correspondence}.
Thus, $\mathcal A_K(\alpha)$ consists of the actions
that are optimal under some recognition distribution
attaining the fit-complexity frontier given $\alpha$.
For $\mathcal S\subseteq A$, let $\Delta(\mathcal S)$
denote the distributions in $\Delta(A)$ supported on
$\mathcal S$.
Starting from $\mathcal S_0=A$, recursively define
$\mathcal S_{j+1}:=
\mathcal S_j\cap
\bigcup_{\alpha\in\Delta(\mathcal S_j)}
\mathcal A_K(\alpha)$, and let
$\mathcal S^\star:=\bigcap_{j\ge0}\mathcal S_j$.

We call a complexity-constrained $c$-robust equilibrium
\emph{pure} if its action distribution is a Dirac
measure and \emph{mixed} otherwise. When $A=\{a,b\}$,
write $\alpha^x:=(1-x)\delta_a+x\delta_b$ for
$x\in[0,1]$, and define
$\mathcal M_K:=
\big\{x\in(0,1):
\exists\eta\in\mathcal H_K(\alpha^x)
\textnormal{ s.t. }
\{a,b\}\subseteq
\operatorname{BR}_{\tau(\alpha^x)}(\eta)\big\}.$
By Proposition \ref{prop:frontier}\textnormal{(i)},
$\mathcal M_K$ is the set of frequencies of action
$b$ in mixed complexity-constrained $c$-robust
equilibria. All the proofs of results in this appendix are in Appendix \ref{app:convergence_proofs}.

\begin{proposition}[Convergence]
\label{prop:action_convergence}
For every policy, the following hold a.s.:
\begin{enumerate}
\item[\textnormal{(i)}]
The sequence $(\mathcal S_j)_{j\ge0}$ stabilizes
after finitely many iterations at a nonempty set
$\mathcal S^\star$, and $a_t\in\mathcal S^\star$
at every sufficiently large date. In particular,
if $\mathcal S^\star=\{a^\star\}$, then
$a_t=a^\star$ at every sufficiently large date.

\item[\textnormal{(ii)}]
Suppose $A=\{a,b\}$. If $\mathcal M_K$ has empty
interior, then $\alpha_t$ converges to the action
distribution of a complexity-constrained $c$-robust
equilibrium.

\item[\textnormal{(iii)}]
Suppose $A=\{a,b\}$ and $\mathcal M_K=\varnothing$.
If $\mathcal A_K(\delta_{a'})=\{a'\}$ 
$\forall a'\in A$ such that
$a'\in\mathcal A_K(\delta_{a'})$, then there is
an action $a^\star$ such that $a_t=a^\star$
at every sufficiently large date.
\end{enumerate}
\end{proposition}

Proposition \ref{prop:action_convergence}\textnormal{(i)}
exploits the fact that an action may be optimal early on but cease to be
optimal as the agent accumulates observations.
For an action to be chosen infinitely often, it must
be optimal under some recognition distribution and
misspecification concern that can arise in the long
run. The first iteration rules out actions for which
this is impossible, regardless of the frequencies
with which the agent chooses the available actions.
Since these actions are chosen only finitely often,
they cannot affect the limiting empirical frequencies.
The next iteration can therefore restrict attention
to data generated by the remaining actions, which
may rule out further actions. Repeating this argument
successively restricts the set of actions that the
agent can choose infinitely often. Since the action set is finite, this procedure terminates, and if only
one action remains, the agent eventually chooses
this action forever.

A simple way through which this mechanism operates is when limited
capacity makes some actions unattractive because models that favor them are too costly to use, while observations from the remaining actions favor models under which the same action is optimal. Formally, suppose $a^\star$ yields a constant payoff $\bar u$
and there is a set $\mathcal S\subseteq A$ containing
$a^\star$ such that
$\max_{\eta\in\mathcal E_K}U(b;\eta)<\bar u$
for every $b\notin\mathcal S$.
Suppose further that
$\mathcal H_K(\delta_a)=\{\nu\}$ for every
$a\in\mathcal S$ and that
$V_{\underline\lambda_{\mathcal S}}(b;\nu)<\bar u$
for every $b\in\mathcal S\setminus\{a^\star\}$,
where $\underline\lambda_{\mathcal S}:=
\min_{a\in\mathcal S}\tau(\delta_a)$.
Then, $\mathcal S_2=\mathcal S^\star=\{a^\star\}$.
Indeed, since $V_\lambda(b;\eta)\le U(b;\eta)$
and $V_\lambda(a^\star;\eta)=\bar u$,
$\mathcal S_1\subseteq\mathcal S$.
For every $\alpha\in\Delta(\mathcal S)$, $\nu$
uniquely minimizes $\E_\eta[D(\cdot;\alpha)]$
over $\mathcal E_K$ because it uniquely minimizes
$\E_\eta[D_a]$ for every $a\in\mathcal S$.
Moreover, concavity of $\tau$ implies
$\tau(\alpha)\ge\underline\lambda_{\mathcal S}$,
so the monotonicity of robust values in $\lambda$
gives $\mathcal A_K(\alpha)=\{a^\star\}$.
Hence, $\mathcal S_2\subseteq\{a^\star\}$, and
part \textnormal{(i)} ensures that $\mathcal S_2$
is nonempty.

These conditions do not require that the same model
best explain the consequences of every action or
that every model favor $a^\star$. A simple case is
when a model with complexity at most $K$ fits the
outcomes of each action in $\mathcal S$ strictly
better than every other model and favors $a^\star$
under expected utility. Other models may predict
the consequences of actions outside $\mathcal S$
better and recommend different actions.
However, since the agent does not choose the
actions outside $\mathcal S$, she does not obtain
the observations that would favor those models.
The next example illustrates how this can lead to the successive elimination
of actions.

\begin{example}[Iterated elimination of actions]\normalfont
Let $A=\{\mathfrak s,\mathfrak r,b\}$ and
$Y=\{0,1\}$, with $u(\mathfrak s,y)=0.5$
and $u(\mathfrak r,y)=u(b,y)=y$.
There are two structured models
$Q=\{\m q^S,\m q^C\}$ with
$k(\m q^S)=0$, $k(\m q^C)=1$, and $K=0.5$.
Their predictions and the true outcome probabilities
are
\[
\begin{array}{c|ccc}
 & q_a^S(1) & q_a^C(1) & p_a^\star(1)\\ \hline
\mathfrak s & 0.5 & 0.6 & 0.5\\
\mathfrak r & 0.4 & 0.7 & 0.4\\
b           & 0.1 & 0.8 & 0.8
\end{array}
\]
Thus, the simple model correctly predicts outcomes
under $\mathfrak s$ and $\mathfrak r$, while the
complex model correctly predicts outcomes under $b$. Since feasibility requires $\eta(\m q^C)\le0.5$,
$U(b;\eta)\le0.45<0.5$, so $b$ cannot be optimal
at any history, regardless of misspecification
concern. However, if all observations were generated
by $b$, Corollary \ref{cor:two_model} would give
$\mathcal H_K(\delta_b)=\{\nu\}$ with
$\nu(\m q^C)=0.5$.
Since $d(\delta_b)=\psi(\delta_b)=0$,
$\tau(\delta_b)=0$ and
$V_0(\mathfrak r;\nu)=0.55>0.5$, so $\mathfrak r\in\mathcal A_K(\delta_b)$.
Also, $\mathfrak s\in\mathcal A_K(\delta_{\mathfrak s})$ because $\m q^S$ uniquely fits best under $\mathfrak s$
and is feasible. Thus,
$\mathcal S_1=\{\mathfrak s,\mathfrak r\}$.

For every $\alpha\in\Delta(\mathcal S_1)$,
$\m q^S$ uniquely minimizes $D(\m q;\alpha)$
and $k(\m q^S)<K$, so Proposition
\ref{thm:independent} gives
$\mathcal H_K(\alpha)=\{\delta_{\m q^S}\}$.
Since
$V_{\tau(\alpha)}(\mathfrak r;\delta_{\m q^S})
\le0.4<0.5$, $\mathcal S_2=\mathcal S^\star=\{\mathfrak s\}$.
By Proposition \ref{prop:action_convergence}(i),
the agent will eventually choose $a_t=\mathfrak s$
a.s., for every $c>0$ and every full-support prior.

Although $b$ is objectively optimal and is strictly
optimal under the complex model when $\lambda=0$,
limited capacity prevents the agent from relying
sufficiently on that model to choose $b$. This
eliminates the observations under which the complex
model fits better, and therefore eliminates
$\mathfrak r$ as well. The agent eventually settles on safety because the capacity constraint rules out the action $b$ whose outcomes would have favored the complex model.\hfill$\bigtriangledown$
\end{example}

\par Proposition \ref{prop:action_convergence}\textnormal{(ii)} establishes convergence of
empirical frequencies without requiring convergence
of recognition or uniqueness of equilibrium.
Nonconvergence would require repeated crossings of
every frequency in some nondegenerate interval,
which would require each of those frequencies
to be supported by a mixed equilibrium.
Proposition \ref{prop:action_convergence}\textnormal{(iii)} distinguishes convergence
to a pure frequency from eventual stabilization:
strict optimality under every fit-optimal recognition
distribution at a pure equilibrium also rules out
infinitely recurring choices with vanishing
empirical frequency.

The following corollary provides sufficient
conditions for part \textnormal{(ii)} in the
risky-safe problems of Definition
\ref{def:risky_safe}, while allowing for an arbitrary
finite model class $Q$. Let $\mathcal V_K$ denote
the set of vertices of the polytope $\mathcal E_K$,
which is determined by the triplet $(Q,k,K)$.

\begin{corollary}[Risky-safe problems]
\label{cor:general_risky_safe_convergence}
Suppose $A=\{\mathfrak s,\mathfrak r\}$ is a
risky-safe problem and:
\begin{enumerate}
\item[\textnormal{(i)}]
For every distinct $\nu,\nu'\in\mathcal V_K$,
$(\E_\nu[D_{\mathfrak s}],\E_\nu[D_{\mathfrak r}])
\ne
(\E_{\nu'}[D_{\mathfrak s}],\E_{\nu'}[D_{\mathfrak r}])$.

\item[\textnormal{(ii)}]
For every $\m q\in Q$,
$D_{\mathfrak r}(\m q)+\psi(\delta_{\mathfrak r})
\ne
D_{\mathfrak s}(\m q)+\psi(\delta_{\mathfrak s})$.
\end{enumerate}
Then $\mathcal M_K$ is finite, and for every policy,
$\alpha_t$ converges a.s. to the action distribution
of a complexity-constrained $c$-robust equilibrium.
\end{corollary}

Condition \textnormal{(i)} requires that distinct
extreme feasible recognition distributions differ
in their statistical accuracy under at least one
action. Condition \textnormal{(ii)} requires that
safe and risky observations differ in their
contributions to safety-adjusted misspecification
concern under every structured model.
Together, these conditions rule out intervals of
indifference between risk and safety. Unlike in
Theorem \ref{thm:complexity_cycles}, the best-fitting
model can depend on the frequency of risky play,
and risky observations need not worsen model fit.

Obtaining similarly tractable conditions for
frequency convergence with three or more actions
is more challenging: the same frequency of an
action can be associated with different distributions
of the remaining actions and hence with different
recognition distributions and misspecification
concerns. Moreover, different actions can be
asymptotically optimal under different fit-optimal
recognition distributions without being jointly
optimal under any one of them. We therefore leave the
analysis of these general dynamics for future work.

\section{Additional details for Section \ref{sec:ba}}
\label{app:ba_definitions}

\subsection{Definitions}

\paragraph{Bayesian benchmark.}
For $\zeta\in\Delta(Q)$, define the Bayesian best-reply correspondence
\begin{equation}\label{eq:app_bayesian_best_reply}
\operatorname{BR}^B_Q(\zeta):=
\arg\max_{a\in A}\sum_{\m q\in Q}\zeta(\m q)
\sum_{y\in Y}q_a(y)u(a,y).
\end{equation}
Fix a rule choosing one maximizing action whenever several actions
maximize the criterion. In the risky--safe example, this rule chooses
$\mathfrak s$ when the two actions have equal value.

\begin{definition}[Self-confirming equilibrium]
\label{def:ba_sce}
A strategy $\sigma\in\Delta(A)$ is a Bayesian
\emph{self-confirming equilibrium} (SCE) of $Q$ if there is
$\zeta\in\Delta(Q)$ such that
$\operatorname{supp}(\sigma)\subseteq\operatorname{BR}^B_Q(\zeta)$
and $q_a=p_a^\star$ for every
$a\in\operatorname{supp}(\sigma)$ and
$\m q\in\operatorname{supp}(\zeta)$.
The belief $\zeta$ supports the SCE. 
\end{definition}

\paragraph{Scoring, updating, and choice.}
Fix $h\in[1,\infty]$ and $\gamma>1$.
Both classes use every observation to compute their likelihoods,
Bayesian posteriors, and scores from \eqref{eq:class_score}.
Let $m_t\in\{Q,Q'\}$ denote the class used to choose $a_{t+1}$,
with $m_0=Q$. For $t\ge1$, set
$B_{h,t}:=S_{h,t}(Q')/S_{h,t}(Q)$ and
\begin{equation}\label{eq:app_switching_rule}
m_t=\begin{cases}
Q',&B_{h,t}>\gamma,\\
Q,&B_{h,t}<1/\gamma,\\
m_{t-1},&1/\gamma\le B_{h,t}\le\gamma.
\end{cases}
\end{equation}
For each $\widetilde Q\in\{Q,Q'\}$, let
$\pi_t^{\widetilde Q}$ denote its Bayesian posterior,
$\eta_t^{\widetilde Q}$ its recognition distribution from
\eqref{eq:dynamic_recognition}, and
$\lambda_t^{\widetilde Q}$ its concern from \eqref{eq:dynamic_rule}.
The Bayesian benchmark selects $a_{t+1}$ from
$\operatorname{BR}^B_{m_t}(\pi_t^{m_t})$; the extension selects it from
$\operatorname{BR}_{\lambda_t^{m_t}}(\eta_t^{m_t})$.
In the extension, both classes use the same capacity $K$ and
sensitivity $c>0$.

\paragraph{Persistence and robustness.}
In the benchmark, an admissible competing class is any finite nonempty
class with full-support predictions and a full-support prior.
In the extension it also has a nonnegative complexity-cost assignment
$k'$ satisfying $\min_{\m q\in Q'}k'(\m q)\le K$.
Robustness is evaluated for the specified action rule and fixed
$h,\gamma$, and, in the extension, fixed $K,c$ and the initial
class's complexity costs.

\begin{definition}[Persistence and robustness]
\label{def:ba_robustness}
The class $Q$ \emph{persists} against $Q'$ at the specified priors
and costs if
$\mathbb{P}\big(\exists T<\infty:\ m_t=Q\,\forall t\ge T\big)>0$.
It is \emph{globally robust} at $\pi_0$ if it persists against every
admissible competing class, every full-support prior on that class,
and, in the extension, every admissible assignment of its costs.

Equip $Y$ with the discrete metric and let
$d_P$ be the resulting Prokhorov metric on distributions.
On structured models, set
$d_D(\m q,\m r):=\max_{a\in A}d_P(q_a,r_a)$,
and define the Hausdorff distance by
$d_H(Q,Q'):=\max\big\{\max_{\m q\in Q}\min_{\m r\in Q'}d_D(\m q,\m r),\ 
\max_{\m r\in Q'}\min_{\m q\in Q}d_D(\m q,\m r)\big\}.$
Associate the prior on $Q$ with the probability measure
$\nu_Q:=\sum_{\m q\in Q}\pi_0(\m q)\delta_{\m q}$,
where $\delta_{\m q}$ is the point mass at $\m q$;
define $\nu_{Q'}$ analogously.
The class $Q$ is \emph{locally robust} at $\pi_0$ if some
$\varepsilon>0$ ensures persistence against every admissible
competitor, prior, and cost assignment satisfying
$d_H(Q,Q')<\varepsilon$ and
$d_P(\nu_Q,\nu_{Q'})<\varepsilon$,
where the last metric is induced by $d_D$.
\end{definition}

The definitions use the criteria in
\citet[Definitions 1--3 and Appendix A.2]{ba26} with the scoring and
action rules specified above.

\section{Omitted Proofs}\label{app:oproof}

\subsection{Proofs for Sections \ref{sec:static}--\ref{sec:dynamic}}
\begin{proof}[Proof of Lemma \ref{prop:constrained_posterior}]
By \eqref{eq:bayes-variational}, \eqref{eq:recognition_problem}
is equivalent to minimizing $R(\eta\Vert\pi_n)$ over
$\mathcal E_K$. Since $\pi_n$ has full support, this
objective is continuous and strictly convex on the
nonempty compact convex set $\mathcal E_K$, so its
minimizer exists and is unique.

If $\E_{\pi_n}[k]\le K$, then $\pi_n\in\mathcal E_K$ and
$R(\pi_n\Vert\pi_n)=0$.  Nonnegativity of relative entropy implies that
$\eta_n^K=\pi_n$.
Suppose instead that $\E_{\pi_n}[k]>K$.  For $\mu\ge0$, define
$\eta^\mu(\m q)
:=
\frac{\pi_n(\m q)e^{-\mu k(\m q)}}
{\sum_{\m r}\pi_n(\m r)e^{-\mu k(\m r)}}$
and let $C(\mu):=\E_{\eta^\mu}[k]$.  Direct differentiation gives
$C'(\mu)
=
-\Var_{\eta^\mu}(k).$
Every $\eta^\mu$ has full support, and $k$ is nonconstant, so
$\Var_{\eta^\mu}(k)>0$.  Thus, $C$ is strictly decreasing.  Moreover,
$C(0)=\E_{\pi_n}[k]>K$, while
$\lim_{\mu\to\infty}C(\mu)=\min_{\m q}k(\m q)<K$.  Hence, there is a
unique $\mu_n>0$ such that $C(\mu_n)=K$.

To verify optimality, for every feasible $\eta$,
$\log\frac{\eta(\m q)}{\pi_n(\m q)}
=
\log\frac{\eta(\m q)}{\eta^{\mu_n}(\m q)}
-\mu_nk(\m q)
-\log\big(\sum_{\m r}\pi_n(\m r)e^{-\mu_nk(\m r)}\big).$
Summing with respect to $\eta$ and subtracting the same expression evaluated at
$\eta^{\mu_n}$ gives
$R(\eta\Vert\pi_n)-R(\eta^{\mu_n}\Vert\pi_n)
=
R(\eta\Vert\eta^{\mu_n})
+
\mu_n\big(K-\E_\eta[k]\big)
\ge0.$
Equality requires $R(\eta\Vert\eta^{\mu_n})=0$, hence
$\eta=\eta^{\mu_n}$.  Therefore,
$\eta_n^K=\eta^{\mu_n}$ is the unique recognition distribution.\end{proof}

\begin{proof}[Proof of Corollary \ref{cor:evidence_complexity}]
For every $\m q$,
$
D_a(\m q)
=
R(p^\star_a\Vert q_a)
=
\sum_y p^\star_a(y)\log p^\star_a(y)
-
\sum_y p^\star_a(y)\log q_a(y),$
so $
\E_{p^\star_a}[\log q_a(Y)]
=
-H(p^\star_a)-D_a(\m q).$ The entropy term is constant across structured models.  Therefore, we have
$
\Cov_{\eta^s}
\left(
k,
\E_{p^\star_a}[\log q_a(Y)]
\right)
=
-\Cov_{\eta^s}(k,D_a).$  The first identity in Theorem \ref{thm:complexity_filters} gives the result. \end{proof}

\begin{proof}[Proof of Corollary \ref{prop:lag}]
Let
$\Delta k:=k(\m q^C)-k(\m q^S)>0$.  Since $Q$ contains  two models and
$k(\m q^C)\ne k(\m q^S)$, every function on $Q$ has a unique representation
$\zeta\mathbf 1_Q+\beta k$.  For $g_{t,T}$, the coefficient on $k$ is
$\beta
=
\frac{g_{t,T}(\m q^C)-g_{t,T}(\m q^S)}{\Delta k}
=
B_{t,T}.$
If $B_{t,T}\ge-\mu_t$, Proposition \ref{prop:recognition_invariance}(ii) gives
$\eta_T=\eta_t$ and $\mu_T=\mu_t+B_{t,T}$.

Suppose now $B_{t,T}<-\mu_t$. Since
$\E_{\eta_t}[k]=K$ and there are only two models,
$\eta_t(\m q^C)
=
\bar x
:=
\frac{K-k(\m q^S)}{\Delta k}.$
The exponential-tilt representation at date $t$ gives
$\frac{\pi_t(\m q^C)}{\pi_t(\m q^S)}
=
\frac{\eta_t(\m q^C)}{\eta_t(\m q^S)}e^{\mu_t\Delta k}.$
Using Bayes rule between $t$ and $T$ and the definition of $B_{t,T}$,
$\frac{\pi_T(\m q^C)}{\pi_T(\m q^S)}
=
\frac{\eta_t(\m q^C)}{\eta_t(\m q^S)}
e^{(\mu_t+B_{t,T})\Delta k}
<
\frac{\eta_t(\m q^C)}{\eta_t(\m q^S)}.$
Hence, $\pi_T(\m q^C)<\bar x=\eta_t(\m q^C)$, so
$\E_{\pi_T}[k]<K$.  Lemma \ref{prop:constrained_posterior} implies that the
constraint is slack at $T$, so $\eta_T=\pi_T$ and $\mu_T=0$.  The strict inequality
above then gives
$\eta_T(\m q^C)<\eta_t(\m q^C)$. \end{proof}

\begin{proof}[Proof of Corollary \ref{cor:frozen_recognition_switch}]
Suppose first that $g_{t,T}=\zeta\mathbf 1_Q+\beta k$
with $\beta\ge-\mu_t$.
Proposition \ref{prop:recognition_invariance}\textnormal{(ii)}
gives $\eta_T=\eta_t$.
Let $S_t:=\max_{\m q\in Q}L_t(\m q)$.
By \eqref{eq:dynamic_rule},
$ct\lambda_t=\log U_t-\log S_t+\psi_t$.
Since $L_T(\m q)=L_t(\m q)e^{\zeta+\beta k(\m q)}$,
$\log S_T=\log S_t+\zeta+\mathcal M_t(\beta)$.
Also, $\log U_T=\log U_t+\Delta^U_{t,T}$.
Thus,
$cT\lambda_T=\log U_T-\log S_T+\psi_T
=ct\lambda_t+\Delta^U_{t,T}-\zeta-\mathcal M_t(\beta)
+\psi_T-\psi_t$,
which proves
$\lambda_T=\widetilde\lambda_{t,T}(\zeta,\beta)$.

Since $\eta_T=\eta_t$ and the best replies at both
histories are unique, the choice rule
\eqref{eq:dynamic_criterion} implies that $a_{t+1}=a$
exactly when $\lambda_t\in\mathcal C_a(\eta_t)$,
and that $a_{T+1}=b$ exactly when
$\widetilde\lambda_{t,T}(\zeta,\beta)
\in\mathcal C_b(\eta_t)$.
Thus, the two actions differ if and only if these
misspecification concerns belong to strict best-reply
regions corresponding to distinct actions.

Conversely, suppose the agent switches actions while
$\eta_T=\eta_t$. Since $\E_{\eta_t}[k]=K$,
Proposition \ref{prop:recognition_invariance}\textnormal{(ii)}
yields unique coefficients $\zeta\in\R$ and
$\beta\ge-\mu_t$ such that
$g_{t,T}=\zeta\mathbf 1_Q+\beta k$.
The calculation above gives
$\lambda_T=\widetilde\lambda_{t,T}(\zeta,\beta)$,
and uniqueness of best replies at the two histories
places $\lambda_t$ and $\lambda_T$ in the strict
best-reply regions of the two distinct chosen actions.
This proves the equivalence.
\end{proof}

\begin{proof}[Proof of Corollary \ref{lem:eu_constrained}]
Work on the probability-one event on which
Theorem \ref{thm:long_run_frontier} holds and
$\alpha_t\to\alpha$.
Since $\mathcal H_K(\alpha)=\{\eta^\star\}$,
part \textnormal{(ii)} of that theorem and compactness
of $\mathcal E_K$ imply $\eta_t\to\eta^\star$.
Since $d(\alpha)=\psi(\alpha)=0$,
part \textnormal{(i)} gives
$\lambda_t\to\tau(\alpha)=0$.
By continuity of $v_\lambda(a;\m q)$ at zero
and finiteness of $Q$, for every $a\in A$,
$V_{\lambda_t}(a;\eta_t)
\to V_0(a;\eta^\star)=U(a;\eta^\star)$. Fix $a\in\operatorname{supp}(\alpha)$.
Since $\alpha(a)>0$, action $a$ is chosen infinitely
often. Choose a sequence $t_n\to\infty$ such that
$a_{t_n+1}=a$.
By \eqref{eq:dynamic_criterion}, for every $b\in A$,
$V_{\lambda_{t_n}}(a;\eta_{t_n})
\ge V_{\lambda_{t_n}}(b;\eta_{t_n})$.
Taking limits gives
$U(a;\eta^\star)\ge U(b;\eta^\star)$
for every $b\in A$.
Since $a$ was arbitrary,
$\operatorname{supp}(\alpha)
\subseteq\arg\max_{b\in A}U(b;\eta^\star)$.
\end{proof}

\begin{proof}[Proof of Corollary \ref{cor:two_model}]

Write $k_S:=k(\m q^S),$ $k_C:=k(\m q^C),$ $D_S:=D(\m q^S;\alpha),$ and $D_C:=D(\m q^C;\alpha).$
Any $\eta\in\Delta(Q)$ is uniquely determined by
$x:=\eta(\m q^C)\in[0,1]$.  Its complexity is
$(1-x)k_S+xk_C,$
so feasibility is equivalent to
$x\le
\frac{K-k_S}{k_C-k_S}.$
Its expected misspecification is
$(1-x)D_S+xD_C
=
D_S+x(D_C-D_S).$
Since $D_C<D_S$, this expression is strictly decreasing in $x$.  Thus, the unique
optimum places the maximal feasible weight on $\m q^C$:
$\eta^\star(\m q^C)
=
\frac{K-k_S}{k_C-k_S}.$

Both models receive strictly positive probability and the capacity constraint binds.  By Proposition \ref{prop:frontier}\textnormal{(i)}, $\mathcal R_K(\alpha)$ is nonempty.  Fix any $\rho\in\mathcal R_K(\alpha)$.  Proposition
\ref{prop:frontier}\textnormal{(iii)} implies that both models minimize the shadow-priced objective.  Hence, $D_S+\rho k_S=D_C+\rho k_C,$ so $\rho
=
\frac{D_S-D_C}{k_C-k_S}
=:\rho^\star.$ Thus, every element of $\mathcal R_K(\alpha)$ equals $\rho^\star$, and nonemptiness implies $\mathcal R_K(\alpha)=\{\rho^\star\}$.  Since $D_C<D_S$,
$d(\alpha)=D_C$, so $\tau^\star=\tau(\alpha)=[D_C+\psi(\alpha)]/c$. \end{proof}

\begin{proof}[Proof of Lemma \ref{lem:delta_monotone}]

Fix $j\in\{S,C\}$ and write
$q=q_{\mathfrak{r}}^j$ and
$u_y=u(\mathfrak{r},y)$.  For $\lambda>0$, define
$Z(\lambda)
:=
\sum_{y\in Y}q(y)e^{-\lambda u_y}$ and $\widehat p_\lambda(y)
:=
\frac{q(y)e^{-\lambda u_y}}{Z(\lambda)}.$
Then, $v_\lambda(\mathfrak{r};\m q^j)
=
-\frac{1}{\lambda}\log Z(\lambda).$
Since $Y$ is finite, this function is differentiable on $\R_{++}$, and
$\frac{Z'(\lambda)}{Z(\lambda)}
=
-\E_{\widehat p_\lambda}[u(\mathfrak{r},y)].$
Therefore,
$\frac{d}{d\lambda}
v_\lambda(\mathfrak{r};\m q^j)=
\frac{1}{\lambda^2}\log Z(\lambda)
+
\frac{1}{\lambda}
\E_{\widehat p_\lambda}[u(\mathfrak{r},y)]=
-\frac{1}{\lambda^2}
R(\widehat p_\lambda\Vert q).$
The last equality follows from
$R(\widehat p_\lambda\Vert q)
=
\sum_y
\widehat p_\lambda(y)
\big[-\lambda u_y-\log Z(\lambda)\big].$
Full support of $q$ and nonconstancy of $u(\mathfrak{r},\cdot)$ imply
$\widehat p_\lambda\neq q$ for every $\lambda>0$.  Hence,
$R(\widehat p_\lambda\Vert q)>0$, so
$v_\lambda(\mathfrak{r};\m q^j)$ is strictly decreasing on $\R_{++}$. At $\lambda=0$, finiteness of $Y$ gives
$Z(\lambda)
=
1-\lambda\E_q[u(\mathfrak{r},y)]
+
O(\lambda^2),$
and therefore
$\lim_{\lambda\downarrow0}
v_\lambda(\mathfrak{r};\m q^j)
=
\E_q[u(\mathfrak{r},y)]
=
v_0(\mathfrak{r};\m q^j).$
Moreover, for every $\lambda>0$, strict Jensen inequality gives
$\log
\E_q[e^{-\lambda u(\mathfrak{r},y)}]
>
-\lambda\E_q[u(\mathfrak{r},y)],$
and hence
$v_\lambda(\mathfrak{r};\m q^j)
<
v_0(\mathfrak{r};\m q^j).$
Thus, $\Delta_j$ is continuous and strictly decreasing on $\R_+$.  Since
$\chi_K\in(0,1)$,
$\Delta_K
=
(1-\chi_K)\Delta_S+\chi_K\Delta_C$
is also continuous and strictly decreasing on $\R_+$. \end{proof}

\begin{proof}[Proof of Corollary \ref{cor:complexity_eliminates_cycles}]

The assumptions in the statement are equivalent to
$\Delta_S(\lambda_{\mathfrak{s}})<0<\Delta_C(\lambda_{\mathfrak{s}})$
and
$\Delta_C(\lambda_{\mathfrak{r}})<0.$
Since $\Delta_S$ is strictly decreasing and
$\lambda_{\mathfrak{r}}>\lambda_{\mathfrak{s}}$,
$\Delta_S(\lambda_{\mathfrak{r}})
<
\Delta_S(\lambda_{\mathfrak{s}})
<
0.$
Hence,
$\Delta_K(\lambda_{\mathfrak{r}})<0$ 
$\forall K\in(k_S,k_C).$ At the lower endpoint,
$\Delta_K(\lambda_{\mathfrak{s}})
=
(1-\chi_K)\Delta_S(\lambda_{\mathfrak{s}})
+
\chi_K\Delta_C(\lambda_{\mathfrak{s}}).$
Because the two terms have opposite signs, there is a unique
$\chi^\ast\in(0,1)$ for which this expression equals zero.  Solving gives
$\chi^\ast
=
\frac{
-\Delta_S(\lambda_{\mathfrak{s}})
}{
\Delta_C(\lambda_{\mathfrak{s}})
-
\Delta_S(\lambda_{\mathfrak{s}})
}
=
\dfrac{
\bar u-v_{\lambda_{\mathfrak{s}}}(\mathfrak{r};\m q^S)
}{
v_{\lambda_{\mathfrak{s}}}(\mathfrak{r};\m q^C)
-
v_{\lambda_{\mathfrak{s}}}(\mathfrak{r};\m q^S)
}.$
Therefore,
$K^\ast
=
k_S+(k_C-k_S)\chi^\ast
\in(k_S,k_C).$
Since
$\chi_K=(K-k_S)/(k_C-k_S)$ is strictly increasing in $K$,
\[
\Delta_K(\lambda_{\mathfrak{s}})
\begin{cases}
<0,&K<K^\ast,\\
=0,&K=K^\ast,\\
>0,&K>K^\ast.
\end{cases}
\]

If $K<K^\ast$, Theorem \ref{thm:complexity_cycles} gives
$\alpha_t(\mathfrak{r})\to0$, and therefore
$\lambda_t\to\lambda_{\mathfrak{s}}$.  Since
$\Delta_K(\lambda_{\mathfrak{s}})<0$ and $\Delta_K$ is continuous,
$\Delta_K(\lambda_t)<0$ for all sufficiently large $t$.  Also,
$\eta_t(\m q^C)=\chi_K$ for all sufficiently large $t$ by the proof of Theorem \ref{thm:complexity_cycles}.
Thus, the safe action is the unique best reply at every sufficiently large date, proving
part \textnormal{(i)}. If $K=K^\ast$, then
$\Delta_K(\lambda_{\mathfrak{s}})=0$ and
$\Delta_K(\lambda_{\mathfrak{r}})<0$.  Theorem
\ref{thm:complexity_cycles} therefore gives
$\alpha_t(\mathfrak{r})\to0$, proving part \textnormal{(ii)}. Finally, if $K>K^\ast$, then
$\Delta_K(\lambda_{\mathfrak{s}})>0>
\Delta_K(\lambda_{\mathfrak{r}}).$
Theorem \ref{thm:complexity_cycles} gives a unique
$x_K\in(0,1)$ with
$\alpha_t(\mathfrak{r})\to x_K$, so the environment exhibits a
$\lambda$-cycle.  This proves part \textnormal{(iii)}. \end{proof}
\subsection{Proofs for Section \ref{sec:capacity_risky_play}}\label{app:olarisky}

\begin{proof}[Proof of Proposition
\ref{prop:capacity_risky_reversal}]
Part (i). At date $0$, the likelihood is constant, so $\pi_0$ equals the prior and $\lambda_0=0$. The safe action has value zero under every model.
The expected risky payoffs under $\m q^P$, $\m q^S$, and $\m q^C$ are respectively $-0.87$, $0.90$, and $1$. The prior expected complexity equals $0.90$.
For $K<0.90$, the capacity constraint therefore binds.
Because $\m q^P$ and $\m q^S$ have equal complexity,
Lemma \ref{prop:constrained_posterior} implies that their recognition odds equal their prior odds. Since $\m q^C$ is the only model with complexity one, binding of the capacity constraint implies that its recognition weight equals $K$. Thus, $\eta_0(\m q^P)=0.9(1-K)$, $\eta_0(\m q^S)=0.1(1-K)$, and $\eta_0(\m q^C)=K$. The risky value is consequently $1.693K-0.693$.
It is strictly negative for $K<\bar K$, zero at $K=\bar K$, and strictly positive for $K>\bar K$. For $K\ge0.90$, the prior is feasible. Its expected risky payoff is $0.831>0$.
Since $\bar K<0.90$, part \textnormal{(i)} follows for every
$K\in(0,1)$.

\par Part (ii) For either action, consider the log-likelihood increment favoring
$\m q^S$ over $\m q^P$. Under the safe action, its expectation is one third of $\log(125/27)$ and is therefore strictly positive.
Under the risky action, multiplying its expected value by one hundred and exponentiating gives $16\cdot10^{99}/94^{40}>1$, so this expectation is also strictly positive. Because there are only two actions, these conditional expected increments are uniformly bounded below by some strictly positive constant.

\par Subtract from each realized log-likelihood increment its conditional
expectation given the previous history. The resulting partial sums form a martingale with uniformly bounded increments: the action at each date is measurable before the outcome is drawn, while the outcome is generated from the corresponding true
distribution. Azuma-Hoeffding implies exponentially summable tail probabilities for deviations of order $t$. Borel-Cantelli therefore implies that the centered sum divided by $t$ converges to zero a.s. It follows that the log posterior odds of $\m q^S$ against $\m q^P$ diverge to positive infinity. Hence, $\pi_t(\m q^P)/\pi_t(\m q^S)\to0$ a.s.

The two models have identical complexity.
When the capacity constraint is slack, recognition equals the posterior; when it binds, the exponential tilt in Lemma \ref{prop:constrained_posterior} multiplies their posterior weights by the same factor. Thus, their recognition odds always equal their posterior odds. Consequently, $\eta_t(\m q^P)/\eta_t(\m q^S)\to0$ a.s.

 We next show that the risky action is chosen infinitely often a.s.
Suppose, to the contrary, that on an event of positive probability only finitely many risky choices occur. On that event, all sufficiently late observations are generated by the
safe action. Both $\m q^S$ and $\m q^C$ coincide with
$p_{\mathfrak s}^\star$, while the recognition weight on $\m q^P$
converges to zero. The best structured model therefore has zero asymptotic
safe-action divergence. Moreover, the maxmin payoff is
zero and the objective expected payoff of the risky
action is $0.79$, so $\psi(\alpha)=0$ for every
$\alpha\in\Delta(A)$. Since
$\alpha_t(\mathfrak s)\to1$ on this event,
Lemma \ref{lem:safety_statistic} gives
$\lambda_t\to\tau(\delta_{\mathfrak s})=0$ on that event.

At $\lambda=0$, the risky expected payoffs under
$\m q^S$ and $\m q^C$ are $0.90$ and $1$, respectively.
By continuity of the robust values in $\lambda$, once the recognition
weight on $\m q^P$ is sufficiently small and $\lambda_t$ sufficiently
close to zero, the risky action is strictly preferred to the safe action. This contradicts eventual exclusive play of the safe action.
Risky play must therefore occur infinitely often a.s.

Safe observations do not change the likelihood ratio of
$\m q^C$ against $\m q^S$, since the two models have identical
safe-action distributions. Under the risky action, the expected log-likelihood increment in favor of $\m q^C$ equals $D_{\mathfrak r}(\m q^S)-D_{\mathfrak r}(\m q^C)$. Multiplying this expectation by one hundred and exponentiating gives $2^{56}3^{99}>1$, so the expectation is strictly positive.

Represent the outcome process using independent
action-specific sequences $(Y_n^a)_{n\ge1}$, where,
for each $a\in A$, the variables $Y_n^a$ are i.i.d.
with distribution $p_a^\star$, and the $n$th play
of action $a$ reveals $Y_n^a$.
This construction induces the prescribed conditional
outcome distributions under every history-dependent
action rule.
By the strong law, the average of the first $n$
risky-action log-likelihood increments in favor of
$\m q^C$ against $\m q^S$ converges a.s. to
$D_{\mathfrak r}(\m q^S)-D_{\mathfrak r}(\m q^C)>0$.
Since risky play occurs infinitely often a.s.,
the same convergence holds along the random number
of risky observations. Safe observations do not
affect this likelihood ratio, so
$\pi_t(\m q^C)/\pi_t(\m q^S)\to\infty$ a.s.
Together with
$\pi_t(\m q^P)/\pi_t(\m q^S)\to0$,
this yields $\pi_t(\m q^C)\to1$ a.s. Fix $K\in(0,1)$. Eventually $\pi_t(\m q^C)>K$. Because $\m q^C$ is the only model with complexity one and the other two models have complexity zero, the capacity constraint then binds and $\eta_t(\m q^C)=K$. Moreover, the recognition odds of $\m q^P$ against $\m q^S$ equal their posterior odds. It follows that $\eta_t(\m q^P)\to0$ and $\eta_t(\m q^S)\to1-K$ a.s.

We next establish the limits of misspecification concern
and risky frequency in part \textnormal{(ii)}.
The preceding likelihood comparisons establish that
$\m q^S$ fits better than $\m q^P$ under both actions,
while $\m q^C$ and $\m q^S$ fit equally well under
the safe action and $\m q^C$ fits strictly better
under the risky action.
Consequently, for every $\alpha\in\Delta(A)$,
$d(\alpha)=D(\m q^C;\alpha)
=\alpha(\mathfrak r)D_{\mathfrak r}(\m q^C)$,
where the second equality uses
$q_{\mathfrak s}^C=p_{\mathfrak s}^\star$.
Since $\psi(\alpha)=0$ for every $\alpha\in\Delta(A)$,
Lemma \ref{lem:safety_statistic} gives
$\lambda_t-
\alpha_t(\mathfrak r)D_{\mathfrak r}(\m q^C)/c
\to0$ a.s.

The robust values are uniformly bounded by the finite range of utility.
The convergence of the recognition distribution therefore implies that
the actual risky-safe value difference differs from
$\Delta_K(\lambda_t)$ by a term converging to zero.
Lemma \ref{lem:delta_monotone} implies that $\Delta_K$ is continuous and strictly decreasing. At zero, $\Delta_K(0)=0.90+0.10K>0$, while its limit as $\lambda$ tends to infinity is $-1$. Thus, $\Delta_K$ has a unique zero $\tau_K$.

For $\m q^C$, the unique positive value of $\lambda$ at which its robust risky value equals zero is $\log3\simeq1.099$.
For $\m q^S$, the corresponding value is
$\log((9+\sqrt{85})/2)\simeq2.209$.
Since $K\in(0,1)$, the zero $\tau_K$ of their strict convex combination lies strictly between these two values.
Direct evaluation gives $D_{\mathfrak r}(\m q^C)/c\simeq2.738$.
Consequently, $x_K:=c\tau_K/D_{\mathfrak r}(\m q^C)$ belongs to $(0,1)$.

Since $\lambda_t-\alpha_t(\mathfrak r)D_{\mathfrak r}(\m q^C)/c\to0$ a.s., the sequence $(\lambda_t)$ is eventually contained in a compact interval. Uniform continuity of $\Delta_K$ on that interval gives $\Delta_K(\lambda_t)-
\Delta_K(\alpha_t(\mathfrak r)
D_{\mathfrak r}(\m q^C)/c)\to0$ a.s. Fix $\varepsilon>0$ sufficiently small that
$x_K-\varepsilon>0$ and $x_K+\varepsilon<1$.
Strict decrease of $\Delta_K$ gives a strictly positive margin favoring the risky action whenever the risky frequency is at most
$x_K-\varepsilon$, and a strictly positive margin favoring the safe action whenever it is at least $x_K+\varepsilon$.
The vanishing approximation errors imply that these choices are indeed
strictly optimal for all sufficiently large dates.

The empirical risky frequency cannot remain forever below
$x_K-\varepsilon$: if it did, every sufficiently late choice would be
risky and the empirical frequency would converge to one, a contradiction. Likewise it cannot remain forever above $x_K+\varepsilon$, since then every sufficiently late choice would be safe and the empirical frequency would converge to zero. Hence, the risky frequency eventually enters
$[x_K-\varepsilon,x_K+\varepsilon]$.
Its absolute one-period change is at most $1/(t+1)$.
After this bound is below $\varepsilon$, any exit through the lower
boundary is followed by risky choices and any exit through the upper
boundary by safe choices; moreover, a single update can overshoot a
boundary by less than $\varepsilon$. Thus, the risky frequency is eventually contained in $[x_K-2\varepsilon,x_K+2\varepsilon]$.
Since $\varepsilon$ is arbitrary, $\alpha_t(\mathfrak r)\to x_K$ a.s.
The preceding relation between $\lambda_t$ and the risky frequency then implies $\lambda_t\to\tau_K$ a.s. This proves part \textnormal{(ii)}.

It remains to prove part \textnormal{(iii)}.
For $\lambda>0$, let $M_S(\lambda)$ and $M_C(\lambda)$ be the exponential moments entering the multiplier criteria for the risky action under $\m q^S$ and $\m q^C$. Since the multiplier value equals the negative logarithm of the relevant moment divided by $\lambda$, $v_\lambda(\mathfrak r;\m q^C)<v_\lambda(\mathfrak r;\m q^S)$ if and only if $M_C(\lambda)>M_S(\lambda)$. Writing $z=e^\lambda$ and multiplying
$M_C(\lambda)-M_S(\lambda)$ by $z^2$ gives
$(z-1)(0.2z^2+0.2z-0.5)$. The positive root of the second factor is
$(\sqrt{11}-1)/2<3$.
Therefore, $v_\lambda(\mathfrak r;\m q^C)
<v_\lambda(\mathfrak r;\m q^S)$ whenever
$\lambda\ge\log3$.
Every $\tau_K$ is strictly larger than $\log3$, so this inequality holds
at every $\tau_K$.

Now fix $K'>K$.
At $\lambda=\tau_K$, increasing recognition weight on $\m q^C$ and
reducing recognition weight on $\m q^S$ strictly lowers the risky-safe
value difference.
Hence, $\Delta_{K'}(\tau_K)<0$.
But $\Delta_{K'}(0)>0$, and
Lemma \ref{lem:delta_monotone} implies that $\Delta_{K'}$ is strictly
decreasing.
Its unique zero must therefore satisfy
$\tau_{K'}<\tau_K$.
Thus, $\tau_K$ is strictly decreasing in $K$.
Since the limiting risky frequency from part \textnormal{(ii)} is
$c\tau_K/D_{\mathfrak r}(\m q^C)$ and the multiplicative factor is
strictly positive and independent of $K$, that frequency is also strictly
decreasing in $K$.
This proves part \textnormal{(iii)}.\end{proof}

\begin{proof}[Proof of Observation \ref{prop:one_simple}]

Consider the two coupled copies defined in the statement.
Suppose first that, at some date, they have chosen the risky action the same number of times. They have then also chosen the safe action the same number of times. Under the action-by-action coupling, both copies have observed  the same first observations from the risky sequence and the same first observations from the safe sequence. Their action-specific data sets are therefore identical. Because likelihoods, the unrestricted likelihood, and the statistics entering \eqref{eq:dynamic_rule} depend on these action-specific samples but not on their chronological ordering, the two copies have the same Bayesian posterior and the same misspecification concern at that date.

Let $p$ denote their common posterior probability on $\m q^C$.
A distribution over the two models is determined by its weight $x$ on
$\m q^C$. Its expected complexity is $k(\m q^S)+x[k(\m q^C)-k(\m q^S)]$. Thus, at capacity $K$, feasibility is equivalent to
$x\le\chi_K$, where $\chi_K=(K-k(\m q^S))/(k(\m q^C)-k(\m q^S))$. The unconstrained minimizer of relative entropy is $x=p$.
Strict convexity of relative entropy therefore implies that the
constrained solution is $x_K=\min\{p,\chi_K\}$. Since $K'>K$, we have $\chi_{K'}>\chi_K$ and hence $x_{K'}\ge x_K$.

We now determine how the risky-safe value difference varies with $x$.
At the common misspecification concern $\lambda$, it equals
$(1-x)\Delta_S(\lambda)+x\Delta_C(\lambda)$. Suppose first that $\bar\lambda_C>\bar\lambda_S$. By Lemma \ref{lem:delta_monotone}, each $\Delta_j$ is continuous and strictly decreasing. Below $\bar\lambda_S$, whenever this interval is nonempty, both models strictly favor the risky action.
Above $\bar\lambda_C$, both strictly favor the safe action.
For misspecification concerns strictly between the two thresholds,
$\m q^S$ strictly favors safety and $\m q^C$ strictly favors risk, so the risky-safe value difference is strictly increasing in $x$.
At either threshold the same weak monotonicity holds.
Consequently, at every common decision problem, the copy with capacity
$K'$ cannot select the safe action when the copy with capacity $K$
selects the risky action. If both are exactly indifferent, the common selection specified in the statement gives them the same action.

Define $D_t:=R_t^{K'}-R_t^K$. Initially $D_0=0$. Assume inductively that $D_t\ge0$. If $D_t\ge1$, one additional choice by each copy can reduce the difference by at most one, so $D_{t+1}$ cannot be negative.
If $D_t=0$, the two copies have equal risky and safe counts and therefore face the common decision problem characterized above.
The higher-capacity copy cannot choose safe while the lower-capacity copy chooses risky. Hence, again $D_{t+1}\ge0$. Induction yields
$R_t^{K'}\ge R_t^K$ for every $t$ on every coupled outcome path.

Suppose next that $\bar\lambda_C<\bar\lambda_S$. Between the two thresholds, $\m q^C$ strictly favors safety and $\m q^S$ strictly favors risk. The risky-safe value difference is therefore strictly decreasing in $x$ on precisely the region in which model recognition can affect the choice; outside that region both models rank the actions in the same direction. Repeating the preceding argument with
$D_t:=R_t^K-R_t^{K'}$ gives $R_t^{K'}\le R_t^K$ for every $t$ on every coupled outcome path.

Finally, suppose $\bar\lambda_C=\bar\lambda_S$. If the common threshold is strictly positive, both models strictly favor risk below it, both strictly favor safety above it, and both are indifferent at the threshold, so the two copies make the same choice whenever their action counts are equal. If the common threshold equals zero, each model weakly favors safety at $\lambda=0$ and strictly favors safety at every positive $\lambda$. At every finite history the Bayesian posterior has full support, and because the capacities lie strictly between the two complexity levels, each recognition weight on $\m q^C$ lies strictly between zero and one. Hence, if either model strictly favors safety at $\lambda=0$, every recognition mixture also strictly favors safety; if both models are indifferent, the common selection gives the same action. Thus, the two copies again make the same choice whenever their action counts are equal.

Starting from equal counts at date zero and applying induction therefore gives $R_t^{K'}=R_t^K$ for every $t$ on every coupled outcome path. This completes the proof. \end{proof}

\subsection{Proofs for Section \ref{sec:ba}}
\label{app:generalized_selection_proofs}

The proofs maintain the risky-safe environment of
Section \ref{sec:ba}, with
$Q=\{\m q^S,\m q^C\}$, $w=\pi_0(\m q^C)$, and
$Q^\star=\{\m p^\star\}$. All predictions and both true
outcome distributions have full support. Write
$\pi_t=\pi_t^Q$ and $\eta_t=\eta_t^Q$. Recall that
$a_h(w)=\min\{hw,1\}$ for finite $h$ and $a_\infty(w)=1$.

\paragraph{Probability space.}
For each $a\in A$, let $(Y_n^a)_{n\ge1}$ be an i.i.d. sequence with
law $p_a^\star$, independent across actions. The $n$th choice of $a$
reveals $Y_n^a$. Define $N_t(a):=\sum_{i=1}^t\mathbf1\{a_i=a\}$.
A policy chooses $a_{t+1}$ as a function of $h_t$. Probabilities are taken under the true outcome distributions
and the specified policy.

\begin{proof}[Proof of Observation \ref{lem:ba_cvar}]
For every feasible $\nu$ in \eqref{eq:class_score}
and every $\zeta\in\R$,
$\sum_{\m q}\nu(\m q)L_t(\m q)
\le\zeta+h\sum_{\m q}\pi_0(\m q)
(L_t(\m q)-\zeta)_+$.
Taking the maximum over $\nu$ and the infimum over
$\zeta$ proves one inequality. For the reverse inequality, choose a likelihood value
$\zeta^\star$ satisfying
$\pi_0\{L_t>\zeta^\star\}\le1/h
\le\pi_0\{L_t\ge\zeta^\star\}$; such a value exists
because $Q$ is finite. Assign weight $h\pi_0(\m q)$
to every model with $L_t(\m q)>\zeta^\star$, zero
to every model with $L_t(\m q)<\zeta^\star$, and
distribute the remaining mass among models with
$L_t(\m q)=\zeta^\star$, respecting their caps.
The inequalities defining $\zeta^\star$ guarantee
feasibility. This distribution attains
$\zeta^\star+h\sum_{\m q}\pi_0(\m q)
(L_t(\m q)-\zeta^\star)_+$, establishing the
reverse inequality and hence \eqref{eq:ba_cvar}.
\end{proof}

\begin{lemma}\label{lem:selection_limits}
Fix $h\in[1,\infty]$ and a policy.
\begin{enumerate}
\item[\textnormal{(i)}]
On $\{N_t(\mathfrak r)\to\infty\}$,
$S_{h,t}(Q)/L_t(\m p^\star)\to0$ almost surely.
For each fixed $\varepsilon\in(0,1)$, set
$\m q^\varepsilon:=(1-\varepsilon)\m q+\varepsilon\m p^\star$
and $Q^\varepsilon:=\{\m q^\varepsilon:\m q\in Q\}$, with
$\pi_0^\varepsilon(\m q^\varepsilon):=\pi_0(\m q)$.
On the same event,
$S_{h,t}(Q^\varepsilon)/S_{h,t}(Q)\to\infty$ almost surely.
\item[\textnormal{(ii)}]
$\lambda_t^{Q^\star}\to0$ almost surely.
On $\{\sup_tN_t(\mathfrak r)<\infty\}$,
$\pi_t(\m q^C)\to1$ and $\lambda_t^Q\to0$ almost surely.
\end{enumerate}
\end{lemma}

\begin{proof}[Proof of Lemma \ref{lem:selection_limits}]
For each fixed action $a$ and $\m q\in Q$, the variables
$\log[q_a(Y_n^a)/p_a^\star(Y_n^a)]$ are i.i.d.
with mean $-R(p_a^\star\Vert q_a)$.
The strong law applies to their partial sums indexed by
$n$, and hence along $n=N_t(a)$ whenever $N_t(a)\to\infty$.
These partial sums tend to $-\infty$ if $q_a\ne p_a^\star$
and are identically zero otherwise. Thus each sequence of partial
sums is bounded above almost surely. Both models are incorrect at
$\mathfrak r$, so $N_t(\mathfrak r)\to\infty$ implies
$L_t(\m q)/L_t(\m p^\star)\to0$ for both models.
The bound $S_{h,t}(Q)\le\max_{\m q\in Q}L_t(\m q)$ proves
the first assertion.

For the perturbed model, the mean of
$\log[q_a^\varepsilon(Y_n^a)/q_a(Y_n^a)]$ is
$R(p_a^\star\Vert q_a)-R(p_a^\star\Vert q_a^\varepsilon)
\ge\varepsilon R(p_a^\star\Vert q_a)$.
The corresponding partial sums tend to $+\infty$ when
$q_a\ne p_a^\star$ and are zero otherwise; each is bounded below
almost surely. Hence
$L_t(\m q^\varepsilon)/L_t(\m q)\to\infty$ for both models
when risk is chosen infinitely often. 
Since $\m q^\varepsilon-\m r^\varepsilon
=(1-\varepsilon)(\m q-\m r)$ and $\varepsilon<1$,
the map $\m q\mapsto\m q^\varepsilon$ is injective.
 For finite $h$,
$\nu^\varepsilon(\m q^\varepsilon):=\nu(\m q)$ satisfies the
scoring constraints for $Q^\varepsilon$ exactly when $\nu$ satisfies
them for $Q$, since $\pi_0^\varepsilon(\m q^\varepsilon)=\pi_0(\m q)$.
For $h=\infty$, each score is the largest model likelihood.
In both cases, $S_{h,t}(Q^\varepsilon)/S_{h,t}(Q)$ is at least $\min_{\m q\in Q}L_t(\m q^\varepsilon)/L_t(\m q)$.

For \textnormal{(ii)}, since the maxmin payoff is
$\bar u$ and the objective expected risky payoff is
$p_R>\bar u$, we have $\psi(\alpha)=0$
for every $\alpha\in\Delta(A)$.
Lemma \ref{lem:safety_statistic}\textnormal{(i)}
applies to the likelihoods of either class computed
along the given action rule, regardless of which class
is adopted at each date. For $Q^\star=\{\m p^\star\}$, the minimum
action-weighted relative entropy is zero for every
$\alpha\in\Delta(A)$. Hence, $\lambda_t^{Q^\star}\to0$ a.s. Suppose now that only finitely many risky observations
occur. Since
$q_{\mathfrak s}^C=p_{\mathfrak s}^\star
\ne q_{\mathfrak s}^S$, the strong law along the
safe-action outcome sequence implies
$L_t(\m q^S)/L_t(\m q^C)\to0$ and hence
$\pi_t(\m q^C)\to1$.
Moreover, $\alpha_t(\mathfrak s)\to1$ and
$0\le d(\alpha_t)
\le D(\m q^C;\alpha_t)
=\alpha_t(\mathfrak r)
R(p_{\mathfrak r}^\star\Vert q_{\mathfrak r}^C)
\to0$.
Since $\psi(\alpha_t)=0$,
Lemma \ref{lem:safety_statistic}\textnormal{(i)}
gives $\lambda_t^Q\to0$.
\end{proof}

\begin{lemma}[Probability of remaining with $Q$]
\label{lem:selection_retention}
Suppose that, after any history consisting only of safe choices,
the action rule under $Q$ selects safety whenever
$\pi_t(\m q^C)\ge\theta$, where $0\le\theta<w$.
Define $e_\theta(w):=(1-w)\theta/[w(1-\theta)]$.
\begin{enumerate}
\item[\textnormal{(i)}]
Against $Q^\star$, if $a_h(w)\ge1/\gamma$, the probability of
never leaving $Q$ and always choosing safety is at least
$1-e_\theta(w)$.
\item[\textnormal{(ii)}]
For finite $h$, against any admissible competitor and prior,
this probability is at least
$1-e_\theta(w)-h/[\gamma a_h(w)]$.
\end{enumerate}
\end{lemma}

\begin{proof}[Proof of Lemma \ref{lem:selection_retention}]
Evaluate likelihoods on
$((\mathfrak s,Y_1^{\mathfrak s}),\ldots,
(\mathfrak s,Y_t^{\mathfrak s}))$ and let
$J_t:=L_t(\m q^S)/L_t(\m p^\star)$ and
$M_t:=\ell_t(Q')/L_t(\m p^\star)$. For each $\m r\in Q'\cup\{\m q^S\}$,
$\E[r_{\mathfrak s}(Y_{t+1}^{\mathfrak s})/
p_{\mathfrak s}^\star(Y_{t+1}^{\mathfrak s})
|\mathcal F_t^{\mathfrak s}]=1$, where $\mathcal F_t^{\mathfrak s}
:=\sigma(Y_1^{\mathfrak s},\ldots,Y_t^{\mathfrak s})$.
Thus, $L_t(\m r)/L_t(\m p^\star)$ is a nonnegative
martingale starting at one. This applies to $J_t$
and to the prior-weighted sum $M_t$.
The martingale maximal inequality therefore gives
$\mathbb{P}(\exists t\ge0:Z_t\ge d)\le1/d$
for $Z\in\{J,M\}$ and $d>0$.

Since $\pi_t(\m q^C)=w/[w+(1-w)J_t]$, for $\theta>0$ the event
$\{J_t<1/e_\theta(w)\ \forall t\ge0\}$ ensures
$\pi_t(\m q^C)>\theta$ at every date and has probability at least
$1-e_\theta(w)$. For $\theta=0$, the posterior condition holds
on every history and $e_0(w)=0$.
Moreover, $L_t(\m q^C)=L_t(\m p^\star)$ and the maximum feasible
scoring weight on $\m q^C$ is $a_h(w)$, so
$S_{h,t}(Q)\ge a_h(w)L_t(\m p^\star)$.
Against $Q^\star$, this gives $B_{h,t}\le1/a_h(w)\le\gamma$,
proving (i). For finite $h$, \eqref{eq:class_score_bounds} gives
$B_{h,t}\le hM_t/a_h(w)$. The maximal inequality and union bound
therefore give probability at least
$1-e_\theta(w)-h/[\gamma a_h(w)]$ that the posterior condition
holds and $M_t<\gamma a_h(w)/h$ for all $t$.
On either event, induction implies $m_t=Q$ and
$a_{t+1}=\mathfrak s$ for every $t\ge0$.
\end{proof}

\begin{proof}[Proof of Observation \ref{prop:ba_unconstrained}]
(i) If $Q$ is eventually used forever and risk is chosen only finitely
often, Lemma \ref{lem:selection_limits}(ii) gives
$\pi_t(\m q^C)\to1$, making risk strictly optimal since $z>\bar u$.
Thus eventual permanent use of $Q$ requires infinitely many risky
observations. Against each $Q^\varepsilon$ in Lemma
\ref{lem:selection_limits}(i), the ratio $S_{h,t}(Q^\varepsilon)/S_{h,t}(Q)$ would tend to infinity,
contradicting permanent use of $Q$.
The perturbations have full support. For each action,
$\tfrac12\sum_{y\in Y}|q_a^\varepsilon(y)-q_a(y)|\le\varepsilon$,
so $d_H(Q,Q^\varepsilon)\le\varepsilon$. Pairing each $\m q$ with
$\m q^\varepsilon$ under the prior gives
$d_P(\nu_Q,\nu_{Q^\varepsilon})\le\varepsilon$.
Thus $Q$ is not locally robust.
Against $Q^\star$, finitely many risky observations would eventually
make risk strictly optimal in both classes. Hence risk is observed
infinitely often and Lemma \ref{lem:selection_limits}(i) forces
permanent adoption of $Q^\star$, which always selects risk.

(ii) Under Bayesian choice, $\pi_t(\m q^C)\ge K_\star$
implies safety. Apply Lemma \ref{lem:selection_retention} with
$\theta=K_\star$. Part (i) gives the claim against $Q^\star$;
part (ii) gives global robustness because
\eqref{eq:ba_global_prior} makes its probability bound positive.
\end{proof}

\begin{proof}[Proof of Proposition \ref{prop:ba_capacity}]
If $x+(z-x)K>\bar u$, eventual permanent use of $Q$ with finitely many
risky observations would imply
$\eta_t(\m q^C)\to K$ and $\lambda_t^Q\to0$, by Lemma
\ref{lem:selection_limits}(ii). Continuity of $v_\lambda$ at zero
would then make risk strictly optimal. Thus permanent use requires
infinitely many risky observations. The perturbations $Q^\varepsilon$ of Lemma \ref{lem:selection_limits}(i),  with
$k(\m q^\varepsilon):=k(\m q)$, have a nonempty feasible
recognition set whenever $Q$ does, so local robustness
fails in both parts.
(i) For $K\le K_\star$, the value $V_{\lambda_t^Q}(\mathfrak r;\eta_t)$ is at most
$x+(z-x)K\le \bar u$ at every history, so safety is always chosen under $Q$. Lemma \ref{lem:selection_retention}(ii), with
$\theta=0$, gives a retention probability of at
least $1-h/[\gamma a_h(w)]>0$, since
$h/a_h(w)=\max\{h,1/w\}<\gamma$. A Bayesian SCE would have to prescribe safety, since neither model
predicts risk correctly. Its supporting belief must concentrate on
$\m q^C$, which instead makes risk strictly preferable. No such SCE exists.

(ii) If $\pi_t(\m q^C)\ge K_\star$ and $K\ge K_\star$,
then $\eta_t(\m q^C)\ge K_\star$. Since $z<x$,
$V_{\lambda_t^Q}(\mathfrak r;\eta_t)
\le x+(z-x)K_\star=\bar u$, so safety is chosen.
Lemma \ref{lem:selection_retention}(ii), with $\theta=K_\star$,
and \eqref{eq:ba_global_prior} establish global robustness.
\end{proof}

\begin{proof}[Proof of Corollary \ref{cor:ba_welfare}]
(i) If risk were chosen finitely often, Lemma
\ref{lem:selection_limits}(ii) and $x+(z-x)K>\bar u$ would eventually
make risk strictly optimal in both classes, a contradiction, so Lemma
\ref{lem:selection_limits}(i) implies  permanent adoption of
$Q^\star$. Since $\lambda_t^{Q^\star}\to0$ and $p_R>\bar u$, risk is
chosen at every sufficiently late date. The strong law of large numbers  gives the average-payoff conclusion.

(ii) While $Q$ is adopted, the value
$V_{\lambda_t^Q}(\mathfrak r;\eta_t)$ is at most
$x+(z-x)K\le \bar u$, so safety is chosen. Define
$J_t:=\prod_{n=1}^t
\frac{q^S_{\mathfrak s}(Y_n^{\mathfrak s})}
 {p^\star_{\mathfrak s}(Y_n^{\mathfrak s})}.$
By construction, $(Y_n^{\mathfrak s})_{n\ge1}$ is i.i.d.
with distribution $p^\star_{\mathfrak s}$. Since
$q^S_{\mathfrak s}\ne p^\star_{\mathfrak s}$, the strong law gives
$J_t\to0$ on an event of probability one. On this event,
the convergence also holds along histories in which safety
is always chosen, where
$J_t=L_t(\m q^S)/L_t(\m p^\star)$.
Writing the scoring weight on $\m q^C$ as $v$ gives
$S_{h,t}(Q)/L_t(\m p^\star)=\max_v\{v+(1-v)J_t\}$,
where $v\in[\max\{0,1-h(1-w)\},a_h(w)]$ for finite $h$ and
$v\in[0,1]$ for $h=\infty$.
This ratio converges to $a_h(w)$ and is always at least $a_h(w)$.
If $a_h(w)\ge1/\gamma$, the switching threshold is never crossed.
If $a_h(w)<1/\gamma$, the ratio $B_{h,t}$ eventually exceeds $\gamma$
on these histories, forcing a finite first switch almost surely.
(iii) Choosing risk at dates $j^2$, $j\ge1$, implies
$N_t(\mathfrak r)\ge\lfloor\sqrt t\rfloor$.
Lemma \ref{lem:selection_limits}(i) forces permanent adoption of
$Q^\star$. Since $\lambda_t^{Q^\star}\to0$ by
Lemma \ref{lem:selection_limits}(ii) and $p_R>\bar u$, risk is strictly
optimal under $Q^\star$ at every sufficiently late date, so it is
chosen at all such dates, including those not in $\{j^2\}$.
\end{proof}

\begin{proof}[Proof of Proposition
\ref{prop:ba_information_rates}]
Let $M_t:=\max_{\m q\in Q}L_t(\m q)$ and
$\underline w:=\min\{w,1-w\}>0$.
For every $h\in[1,\infty]$,
$\underline w M_t\le\ell_t(Q)\le
S_{h,t}(Q)\le M_t$.
Since $S_{h,t}(Q^\star)=L_t(\m p^\star)$,
$\log B_{h,t}$ belongs to the interval
$[\log(L_t(\m p^\star)/M_t),
\log(L_t(\m p^\star)/M_t)+\log(1/\underline w)]$.
This interval is independent of $h$, proving the
first assertion.

For each $\m q\in Q$ and $a\in A$, the
action-specific strong law gives
$\frac{1}{n}\sum_{j=1}^n
\log[q_a(Y_j^a)/p_a^\star(Y_j^a)]
\to-R(p_a^\star\Vert q_a)$ a.s.
If $N_t(a)\to\infty$, multiply this convergence
by $N_t(a)/t$; if $N_t(a)$ is bounded, its
log-likelihood contribution divided by $t$
vanishes. Since observation counts are
nondecreasing, these cases exhaust all
possibilities. Thus,
$\frac{1}{t}\log[L_t(\m q)/L_t(\m p^\star)]$
converges a.s. to
$-(1-\xi)R(p_{\mathfrak s}^\star\Vert q_{\mathfrak s})
-\xi R(p_{\mathfrak r}^\star\Vert q_{\mathfrak r})$.
Finiteness of $Q$ permits taking the maximum
over $\m q$. The preceding uniform bounds on
$\log B_{h,t}$ then yield
\eqref{eq:ba_information_rate}.

Finally, both models have strictly positive
risky-action KL divergence, while
$R(p_{\mathfrak s}^\star\Vert
q_{\mathfrak s}^C)=0$. 
Nonnegativity of relative entropy therefore gives
$I(\xi)>0$ for $\xi>0$ and $I(0)=0$, where $I(\xi)$ denotes the limit in \eqref{eq:ba_information_rate}.
\end{proof}

\begin{proof}[Proof of Corollary
\ref{cor:ba_information_rates}]
Let $M_t:=\max_{\m q\in Q}L_t(\m q)$.
On $\{N_t(\mathfrak r)\to\infty\}$,
Lemma \ref{lem:selection_limits}(i), applied
with $h=\infty$, gives
$M_t/L_t(\m p^\star)\to0$ almost surely.
Since $S_{h,t}(Q)\le M_t$ and
$S_{h,t}(Q^\star)=L_t(\m p^\star)$,
$B_{h,t}\to\infty$, uniformly over $h$.
Hence, $B_{h,t}>\gamma$ eventually, so the
switching rule implies permanent adoption
of $Q^\star$. If also $N_t(\mathfrak r)/t\to0$, the
action-specific strong laws imply
$t^{-1}\log[L_t(\m q)/L_t(\m p^\star)]
\to-R(p^\star_{\mathfrak s}\Vert
q_{\mathfrak s})$ for each $\m q\in Q$.
Since $Q$ is finite and
$q^C_{\mathfrak s}=p^\star_{\mathfrak s}$,
$t^{-1}\log[M_t/L_t(\m p^\star)]\to0$.
Finally,
$\min\{w,1-w\}M_t\le S_{h,t}(Q)\le M_t$
implies $t^{-1}\log B_{h,t}\to0$,
uniformly over $h$, as claimed.
\end{proof}
\subsection{Proofs for Appendix \ref{app:safety}}\label{app:sproof}

\begin{lemma}\label{lem:safety_statistic}
For every history-dependent action rule, the following
hold a.s.:
\begin{enumerate}
\item[\textnormal{(i)}]
$\frac{1}{t}\operatorname{LLR}(h_t;Q)-d(\alpha_t)\to0$
and $\frac{1}{t}\psi_t-\psi(\alpha_t)\to0$.

\item[\textnormal{(ii)}]
$\E_{\eta_t}[D(\cdot;\alpha_t)]-F_K(\alpha_t)\to0$.

\item[\textnormal{(iii)}]
$\frac{1}{t}\sum_{s=1}^tu(a_s,y_s)
-\sum_{a\in A}\alpha_t(a)
\E_{p^\star_a}[u(a,y)]\to0$.
\end{enumerate}
In particular, under S-LLR,
$\lambda_t-\tau(\alpha_t)\to0$ a.s.
\end{lemma}

\begin{proof}[Proof of Lemma \ref{lem:safety_statistic}]
Let $\mathcal F_t$ be the sigma-field generated by
the realized history through date $t$, and write
$N_t(a,y):=\sum_{s=1}^t
\mathbbm 1\{a_s=a,y_s=y\}$ and
$N_t(a):=\sum_yN_t(a,y)$.
For every $(a,y)\in A\times Y$,
$M_t^{a,y}:=N_t(a,y)-N_t(a)p^\star_a(y)$
is a martingale with increments bounded in absolute
value by one: the action $a_s$ is
$\mathcal F_{s-1}$-measurable and the conditional
distribution of $y_s$ is $p^\star_{a_s}$.
Azuma-Hoeffding therefore gives
$\mathbb P(|M_t^{a,y}|/t>\varepsilon)
\le2e^{-t\varepsilon^2/2}$ for every $\varepsilon>0$.
Applying Borel-Cantelli to $\varepsilon=1/m$,
$m\in\mathbb N$, and using finiteness of $A$ and $Y$,
there is an event of probability one on which,
simultaneously for every $(a,y)\in A\times Y$,
\begin{equation}\label{eq:safety_empirical}
\frac{N_t(a,y)}{t}
-\alpha_t(a)p^\star_a(y)\longrightarrow0.
\end{equation}
Fix a path in this event. Part \textnormal{(i)}: since all structured models
have full support and $Q$ is finite,
\eqref{eq:safety_empirical} implies
$\frac{1}{t}\log L_t(\m q)
=\sum_{a,y}N_t(a,y)\frac{1}{t}\log q_a(y)
=-\sum_a\alpha_t(a)H(p^\star_a)
-D(\m q;\alpha_t)+o(1)$ uniformly over
$\m q\in Q$.
We get
$\frac{1}{t}\log\max_{\m q\in Q}L_t(\m q)
=-\sum_a\alpha_t(a)H(p^\star_a)
-d(\alpha_t)+o(1)$ after taking the maximum over $\m q\in Q$. For every action $a$ with $N_t(a)>0$, nonnegativity of
relative entropy implies that its unrestricted likelihood
is maximized at $p_a(y)=N_t(a,y)/N_t(a)$.
Thus,
$\frac{1}{t}\log U_t
=\sum_{a,y}[N_t(a,y)/t]\log[N_t(a,y)/t]
-\sum_a\alpha_t(a)\log\alpha_t(a).$
Uniform continuity of $x\mapsto x\log x$ on $[0,1]$
and \eqref{eq:safety_empirical} imply that the
first sum differs from
$\sum_{a,y}\alpha_t(a)p^\star_a(y)
\log[\alpha_t(a)p^\star_a(y)]$ by $o(1)$. Thus,
$\frac{1}{t}\log U_t=-\sum_a\alpha_t(a)H(p^\star_a)+o(1)$.
Subtracting the maximized structured log likelihood
gives $\frac{1}{t}\operatorname{LLR}(h_t;Q)
=d(\alpha_t)+o(1)$.

If $B>0$, the definition of $\psi_t$ implies
$\frac{1}{t}\psi_t
=\frac{1}{B}\sum_a\big(\alpha_t(a)\underline u
-\sum_yN_t(a,y)t^{-1}u(a,y)\big)_+$.
Since the positive-part function is Lipschitz
continuous, \eqref{eq:safety_empirical} implies
$\frac{1}{t}\psi_t-\psi(\alpha_t)\to0$.
If $B=0$, both adjustments are identically zero.
This proves part \textnormal{(i)}.

For part \textnormal{(ii)}, the recognition
distribution maximizes
$t^{-1}\E_\eta[\log L_t(\cdot)]
-t^{-1}R(\eta\Vert\pi_0)$ over $\mathcal E_K$.
Since $\pi_0$ has full support,
$0\le R(\eta\Vert\pi_0)
\le\log(1/\min_{\m q\in Q}\pi_0(\m q))$
for every $\eta\in\Delta(Q)$.
The preceding uniform likelihood approximation
therefore implies that the recognition objective
equals $-\sum_a\alpha_t(a)H(p^\star_a)
-\E_\eta[D(\cdot;\alpha_t)]+e_t(\eta)$,
where $\sup_{\eta\in\mathcal E_K}|e_t(\eta)|\to0$.
Choose any $\widetilde\eta_t\in\mathcal H_K(\alpha_t)$.
Optimality of $\eta_t$ gives
$0\le\E_{\eta_t}[D(\cdot;\alpha_t)]-F_K(\alpha_t)
\le2\sup_{\eta\in\mathcal E_K}|e_t(\eta)|\to0$.

Finally,
$t^{-1}\sum_{s=1}^tu(a_s,y_s)
=\sum_{a,y}N_t(a,y)t^{-1}u(a,y)$,
so \eqref{eq:safety_empirical} also proves
part \textnormal{(iii)}.
The last assertion follows from part \textnormal{(i)}
and \eqref{eq:dynamic_rule}.
\end{proof}

\begin{proof}[Proof of Theorem \ref{prop:safety_normative}]
Let $n:=|A|$ and
$u_a^\star:=\E_{p^\star_a}[u(a,y)]$.
Choose $a^0\in\arg\max_{a\in A}\min_yu(a,y)$.
For any $\lambda\ge0$ and $\eta\in\Delta(Q)$,
$V_\lambda(a^0;\eta)\ge\underline u$ and
$V_\lambda(a;\eta)\le U(a;\eta)$ for every $a$:
the first inequality follows from nonnegativity
of relative entropy, and the second follows by
evaluating each minimization at $p=q_a$.
Both inequalities also hold at $\lambda=0$.

We first record a property of empirical action
frequencies. Fix $a\in A$ and $r\in(0,1)$.
Suppose that, for every sufficiently large $t$,
\begin{equation}\label{eq:safety_frequency_exclusion}
\alpha_t(a)\ge r\quad\Longrightarrow\quad a_{t+1}\ne a.
\end{equation}
Then, $\limsup_{t\to\infty}\alpha_t(a)\le r$.
Indeed, $\alpha_{t+1}(a)=\alpha_t(a)
+[\mathbbm 1\{a_{t+1}=a\}-\alpha_t(a)]/(t+1)$.
If $\alpha_t(a)\ge r$ at every sufficiently large
date, \eqref{eq:safety_frequency_exclusion} implies
that $a$ is chosen only finitely often, contradicting
$\alpha_t(a)\ge r>0$. Hence, $\alpha_t(a)<r$ at
arbitrarily large dates. Starting at any such date
$T$ after \eqref{eq:safety_frequency_exclusion}
holds, induction gives
$\alpha_t(a)\le r+1/(T+1)$ for every $t\ge T$:
whenever the frequency is at least $r$ it decreases,
and whenever it is below $r$ it increases by at most
$1/(T+1)$. Letting $T\to\infty$ proves the claim.

We also use the following bound. By Pinsker's
inequality and the range of the utility function,
$|\E_{q_a}[u(a,y)]-u_a^\star|
\le B\sqrt{D_a(\m q)/2}$.
Taking expectations under $\eta$ and applying
Jensen's inequality gives, for every $a\in A$
and $\eta\in\Delta(Q)$,
\begin{equation}\label{eq:safety_payoff_accuracy}
|U(a;\eta)-u_a^\star|
\le B\sqrt{\E_\eta[D_a]/2}.
\end{equation}

\par\noindent
\textnormal{(i).}
Fix $\varepsilon>0$. If $B=0$, every action yields
$\underline u$, so safety holds for every $c>0$. Suppose
henceforth that $B>0$.

We first establish safety. Define
$\underline q:=\min_{\m q\in Q,a\in A,y\in Y}
q_a(y)>0$ and
$\bar\lambda:=4[1+\log(1/\underline q)]/\varepsilon$.
For any $a\in A$, choose $y_a$ minimizing $u(a,\cdot)$.
Evaluating the minimization defining $v_\lambda$
at the Dirac distribution on $y_a$ gives
$v_\lambda(a;\m q)
\le\min_yu(a,y)+\lambda^{-1}\log[1/q_a(y_a)]
\le\min_yu(a,y)+\lambda^{-1}\log(1/\underline q)$
for every $\lambda>0$.
Consequently, if
$u_a^\star\le\underline u-\varepsilon/2$
and $\lambda\ge\bar\lambda$, then
$V_\lambda(a;\eta)
\le u_a^\star+\varepsilon/4
\le\underline u-\varepsilon/4
<V_\lambda(a^0;\eta)$
for every $\eta\in\Delta(Q)$.

Let $r:=\min\{1/(2n),\varepsilon/(4nB)\}$
and $\bar c:=r\varepsilon/(4B\bar\lambda)>0$.
Fix $c\in(0,\bar c]$, a true model, and an
S-LLR-optimal policy.
Work on the probability-one event of
Lemma \ref{lem:safety_statistic}.
If $u_a^\star\le\underline u-\varepsilon/2$
and $\alpha_t(a)\ge r$, then
$\psi(\alpha_t)\ge r\varepsilon/(2B)$.
By Lemma \ref{lem:safety_statistic}\textnormal{(i)}
and nonnegativity of $d(\alpha_t)$, at every
sufficiently large such date,
$c\lambda_t\ge r\varepsilon/(4B)$.
Hence, $\lambda_t\ge\bar\lambda$, and the preceding
strict value comparison implies $a_{t+1}\ne a$.
By \eqref{eq:safety_frequency_exclusion},
$\limsup_t\alpha_t(a)\le r$ $\forall a\in A^-:=\{a\in A:
u_a^\star\le\underline u-\varepsilon/2\}$.

Since $\underline u-u_a^\star\le B$ for every
$a$, and $\underline u-u_a^\star<\varepsilon/2$
for $a\notin A^-$, we obtain
$\limsup_t\sum_a\alpha_t(a)(\underline u-u_a^\star)
\le\varepsilon/2+Bnr\le3\varepsilon/4$.
Lemma \ref{lem:safety_statistic}\textnormal{(iii)}
then yields
$\liminf_t t^{-1}\sum_{s=1}^tu(a_s,y_s)
\ge\underline u-\varepsilon$.
The choice of $\bar c$ is independent of the
true model, capacity, complexity costs, prior,
and policy. This completes the proof of part \textnormal{(i)}.

\par\noindent
\textnormal{(ii).}
Consider the following decision problem.
Let $A=\{\mathfrak s,\mathfrak r\}$,
$Y=\{0,1\}$, $u(\mathfrak s,y)=1/2$,
and $u(\mathfrak r,y)=y$.
Let $Q=\{\m q^S,\m q^C\}$, where
$q_{\mathfrak s}^S=\operatorname{Bern}(0.4)$,
$q_{\mathfrak s}^C=\operatorname{Bern}(0.6)$,
and $q_{\mathfrak r}^S=q_{\mathfrak r}^C
=\operatorname{Bern}(0.8)$.
Assign costs $k(\m q^S)=0$, $k(\m q^C)=1$,
let $K=1/2$, and take any full-support prior.
Fix $\varepsilon=0.1$. Since both models agree on the risky outcome
distribution, the value of the risky action is
independent of recognition. It equals $0.8$
at $\lambda=0$ and converges to zero as
$\lambda\to\infty$. The value of the safe action
is $1/2$ for every $\lambda\ge0$ and every
recognition distribution.

Suppose first that
$\Lambda(h_t)/\{[\operatorname{LLR}(h_t;Q)+\psi_t]/t\}
\to0$ for every sequence of histories.
The denominator is uniformly bounded over all dates and
histories: $t^{-1}\operatorname{LLR}(h_t;Q)\le\log 5$,
since every structured model assigns probability at least
$0.2$ to every outcome under either action, and
$0\le t^{-1}\psi_t\le1$.
Under the stated ratio convention, the convergence
therefore implies $\Lambda(h_t)\to0$ along every
realized infinite history. By continuity at zero,
risk is strictly optimal at every sufficiently
large date under every $\Lambda$-optimal policy.

Take the true model to satisfy
$p^\star_{\mathfrak s}=\operatorname{Bern}(0.5)$
and $p^\star_{\mathfrak r}=\operatorname{Bern}(0.2)$.
Since risk is eventually chosen exclusively,
Lemma \ref{lem:safety_statistic}\textnormal{(iii)}
implies that the average realized payoff converges
to $0.2$ a.s. Since $0.2<1/2-\varepsilon$,
$\Lambda$ is not $\varepsilon$-safe.

Suppose next that
$\{[\operatorname{LLR}(h_t;Q)+\psi_t]/t\}/\Lambda(h_t)
\to0$ for every sequence of histories.
Fix an arbitrary $\delta>0$.
By continuity of relative entropy at full-support
distributions, choose $e\in(0,0.1)$ sufficiently
small that the true model
$p^\star_{\mathfrak s}=\operatorname{Bern}(0.4+e)$
and $p^\star_{\mathfrak r}=\operatorname{Bern}(0.8-e)$
satisfies $\max_aD_a(\m q^S)<\min\{\delta,\varepsilon\}$.
Since $\delta_{\m q^S}\in\mathcal E_K$,
the antecedent in Definition
\ref{def:epsilon_self_confirming} holds.

The true outcome distribution differs from both
structured models under both actions. Hence
$m:=\min_{\m q\in Q,a\in A}D_a(\m q)>0$,
so $d(\alpha)\ge m$ for every $\alpha\in\Delta(A)$.
By Lemma \ref{lem:safety_statistic}\textnormal{(i)}
and $\psi_t\ge0$,
$[\operatorname{LLR}(h_t;Q)+\psi_t]/t
\ge m/2$ at every sufficiently large date a.s.
The assumed ratio convergence thus implies
$\Lambda(h_t)\to\infty$ a.s.
Since the robust risky value converges to zero
while the safe value remains $1/2$, safety is
strictly optimal at every sufficiently large date.
The average realized payoff consequently converges
to $1/2$ a.s.

However, under every recognition distribution,
the risky expected payoff is $0.8$ and the safe
expected payoff is $1/2$.
Thus, the safe action is not an
$\varepsilon$-best reply under any recognition
distribution. The risky action belongs to
$\operatorname{SCE}_{\varepsilon,K}(\m p^\star)$
because it is optimal under $\delta_{\m q^S}$
and $D_{\mathfrak r}(\m q^S)<\varepsilon$.
Therefore,
$\operatorname{SCE}_{\varepsilon,K}(\m p^\star)
=\{\mathfrak r\}$ and
$G_{\varepsilon,K}(\m p^\star)=0.8-e$.
Since $e<0.1$,
$G_{\varepsilon,K}(\m p^\star)-\varepsilon
>0.6>1/2$.
As $\delta>0$ was arbitrary, $\Lambda$ is not
$\varepsilon$-self-confirming under cognitively
feasible correct specification.
Both failures have been established for the same
decision problem and the same $\varepsilon$,
which proves part \textnormal{(ii)}. The same examples establish the conclusion when the two
rate conditions use $\operatorname{LLR}(h_t;Q)/t$
instead of $[\operatorname{LLR}(h_t;Q)+\psi_t]/t$:
the former statistic is uniformly bounded in the first
example and eventually bounded below by $m/2$ a.s. in
the second.

\par\noindent
\textnormal{(iii).}
Suppose $\m p^\star\in Q$ and
$k(\m p^\star)\le K$.
Since $\delta_{\m p^\star}\in\mathcal E_K$,
$d(\alpha)=F_K(\alpha)=0$ for every
$\alpha\in\Delta(A)$.
Lemma \ref{lem:safety_statistic}\textnormal{(ii)}
implies $\E_{\eta_t}[D(\cdot;\alpha_t)]\to0$ a.s. Fix a path in the probability-one event of
Lemma \ref{lem:safety_statistic}.
If $B=0$, the safety adjustment is zero.
If $B>0$, fix $a\in A$ with
$u_a^\star<\underline u$ and $r\in(0,1)$.
Whenever $\alpha_t(a)\ge r$,
$\E_{\eta_t}[D_a]\le
r^{-1}\E_{\eta_t}[D(\cdot;\alpha_t)]$,
whose right-hand side converges to zero.
Equation \eqref{eq:safety_payoff_accuracy}
therefore implies that, at every sufficiently
large such date,
$V_{\lambda_t}(a;\eta_t)\le U(a;\eta_t)
<\underline u\le V_{\lambda_t}(a^0;\eta_t)$.
Hence, $a_{t+1}\ne a$, so
\eqref{eq:safety_frequency_exclusion} yields
$\limsup_t\alpha_t(a)\le r$.
As $r>0$ is arbitrary, $\alpha_t(a)\to0$
for every action with $u_a^\star<\underline u$.
Finiteness of $A$ implies $\psi(\alpha_t)\to0$.
In either case, Lemma
\ref{lem:safety_statistic}\textnormal{(i)}
and $d(\alpha_t)=0$ give $\lambda_t\to0$.

Finally, fix $a$ with $\limsup_t\alpha_t(a)>0$
and choose $r$ strictly between zero and this
limit superior. The frequency-exclusion argument
implies that there are infinitely many dates $t$
with $\alpha_t(a)\ge r$ and $a_{t+1}=a$.
Choose a subsequence of these dates along which
$\eta_t\to\eta\in\mathcal E_K$, which exists by
compactness of $\mathcal E_K$.
Along this subsequence,
$\E_{\eta_t}[D_a]\le
r^{-1}\E_{\eta_t}[D(\cdot;\alpha_t)]\to0$.
Continuity gives $\E_\eta[D_a]=0$.
For every $b\in A$, optimality implies
$V_{\lambda_t}(a;\eta_t)\ge
V_{\lambda_t}(b;\eta_t)$ along the same subsequence.
Since $\lambda_t\to0$, continuity gives
$U(a;\eta)\ge U(b;\eta)$ for every $b\in A$.
Thus, $a\in\operatorname{SCE}_{0,K}(\m p^\star)$,
as desired.

\par\noindent
\textnormal{(iv).}
Fix $\varepsilon>0$ and $c>0$. If $B=0$, choose
$\delta=\varepsilon$. Whenever
$\min_{\eta\in\mathcal E_K}\max_a\E_\eta[D_a]<\delta$,
a feasible recognition distribution attaining this
minimum establishes that every action belongs to
$\operatorname{SCE}_{\varepsilon,K}(\m p^\star)$.
Since every action yields $\underline u$, both
claims follow. Suppose henceforth that $B>0$. By continuity of $v_\lambda$ at zero and finiteness
of $A$ and $Q$, there exists $\lambda^0>0$ such that
$0\le U(a;\eta)-V_\lambda(a;\eta)
\le\varepsilon/2$ for every
$a\in A$, $\eta\in\Delta(Q)$, and
$\lambda\in[0,\lambda^0]$.
Choose $\rho\in(0,\varepsilon)$ sufficiently small that
$(\rho+\sqrt{\rho/2})/c<\lambda^0$. Define
$g(\m p):=\min_{\eta\in\mathcal E_K}
\max_{a\in A}\E_\eta[R(p_a\Vert q_a)]$.
This function is continuous on $\Delta(Y)^A$,
and its zero set is
$Q_K:=\{\m q\in Q:k(\m q)\le K\}$.
Indeed, if $g(\m p)=0$, every model receiving positive
weight under a minimizing distribution equals $\m p$,
so feasibility implies $k(\m p)\le K$. Finiteness of $A$ and $Q$ and full support imply
that each $\m q^0\in Q_K$ has a neighborhood
$\mathcal N(\m q^0)$ on which
$\max_aR(p_a\Vert q_a^0)<\rho$ and
$R(p_a\Vert q_a)>R(p_a\Vert q_a^0)$ whenever
$q_a\ne q_a^0$.
By compactness and continuity, there exists $\delta>0$
such that $g(\m p)<\delta$ implies
$\m p\in\bigcup_{\m q^0\in Q_K}\mathcal N(\m q^0)$.
This choice is independent of the true model and policy.

Suppose $g(\m p^\star)<\delta$, choose
$\m q^0\in Q_K$ with
$\m p^\star\in\mathcal N(\m q^0)$,
and fix an S-LLR-optimal policy.
Construct outcomes from independent action-specific
i.i.d. sequences with laws $p_a^\star$, revealing
the next unused outcome whenever $a$ is chosen.
Work on the probability-one event of their
log-likelihood-ratio strong laws and
Lemma \ref{lem:safety_statistic}.
Let $I$ be the set of actions chosen infinitely often. For every $\m q\in Q$ and action $a$ with
$q_a\ne q_a^0$,
$\E_{p_a^\star}[\log(q_a(Y)/q_a^0(Y))]<0$.
Thus, each infinitely sampled action on which
$\m q$ and $\m q^0$ disagree contributes a term
tending to $-\infty$ to
$\log[L_t(\m q)/L_t(\m q^0)]$.
Actions on which they agree contribute zero,
and finitely sampled actions contribute eventually
constant terms.
Consequently, $L_t(\m q)/L_t(\m q^0)$ converges
to a finite limit for every $\m q$, and this limit
is zero whenever $q_a\ne q_a^0$ for some $a\in I$.
It follows that $\pi_t(\m q^0)$ remains bounded
away from zero, while the posterior probability
of every such disagreeing model vanishes.

Since $\delta_{\m q^0}\in\mathcal E_K$,
optimality of recognition gives
$R(\eta_t\Vert\pi_t)
\le R(\delta_{\m q^0}\Vert\pi_t)
=\log[1/\pi_t(\m q^0)]$,
whose right-hand side is bounded.
The inequality
$R(\eta_t\Vert\pi_t)
\ge\eta_t(\m q)\log[1/\pi_t(\m q)]-\log|Q|$
therefore implies $\eta_t(\m q)\to0$
for every model that disagrees with $\m q^0$
on an action in $I$.
For every $a\in I$, we consequently have
$\E_{\eta_t}[D_a]\to D_a(\m q^0)<\rho$
and
$U(a;\eta_t)\to\E_{q_a^0}[u(a,y)]$.

Whenever $a_{t+1}=a$,
$U(a;\eta_t)\ge V_{\lambda_t}(a;\eta_t)
\ge\underline u$.
Since each $a\in I$ is chosen infinitely often,
$\E_{q_a^0}[u(a,y)]\ge\underline u$.
Equation \eqref{eq:safety_payoff_accuracy},
applied to $\delta_{\m q^0}$, then gives
$u_a^\star\ge\underline u-B\sqrt{\rho/2}$
for every $a\in I$.
Actions outside $I$ are chosen only finitely often,
so $\limsup_t\psi(\alpha_t)\le\sqrt{\rho/2}$.
Also,
$d(\alpha_t)\le D(\m q^0;\alpha_t)<\rho$.
Lemma \ref{lem:safety_statistic}\textnormal{(i)}
therefore yields
$\limsup_t\lambda_t
\le(\rho+\sqrt{\rho/2})/c<\lambda^0$.
Hence, $\lambda_t\le\lambda^0$ at every sufficiently
large date.

Suppose now that $a_{t+1}=a\in I$ at a sufficiently
large date. Then $\E_{\eta_t}[D_a]<\varepsilon$.
For every $b\in A$, optimality and the choice of
$\lambda^0$ imply
$U(a;\eta_t)\ge V_{\lambda_t}(a;\eta_t)
\ge V_{\lambda_t}(b;\eta_t)
\ge U(b;\eta_t)-\varepsilon/2$.
Since $\eta_t\in\mathcal E_K$, this establishes
$a\in\operatorname{SCE}_{\varepsilon,K}(\m p^\star)$. Since $A$ is finite, $I$ is nonempty and all
sufficiently late choices belong to $I$.
Thus, $\operatorname{SCE}_{\varepsilon,K}(\m p^\star)$
is nonempty and contains every sufficiently late choice.
Lemma \ref{lem:safety_statistic}\textnormal{(iii)}
then gives
$\liminf_t t^{-1}\sum_{s=1}^tu(a_s,y_s)
\ge G_{\varepsilon,K}(\m p^\star)$,
which implies $\varepsilon$-self-confirmation
and completes the proof of part \textnormal{(iv)}.
\end{proof}

\begin{proof}[Proof of Observation \ref{lem:convergence}]
By Lemma \ref{lem:safety_statistic}\textnormal{(i)},
$\psi_t/t-\psi(\alpha_t)\to0$ a.s.
Continuity of $\psi$ and $\alpha_t\to\alpha$ therefore
give $\psi_t/t\to\psi(\alpha)$.
The second conclusion follows from
$\lambda_t=c^{-1}[\operatorname{LLR}(h_t;Q)/t+\psi_t/t]$
and $\tau(\alpha)=c^{-1}[d(\alpha)+\psi(\alpha)]$.
\end{proof}

\subsection{Proofs for Appendix
\ref{app:action_convergence}}
\label{app:convergence_proofs}

\begin{lemma}\label{lem:convergence_best_reply}
For every policy, there is an event of probability
one such that, on every realized path in that event,
if $t_\ell\to\infty$, $\alpha_{t_\ell}\to\alpha$,
and $a_{t_\ell+1}=a$ for every $\ell$, then
$a\in\mathcal A_K(\alpha)$. In particular, if
$\alpha_t\to\alpha$, then $a_t\in\mathcal A_K(\alpha)$
at every sufficiently large date.
\end{lemma}

\begin{proof}[Proof of Lemma
\ref{lem:convergence_best_reply}]
Fix a policy and work on the probability-one event
of Lemma \ref{lem:safety_statistic}.
Suppose $t_\ell\to\infty$, $\alpha_{t_\ell}\to\alpha$,
and $a_{t_\ell+1}=a$ for every $\ell$.
By compactness of $\mathcal E_K$, pass to a
subsequence along which $\eta_{t_\ell}\to\eta$
for some $\eta\in\mathcal E_K$.
Since $d$ is the minimum of finitely many continuous
functions and $\psi$ is linear, $\tau$ is continuous.
Moreover, $F_K$ is continuous because
$|F_K(\alpha)-F_K(\alpha')|
\le\max_{\m q\in Q}
|D(\m q;\alpha)-D(\m q;\alpha')|$.
By Lemma \ref{lem:safety_statistic},
$\lambda_{t_\ell}\to\tau(\alpha)$ and
$\E_\eta[D(\cdot;\alpha)]=F_K(\alpha)$,
so $\eta\in\mathcal H_K(\alpha)$.
By \eqref{eq:dynamic_criterion},
$V_{\lambda_{t_\ell}}(a;\eta_{t_\ell})
\ge V_{\lambda_{t_\ell}}(b;\eta_{t_\ell})$
for every $b\in A$.
Continuity of $V_\lambda(b;\eta)$ in $(\lambda,\eta)$
implies $a\in\operatorname{BR}_{\tau(\alpha)}(\eta)$,
so $a\in\mathcal A_K(\alpha)$. Now suppose $\alpha_t\to\alpha$.
If an action $a$ is chosen infinitely often,
there is a sequence $t_\ell\to\infty$ such that
$a_{t_\ell+1}=a$ for every $\ell$.
By the preceding argument, $a\in\mathcal A_K(\alpha)$.
Thus, every action outside $\mathcal A_K(\alpha)$
is chosen only finitely often. Since $A$ is finite,
$a_t\in\mathcal A_K(\alpha)$ at every sufficiently
large date.
\end{proof}

\begin{proof}[Proof of Proposition
\ref{prop:action_convergence}]
Fix a policy and work on the probability-one event
of Lemma \ref{lem:convergence_best_reply}.

\par\noindent
\textnormal{(i).}
Let $\mathcal I$ be the nonempty set of actions
chosen infinitely often. We show by induction that
$\mathcal I\subseteq\mathcal S_j$ for every $j$.
The claim is immediate for $\mathcal S_0=A$.
Suppose it holds for $\mathcal S_j$ and fix
$a\in\mathcal I$. Choose a sequence $t_\ell\to\infty$
such that $a_{t_\ell+1}=a$ for every $\ell$.
By compactness of $\Delta(A)$, pass to a subsequence
along which $\alpha_{t_\ell}\to\alpha$.
By the induction hypothesis, every action outside
$\mathcal S_j$ is chosen only finitely often,
so $\alpha\in\Delta(\mathcal S_j)$.
Lemma \ref{lem:convergence_best_reply} implies
$a\in\mathcal A_K(\alpha)$. Since also
$a\in\mathcal S_j$, we have $a\in\mathcal S_{j+1}$.
Thus, $\mathcal I\subseteq\mathcal S_{j+1}$,
completing the induction.

Since $\mathcal I$ is nonempty, every $\mathcal S_j$
is nonempty. Since $\mathcal S_{j+1}\subseteq\mathcal S_j$
and $A$ is finite, the sequence stabilizes after
finitely many iterations at $\mathcal S^\star$.
Since $\mathcal I\subseteq\mathcal S^\star$,
every action outside $\mathcal S^\star$ is chosen
only finitely often. Finiteness of $A$ therefore
gives $a_t\in\mathcal S^\star$ at every sufficiently
large date. The conclusion for
$\mathcal S^\star=\{a^\star\}$ follows immediately.

\par\noindent
\textnormal{(ii).}
Suppose $A=\{a,b\}$. We first note that, for every
$x\in(0,1)$,
\begin{equation}\label{eq:mixed_frequency_equivalence}
x\in\mathcal M_K
\quad\Longleftrightarrow\quad
\{a,b\}\subseteq\mathcal A_K(\alpha^x).
\end{equation}
Necessity follows from the definition of
$\mathcal A_K(\alpha^x)$.
For sufficiency, suppose
$\{a,b\}\subseteq\mathcal A_K(\alpha^x)$.
There exist $\eta^a,\eta^b\in\mathcal H_K(\alpha^x)$
such that
$a\in\operatorname{BR}_{\tau(\alpha^x)}(\eta^a)$
and
$b\in\operatorname{BR}_{\tau(\alpha^x)}(\eta^b)$.
Thus, the value difference
$V_{\tau(\alpha^x)}(b;\eta)
-V_{\tau(\alpha^x)}(a;\eta)$
is nonpositive at $\eta^a$ and nonnegative at
$\eta^b$. Since this difference is linear in
$\eta$ and $\mathcal H_K(\alpha^x)$ is convex,
there exists a convex combination
$\eta\in\mathcal H_K(\alpha^x)$ of $\eta^a$
and $\eta^b$ at which the difference vanishes.
Since there are only two actions,
$\{a,b\}\subseteq\operatorname{BR}_{\tau(\alpha^x)}(\eta)$,
so $x\in\mathcal M_K$.

Let $x_t:=\alpha_t(b)$.
Suppose, toward a contradiction, that
$\underline x:=\liminf_t x_t<
\limsup_t x_t=:\overline x$.
Fix any $x\in(\underline x,\overline x)$.
There are infinitely many upward crossings
$x_t<x\le x_{t+1}$ and infinitely many downward
crossings $x_{t+1}<x\le x_t$.
Since
$x_{t+1}-x_t=
[\mathbbm{1}\{a_{t+1}=b\}-x_t]/(t+1)$,
an upward crossing requires $a_{t+1}=b$ and a
downward crossing requires $a_{t+1}=a$.
Moreover, $|x_{t+1}-x_t|\le1/(t+1)$.
Hence, there are sequences $t_\ell,s_\ell\to\infty$
such that $x_{t_\ell}\to x$, $x_{s_\ell}\to x$,
$a_{t_\ell+1}=b$, and $a_{s_\ell+1}=a$
for every $\ell$.
Since $\alpha_{t_\ell}\to\alpha^x$ and
$\alpha_{s_\ell}\to\alpha^x$, Lemma
\ref{lem:convergence_best_reply} implies
$\{a,b\}\subseteq\mathcal A_K(\alpha^x)$.
By \eqref{eq:mixed_frequency_equivalence},
$x\in\mathcal M_K$.
Since $x$ was arbitrary,
$(\underline x,\overline x)\subseteq\mathcal M_K$,
contradicting the empty-interior condition.
Thus, $x_t$ converges, and hence $\alpha_t$ converges.

Let $x^\star$ denote the limit of $x_t$.
If $x^\star\in(0,1)$, both actions are chosen
infinitely often, so Lemma
\ref{lem:convergence_best_reply} implies
$\{a,b\}\subseteq\mathcal A_K(\alpha^{x^\star})$.
By \eqref{eq:mixed_frequency_equivalence},
$x^\star\in\mathcal M_K$, so there exists
$\eta^\star\in\mathcal H_K(\alpha^{x^\star})$
such that
$\operatorname{supp}(\alpha^{x^\star})
\subseteq
\operatorname{BR}_{\tau(\alpha^{x^\star})}(\eta^\star)$.
If $x^\star\in\{0,1\}$, the action receiving unit
limiting frequency is chosen infinitely often.
By Lemma \ref{lem:convergence_best_reply}, it
belongs to $\mathcal A_K(\alpha^{x^\star})$.
Thus, in this case too, there exists an
$\eta^\star\in\mathcal H_K(\alpha^{x^\star})$
satisfying the preceding inclusion.
By Proposition \ref{prop:frontier}\textnormal{(i)},
$\mathcal R_K(\alpha^{x^\star})$ is nonempty.
Choosing any $\rho^\star\in\mathcal R_K(\alpha^{x^\star})$,
the tuple
$(\alpha^{x^\star},\eta^\star,
\tau(\alpha^{x^\star}),\rho^\star)$
satisfies Definition \ref{def:ccre}.

\par\noindent
\textnormal{(iii).}
Since $\mathcal M_K=\varnothing$, part
\textnormal{(ii)} implies that $\alpha_t$ converges
to the action distribution of a pure equilibrium,
say $\delta_{a^\star}$.
By Definition \ref{def:ccre},
$a^\star\in\mathcal A_K(\delta_{a^\star})$.
By hypothesis,
$\mathcal A_K(\delta_{a^\star})=\{a^\star\}$.
Lemma \ref{lem:convergence_best_reply} therefore
implies $a_t=a^\star$ at every sufficiently
large date.
\end{proof}

\begin{proof}[Proof of Corollary
\ref{cor:general_risky_safe_convergence}]
Write $\alpha^x:=(1-x)\delta_{\mathfrak s}
+x\delta_{\mathfrak r}$ for $x\in[0,1]$.
For every $\nu\in\mathcal V_K$, the function
$f_\nu(x):=\E_\nu[D(\cdot;\alpha^x)]$ is affine
in $x$. By condition \textnormal{(i)},
$f_\nu$ and $f_{\nu'}$ are distinct affine
functions whenever $\nu\ne\nu'$.
Thus, any two of them coincide at at most one
$x\in[0,1]$. Since $\mathcal V_K$ is finite,
there are only finitely many such intersection
points. Moreover, for each $\m q\in Q$, the function
$D(\m q;\alpha^x)+\psi(\alpha^x)$ is affine
in $x$, with slope
$D_{\mathfrak r}(\m q)+\psi(\delta_{\mathfrak r})
-D_{\mathfrak s}(\m q)-\psi(\delta_{\mathfrak s})$,
which is nonzero by condition \textnormal{(ii)}.
Since
$\tau(\alpha^x)=c^{-1}\min_{\m q\in Q}
[D(\m q;\alpha^x)+\psi(\alpha^x)]$,
$\tau(\alpha^x)$ is piecewise affine, with
finitely many pieces and nonzero slope on each
open affine piece.

Let $J$ consist of $0$, $1$, all intersection
points of the functions $(f_\nu)_{\nu\in\mathcal V_K}$
in $[0,1]$, and all breakpoints of $\tau(\alpha^x)$.
The set $J$ is finite. On each open interval
between consecutive points of $J$, the ordering
of the functions $(f_\nu)_{\nu\in\mathcal V_K}$
is constant and there is a unique minimizing
vertex, say $\nu^\star$.
Since every element of $\mathcal E_K$ is a
convex combination of its vertices and the
objective defining $\mathcal H_K(\alpha^x)$
is linear, uniqueness of the minimizing vertex
implies $\mathcal H_K(\alpha^x)=\{\nu^\star\}$
throughout the interval.

By the risky-safe payoff structure and Assumption
\ref{ass:full_support}, the function
$V_\lambda(\mathfrak r;\nu^\star)-\bar u$
is strictly decreasing in $\lambda$.
Since $\tau(\alpha^x)$ is affine with nonzero
slope on the interval, the function
$V_{\tau(\alpha^x)}(\mathfrak r;\nu^\star)-\bar u$
is strictly monotone in $x$ there.
Consequently, it vanishes at most once on the
interval. Since the recognition distribution
attaining the frontier is unique there, any
$x\in\mathcal M_K$ in the interval must satisfy
this equality.

Thus, $\mathcal M_K$ contains at most one point
in each of the finitely many open intervals
between consecutive points of $J$, together with
at most finitely many additional points in $J$.
Hence, $\mathcal M_K$ is finite.
The convergence conclusion follows from
Proposition \ref{prop:action_convergence}\textnormal{(ii)}.
\end{proof}
\end{document}